\documentclass{jfm}

\usepackage{subfigure}
\usepackage{amsmath}
\usepackage{graphicx}
\usepackage{psfrag}
\usepackage{natbib}
\usepackage{color}
\usepackage{soul}

\def\drawline#1#2{\raise 2.5pt\vbox{\hrule width #1pt height #2pt}}

\def\trian{\raise 1.25pt\hbox{$\scriptscriptstyle\triangle$}\nobreak\ }

\def\dl{$\Delta$}

\shortauthor{A. Patel, B. J. Boersma and R. Pecnik}

\title[The influence of near-wall density and viscosity gradients on turbulence]
{The influence of near-wall density and viscosity gradients on turbulence in channel flows}

\author[A. Patel, B. J. Boersma and R. Pecnik]%
{Ashish Patel
  \thanks{Email address for correspondence: a.patel@tudelft.nl},\ns
Bendiks J. Boersma
and Rene Pecnik\thanks{Email address for correspondence: r.pecnik@tudelft.nl},\ns}

\affiliation{Process and Energy Department, Delft University of Technology, \\Leeghwaterstraat 39, 2628 CB Delft, The Netherlands \\[\affilskip]}

\begin{document}

\maketitle

\begin{abstract}
The influence of near-wall density and viscosity gradients on near-wall turbulence in a channel are studied by means of Direct Numerical Simulation (DNS) of the low-Mach number approximation of the Navier--Stokes equations. Different constitutive relations for density $\rho$ and viscosity $\mu$ as a function of temperature are used in order to mimic a wide range of fluid behaviours and to develop a generalised framework for studying turbulence modulations in variable property flows. Instead of scaling the velocity solely based on local density, as done for the van Driest transformation, we derive an extension of the scaling that is based on gradients of the semi-local Reynolds number, defined as $Re_\tau^\star \equiv Re_\tau\sqrt{(\overline{\rho}/{\overline{\rho}_w})}/(\overline{\mu}/{\overline{\mu}_w})$ (bar and subscript $w$ denote Reynolds averaging and wall value, respectively, while $Re_\tau$ is the friction Reynolds number based on wall values). This extension of the van Driest transformation is able to collapse velocity profiles for flows with near-wall property gradients as a function of the semi-local wall coordinate. However, flow quantities like mixing length, turbulence anisotropy and turbulent vorticity fluctuations do not show a universal scaling very close to the wall. This is attributed to turbulence modulations, which play a crucial role in the evolution of turbulent structures and turbulence energy transfer. We therefore investigate the characteristics of streamwise velocity streaks and quasi-streamwise vortices and found that, similar to turbulence statistics, the turbulent structures are also strongly governed by $Re_\tau^\star$ profiles and that their dependence on individual density and viscosity profiles is minor. Flows with near-wall gradients in $Re_\tau^\star$ ($d {Re_\tau^\star}/dy \neq 0$) showed significant changes in inclination and tilting angles of quasi-streamwise vortices. These structural changes are responsible for the observed modulation of the Reynolds stress generation mechanism and the inter-component energy transfer in flows with strong near-wall $Re_\tau^\star$ gradients. 
\end{abstract} 


\section{Introduction}
Turbulent fluid flows with large near-wall gradients of thermo-physical properties can occur in a wide range of engineering applications, where strong wall heat transfer plays an important role. The mean velocity and near-wall turbulent structures are greatly affected by near-wall gradients of, e.g.,  density and viscosity. However, the majority of the studies on turbulent flows structures has been performed for constant property flows. 

The existence of near-wall turbulent structures has been known for several decades and their structural features have been investigated in great detail. Although geometrically thin, the near-wall region that is formed by the viscous and the buffer layers is responsible for a large fraction of the velocity drop \citep{jimenez2007we}. This shear-dominated near-wall region accounts for a significant amount of turbulence production across the boundary layer and thus plays a crucial role in skin friction and heat transfer. The dominant structures of the near-wall region are streamwise velocity streaks and quasi-streamwise vortices. Streaks \citep{kline1967structure} are spanwise modulation of streamwise velocity and consist of low- and high-speed streaks with a characteristic spacing of about 100 wall units in spanwise direction \citep{smith1983characteristics}. The quasi-streamwise vortices, primarily found in the buffer layer, are slightly inclined away from the wall and tilted in spanwise directions \citep{jeong1997coherent}. Additionally, it is known that the dynamics of near-wall turbulence can be maintained autonomously by a self-sustaining process, where the streaks and the vortices continue regenerating each other independently of the outer flow \citep{jimenez1999autonomous}. {According to \citet{hamilton1995regeneration}, the sustainment of near-wall turbulence involves a closed-loop mechanism whereby the generation of streaks are induced by quasi-streamwise vortices that are in turn created by streak instability. An alternate description of the self-sustaining mechanism has been provided by \citet{chernyshenko2005mechanism} in which the physical mechanism of streak formation is related to combined action of wall-normal motions, mean shear and viscous diffusion. While there are differences in the interpretation of the self-sustaining process, a close relation between low-speed streaks and quasi-streamwise vortices is well-established \citep{robinson1991coherent,jeong1997coherent,jimenez1999autonomous,kim2000linear,schoppa2002coherent} and also supported by self-sustaining mechanisms described by \citet{hamilton1995regeneration} and \citet{chernyshenko2005mechanism}.}

The near-wall coherent structures for variable property wall turbulence has been far less explored in the past. Recently \citet{lee2013effect} and \citet{zonta2012modulation} have investigated incompressible flows with near-wall viscosity gradients and they found that the turbulent structures are affected by these gradients. Further knowledge on variable property effects on turbulent structures has been provided by experimental and numerical studies of supersonic flows with adiabatic walls  \citep[e.g.,][]{spina1987organized,ringuette2008coherent,elsinga2010three,pirozzoli2008characterization,lagha2011near}, and cooled or heated walls  \citep[e.g.,][]{coleman1995numerical,lechner,foysi2004,morinishi2004,duan2010,lagha2011,shadloo2015statistical,modesti2016reynolds}. 
The experimental studies on supersonic boundary layer structures are limited to investigate large-scale motions in the outer region of the flow, while numerical studies additionally allow a detailed investigation of near-wall structures. For supersonic boundary layers with adiabatic walls, the dynamics of near-wall turbulent structures is found to be similar to constant property cases and its characteristics scale well with the classical wall based scaling \citep{ringuette2008coherent,pirozzoli2008characterization,lagha2011}. For cooled walls the near-wall streaks extend in streamwise direction \citep{coleman1995numerical,duan2010,lagha2011}, while they shorten for heated walls \citep{lagha2011}. These streak modifications were quantified in terms of wall based viscous units. The semi-local scaling \citep{huang1995compressible}, which uses the wall shear stress $\tau_w$ and local properties to define near-wall scales, has shown to be effective to account for changes in streak length in the buffer layer \citep{patel2015semi,morinishi2004}. Even though, the semi-local scaling is able to account for some of the differences seen between non-adiabatic variable- and constant property flows, it has not been able to provide a universal scaling law. For instance, in a supersonic channel flow with cold isothermal walls, an increase in turbulence-to-mean time scale ratio was reported by \citet{coleman1995numerical} and a reduced pressure-strain rate resulting in an increased turbulence anisotropy was noted by \citet{foysi2004}. 
In a more recent study \citet{modesti2016reynolds} performed DNS at different Reynolds and Mach numbers for compressible isothermal channel flows and they noted an increase of streamwise turbulence intensity with Mach number as a dominant variable property effect.  

The van Driest transformed mean velocity $\overline{u}^{\mathrm{vD}}=\int^{\overline{u}/{u_\tau}}_0\sqrt{{\overline{\rho}}/{\overline{\rho}_w}} d \left({\overline{u}}/{u_\tau}\right)$, when plotted as a function of $y^+=y Re_\tau/h$, has been successful in collapsing velocity profiles of supersonic flows over adiabatic walls with velocity profiles from constant property flows  \citep{guarini2000direct,maeder2001direct,pirozzoli2011turbulence,
lagha2011,duan2011direct}. Above, $\overline{u}$ and $u_\tau$ are the streamwise and friction velocity, respectively, while $y$ is the wall-normal coordinate and $h$ is the half-channel height or boundary layer thickness. The success of the density weighted scaling can be attributed to Morkovin's hypothesis, which assumes that the characteristic time and length scales governing turbulent transport are not affected by changes in properties \citep{smits2006turbulent}. $\overline{u}^{\mathrm{vD}}$ when plotted as a function of $y^+$ for flows with strong near-wall gradients in density and viscosity, shows deviations if compared with profiles from constant property flows. For cooled walls, shrinking of the viscous sub-layer \citep{duan2010} and an outward shift of the log-law region with an increase in additive constant \citep{maeder2000numerical} has been noted. {Recently, \citet{trettel2016} derived a transformation that considers wall-normal density and viscosity gradients. They successfully applied their velocity transformation for supersonic channel flows with isothermally cooled walls. \citet{modesti2016reynolds} conducted a detailed comparison of different compressibility transformations and they  concluded that the transformation introduced by \citet{trettel2016} performs best for collapsing velocity profiles for several different compressible channel flows. However, an explanation that correlates the changes in mean velocity and the observed turbulence modulations is still missing. }

In our recent work \citep{patel2015semi} we studied in detail the effect of variable density $\rho$ and viscosity $\mu$ on near-wall turbulence and on scaling of turbulence statistics. 
Similar to constant property turbulent channel flows, where turbulence statistics and the mean streamwise velocity $\overline{u}$ can be expressed as a function of wall-normal distance and friction Reynolds number $Re_\tau$, we showed that turbulence statistics and the van Driest transformed mean streamwise velocity $\overline{u}^{\mathrm{vD}}$ for variable property turbulent flows can solely be expressed as a function of wall-normal distance and semi-local Reynolds number $Re_\tau^\star \equiv Re_\tau\sqrt{(\overline{\rho}/{\overline{\rho}_w})}/(\overline{\mu}/{\overline{\mu}_w})$. Unlike for constant property flows, where turbulence statistics weakly depend on $Re_\tau$,  in variable property flows the turbulence statistics show a strong dependence on $Re_\tau^\star$. For a case with $d {Re_\tau^\star}/dy <0$ in the near-wall region (decreasing Reynolds number away from the wall), the streamwise normal Reynolds stress anisotropy increases, which can be associated with strengthening of large scale low-speed streaks in the buffer layer. The reverse was observed for $Re_\tau^\star$ profiles with $d {Re_\tau^\star}/dy >0$. However, there are also few similarities that can be observed for cases with different $Re_\tau^\star$ profiles. For example the peak locations of the streamwise Reynolds stress are at a similar semi-local wall distance ($y^\star=y Re_\tau^\star/h$), but their peak values differ. Another similarity is that the streamwise and the spanwise non-dimensional length (based on semi-local scales) of turbulent structures is similar at $y^\star \approx 15$. 

In the present work, our focus is on identifying the effects of near-wall property gradients on mean velocity scaling, near-wall turbulence statistics and turbulent structures. We use the same DNS database that was also used in our previous study \citep{patel2015semi} and supplement it with few additional simulations. First, we derive a velocity scaling that extends the van Driest transformation to account for gradients in $Re_\tau^\star$ and that is able to provide a collapse of velocity profiles for turbulent flows with strong density and viscosity variations. We then characterise the modification of turbulent structures and show that similar to turbulence statistics, near-wall turbulent structures are also strongly governed by $Re_\tau^\star$ profiles. We investigate the physical mechanism that results in structural changes and consequently in modulated turbulence statistics, which explains the long-standing open question on how  turbulence anisotropy is affected by wall heating or cooling~\citep{lechner,foysi2004,modesti2016reynolds}. Even though the DNS have been performed by solving the low Mach-number approximation of the Navier--Stokes equations, the conclusions are also applicable to high-speed flows. This is because in the near-wall region of a supersonic flow most of the near-wall density and temperature fluctuations are the result of solenoidal 'passive mixing' by turbulence and density fluctuations show little correlation with pressure fluctuations \citep{coleman1995numerical,lechner}.

\section{Simulation Details}
Direct numerical simulations of fully developed turbulent channel flows, driven by a constant streamwise pressure gradient, are performed using the low Mach number approximation of the Navier--Stokes equations. In the low Mach number limit, the density and transport properties can be evaluated as a function of temperature only, independent of pressure fluctuations \citep{majda1985derivation,nemati}. Different constitutive relations for density and viscosity are used. In order to achieve variations in temperature $T$, and consequently in density $\rho$ and viscosity $\mu$, the flow is uniformly heated with a volumetric heat source, while the temperature at both channel walls is kept constant. {This allows a wall heat flux and ensures that the flow is in thermal equilibrium.} Other transport properties like thermal conductivity $\kappa$ and specific heat $c_p$ are constant in all simulations. The Prandtl number based on wall quantities is taken to be unity.

The DNS code discretises the spatial derivatives in wall-normal direction using a sixth order  staggered compact  finite  difference  scheme \citep{lele1992compact,boersma20116th} and the derivatives in spanwise and streamwise directions are computed using a Fourier expansion with periodic boundary conditions. The time integration is done with a second order Adams-Bashforth method and a pressure correction scheme, based on the projection method \citep{mcmurtry1986direct}, is used to ensure mass conservation. Additional details of the governing equations can be found in \citet{patel2015semi}. 

\begin{table}
  \begin{center}
\def~{\hphantom{0}}
  \begin{tabular}{lllllll}
Case         &$\rho/\rho_w$&$\mu/\mu_w$       &$Re_\tau$ &${Re_\tau^\star}_c$     &$N_x \times N_y \times N_z$ &$L_x \times L_y \times L_z$\\ [3pt]
CP395        		&   1  & 1          		&395 &395    &$240 \times 264 \times 240$ &$2\pi h \times 2h \times \pi h$   \\
CRe$^\star_\tau$ 		&$(T/T_w)^{-1}$ &$(T/T_w)^{-0.5}$		&395 &395    &$240 \times 264 \times 240$ &$2\pi h \times 2h \times \pi h$   \\
SRe$^\star_{\tau GL}$   &   1  &$(T/T_w)^{1.2}$    	&395 &152    &$360 \times 264 \times 360$ &$5\pi h \times 2h \times 2\pi h$   \\
GL          		    &$(T/T_w)^{-1}$ & $(T/T_w)^{0.7}$  		&395 &142    &$360 \times 264 \times 360$ &$5\pi h \times 2h \times 2\pi h$   \\
LL           		&   1  &$(T/T_w)^{-1}$	    		&150 &543    &$360 \times 264 \times 360$ &$3\pi h \times 2h \times 1.5\pi h$   \\
SRe$^\star_{\tau LL}$   &$(T/T_w)^{0.6}$  &$(T/T_w)^{-0.75}$  &150 &535  &$360 \times 264 \times 360$ &$3\pi h \times 2h \times 1.5\pi h$     \\
CP150	     		&1	   &1	      		&150 &150    &$192 \times 168 \times 168$ &$5\pi h \times 2h \times 2\pi h$   \\
CP550 				&   1  &1			 	&550 &550    &$312 \times 312 \times 312$ &$2\pi h \times 2h \times \pi h$   \\
\end{tabular}
\caption{Simulation parameters for all cases. {CP395 - Constant Property case with $Re_\tau =395$; CRe$^\star_\tau$ - variable property case with Constant $Re_\tau^\star$ ($=395$) across the channel; GL - case with Gas-Like property variations; SRe$^\star_{\tau GL}$- variable property case with $Re_\tau^\star$ Similar to case GL; LL - case with Liquid-Like property variations; SRe$^\star_{\tau LL}$- variable property case with $Re_\tau^\star$ Similar to case LL; CP150 - Constant Property case with $Re_\tau =150$; CP550 - Constant Property case with $Re_\tau =550$.}}
\label{tab:flow1}
  \end{center}
\end{table}
Eight cases have been simulated, which is summarised in table \ref{tab:flow1}. The simulations consist of three constant property (CP395, CP150, CP550) and five variable property cases (CRe$^\star_\tau$, SRe$^\star_{\tau GL}$, GL, LL, SRe$^\star_{\tau LL}$). {The acronym CP refers to a Constant Property case; GL and LL refer to Gas-Like and Liquid-Like property variations, respectively; CRe$^\star_\tau$ refers to a variable property case with Constant $Re_\tau^\star$ ($=395$) across the whole channel; SRe$^\star_{\tau GL}$ refers to a variable property case that has a Similar $Re_\tau^\star$ distribution as case GL; and SRe$^\star_{\tau LL}$ refers to a variable property case with a $Re_\tau^\star$ distribution Similar to case LL.}  The second and third columns show the functional relations for $\rho/\rho_w$ and $\mu/\mu_w$ as a function of $T/T_w$. 
The next two columns report the wall based friction Reynolds number $Re_\tau$, and the semi-local Reynolds number at the channel centre ${Re_\tau^\star}_c = Re_\tau\sqrt{({\overline{\rho}_c}/{\overline{\rho}_w})}/({\overline{\mu}_c}/{\overline{\mu}_w})$ (subscript $c$ denotes the value at channel centre). Note, at the wall ${Re_\tau^\star}_w=Re_\tau$. The last two columns show the number of mesh points $N$ and the length of the domain $L$ along streamwise $x$, wall-normal $y$ and spanwise $z$ directions. The velocity components along $x$, $y$ and $z$ directions are denoted as $u$, $v$ and $w$, respectively. 

\begin{figure}
  \centering
  \subfigure{\label{fig:rho}
    \psfrag{N}[c][][0.9]{(a)}
  	\psfrag{Y}[c][][0.9]{$\overline{\rho}/\overline{\rho}_w$}
  	\psfrag{X}[c][][0.9]{$y/h$}
  	\includegraphics[width=0.5\textwidth]{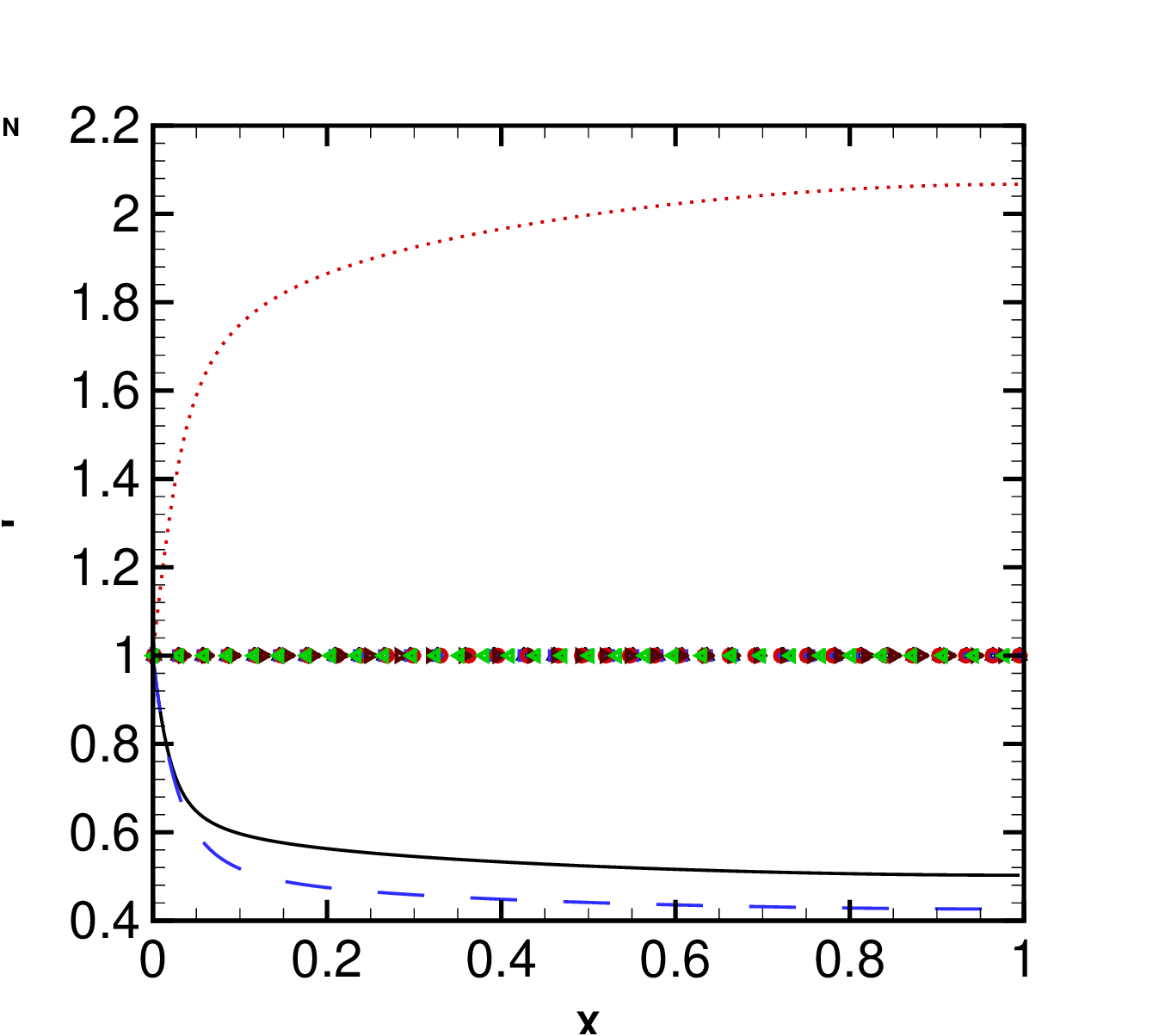}}~~~
  \subfigure{\label{fig:mu}
    \psfrag{N}[c][][0.9]{(b)}
  	\psfrag{Y}[c][][0.9]{$\overline{\mu}/\overline{\mu}_w$}
  	\psfrag{X}[c][][0.9]{$y/h$}		
  	\includegraphics[width=0.5\textwidth]{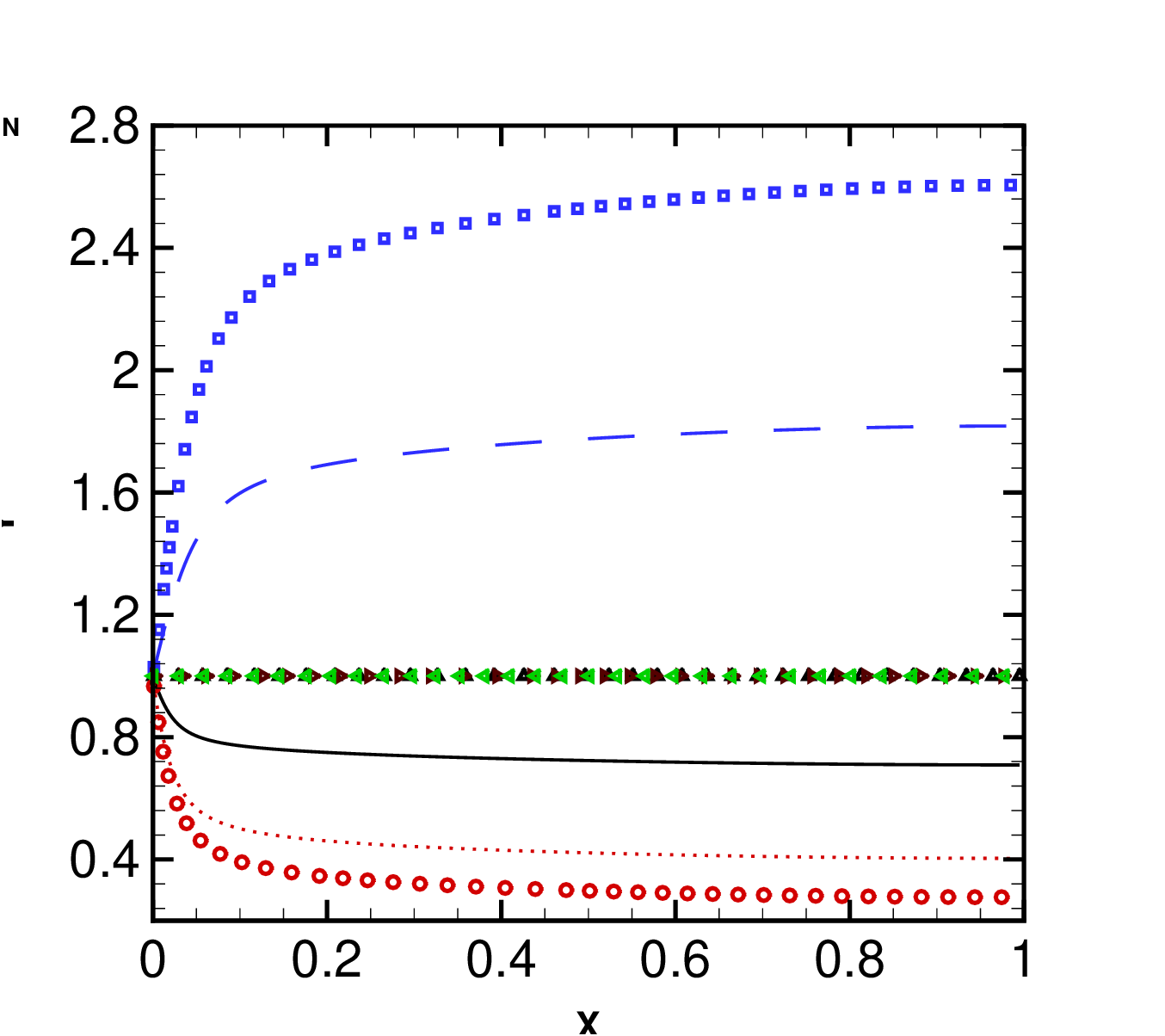}}
  \subfigure{\label{fig:ret}
    \psfrag{N}[c][][0.9]{(c)}
  	\psfrag{Y}[c][][0.9]{$Re_\tau^\star$}
  	\psfrag{X}[c][][0.9]{$y/h$}
  	\psfrag{A}[l][][0.65]{CP395}
  	\psfrag{B}[l][][0.65]{CRe$^\star_\tau$}
  	\psfrag{C}[l][][0.65]{SRe$^\star_{\tau GL}$}
  	\psfrag{D}[l][][0.65]{GL}
  	\psfrag{E}[l][][0.65]{LL}
  	\psfrag{F}[l][][0.65]{SRe$^\star_{\tau LL}$}
  	\psfrag{G}[l][][0.65]{CP150}
  	\psfrag{H}[l][][0.65]{CP550}
  	\includegraphics[width=0.5\textwidth]{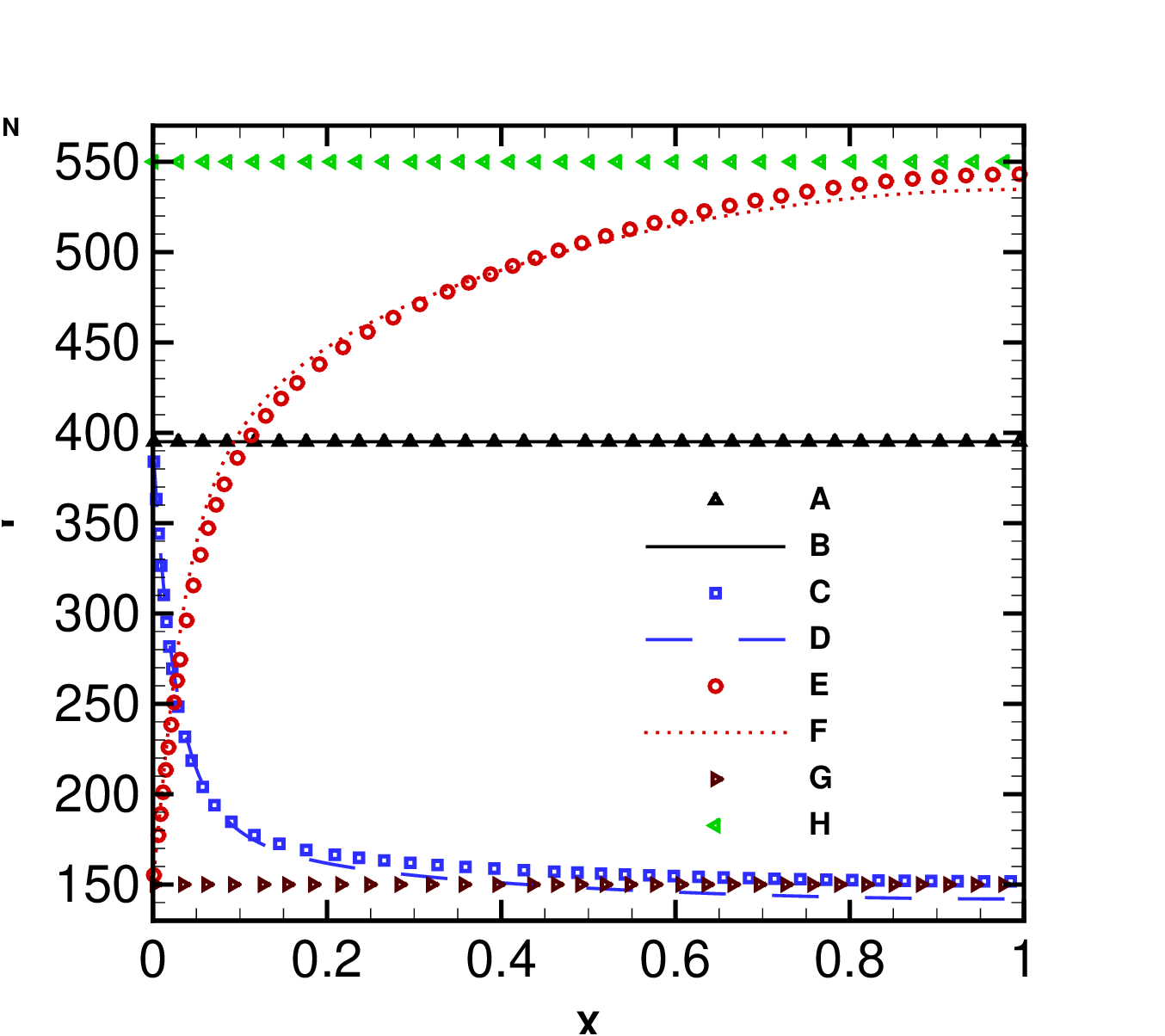}}
   \caption{{(Colour online)} (a) Density $\overline{\rho}/\overline{\rho}_w$, (b) viscosity $\overline{\mu}/\overline{\mu}_w$, and (c) semi-local Reynolds number $Re_\tau^\star$.}   
  \label{fig:prop_ret}
  \vspace*{-1em}
\end{figure} 
Figure~\ref{fig:prop_ret} shows the distributions of density, viscosity and $Re_\tau^\star$ for all cases. Cases with variable density are shown as lines and cases with constant density are shown as symbols. 
As can be seen, we use different combinations of $\rho$ and $\mu$ to obtain quasi-similar $Re_\tau^\star$ profiles. Note, the strongest gradients in $Re_\tau^\star$ occur at the wall. 
The cases forming the first pair -- CP395 and CRe$^\star_\tau$ {(black symbols and line, respectively {- colour online})} -- have constant $Re_\tau^\star$ in wall-normal direction {($dRe_\tau^\star/dy=0$)}, even though for case CRe$^\star_\tau$ the density and viscosity are varying and decrease away from the wall. For the second pair -- SRe$^\star_{\tau GL}$ and GL {(blue symbols and line, respectively {- colour online})} -- the profile for $Re_\tau^\star$ decreases from 395 at the wall to approximately 150 at the channel centre. Later, we will denote these cases as $dRe_\tau^\star/dy<0$. For the pair -- $LL$ and SRe$^\star_{\tau LL}$ (red symbols and line, respectively {- colour online}) -- $Re_\tau^\star$ increases from 150 at the wall to approximately 540 at the channel centre. These cases will be denoted as $dRe_\tau^\star/dy>0$. The simulations CP150 (brown symbol {- colour online}) and CP550 (green symbol {- colour online}) are cases that bound the $Re_\tau^\star$ profiles of the variable property cases in order to investigate and distinguish any Reynolds number effects with respect to effects caused by property gradients. Considerable variations in properties have been obtained for all variable property cases. However, the relative property fluctuations ${{\rho^\prime}_{rms}}/\overline{\rho}$ and ${{\mu^\prime}_{rms}}/\overline{\mu}$ (prime denotes Reynolds average fluctuations, and the subscript $rms$ indicates the root mean square value) are still less than 0.15 for all variable property cases.

\begin{figure}
  \centering
  \subfigure{\label{fig:kolos}
    \psfrag{N}[c][][0.9]{(a)}
  	\psfrag{Y}[c][][0.9]{$\eta/h$}
  	\psfrag{X}[c][][0.9]{$y/h$}
  	\psfrag{A}[l][][0.65]{CP395}
  	\psfrag{B}[l][][0.65]{CRe$^\star_\tau$}
  	\psfrag{C}[l][][0.65]{SRe$^\star_{\tau GL}$}
  	\psfrag{D}[l][][0.65]{GL}
  	\psfrag{E}[l][][0.65]{LL}
  	\psfrag{F}[l][][0.65]{SRe$^\star_{\tau LL}$}
  	\psfrag{G}[l][][0.65]{CP150}
  	\psfrag{H}[l][][0.65]{CP550}
  	\includegraphics[width=0.5\textwidth]{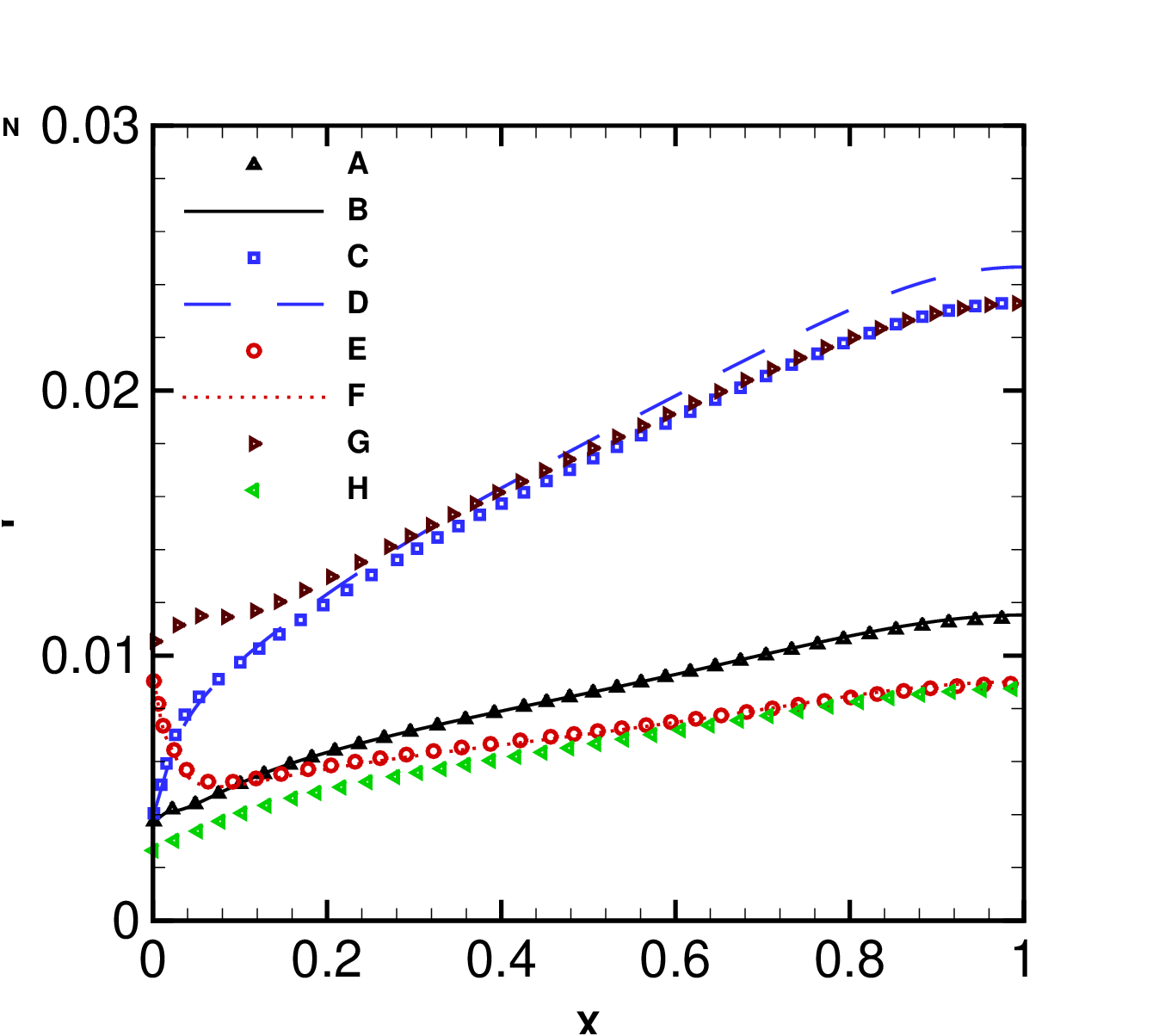}}~~~
  \subfigure{\label{fig:kolvs}
    \psfrag{N}[c][][0.9]{(b)}
  	\psfrag{Y}[c][][0.9]{$\eta Re_\tau/h$}
  	\psfrag{X}[c][][0.9]{$y^+$}		
  	\includegraphics[width=0.5\textwidth]{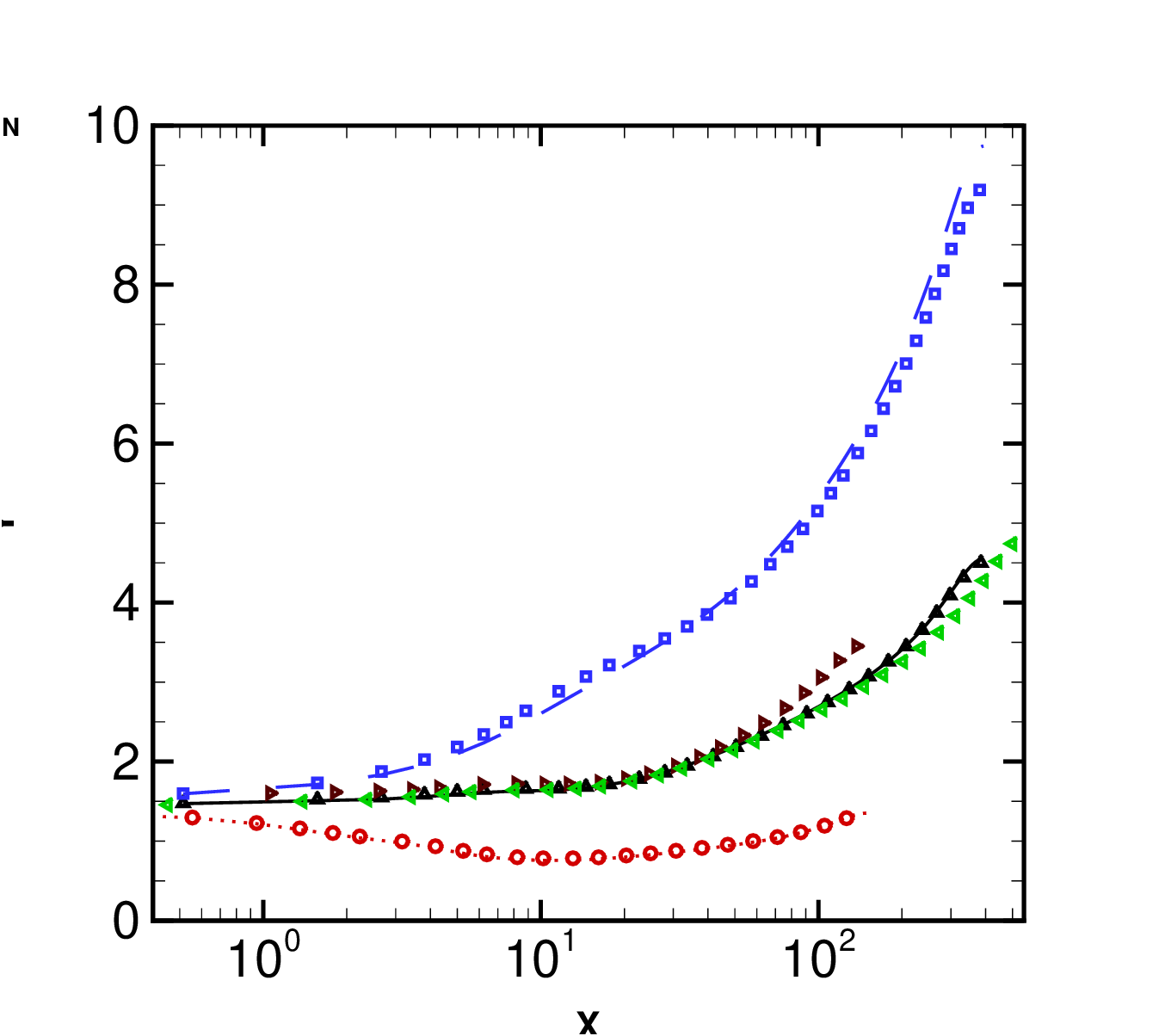}}	
  \subfigure{\label{fig:kolsl}
    \psfrag{N}[c][][0.9]{(c)}
  	\psfrag{Y}[c][][0.9]{$\eta Re_\tau^\star/h$}
  	\psfrag{X}[c][][0.9]{$y^\star$}		
  	\includegraphics[width=0.5\textwidth]{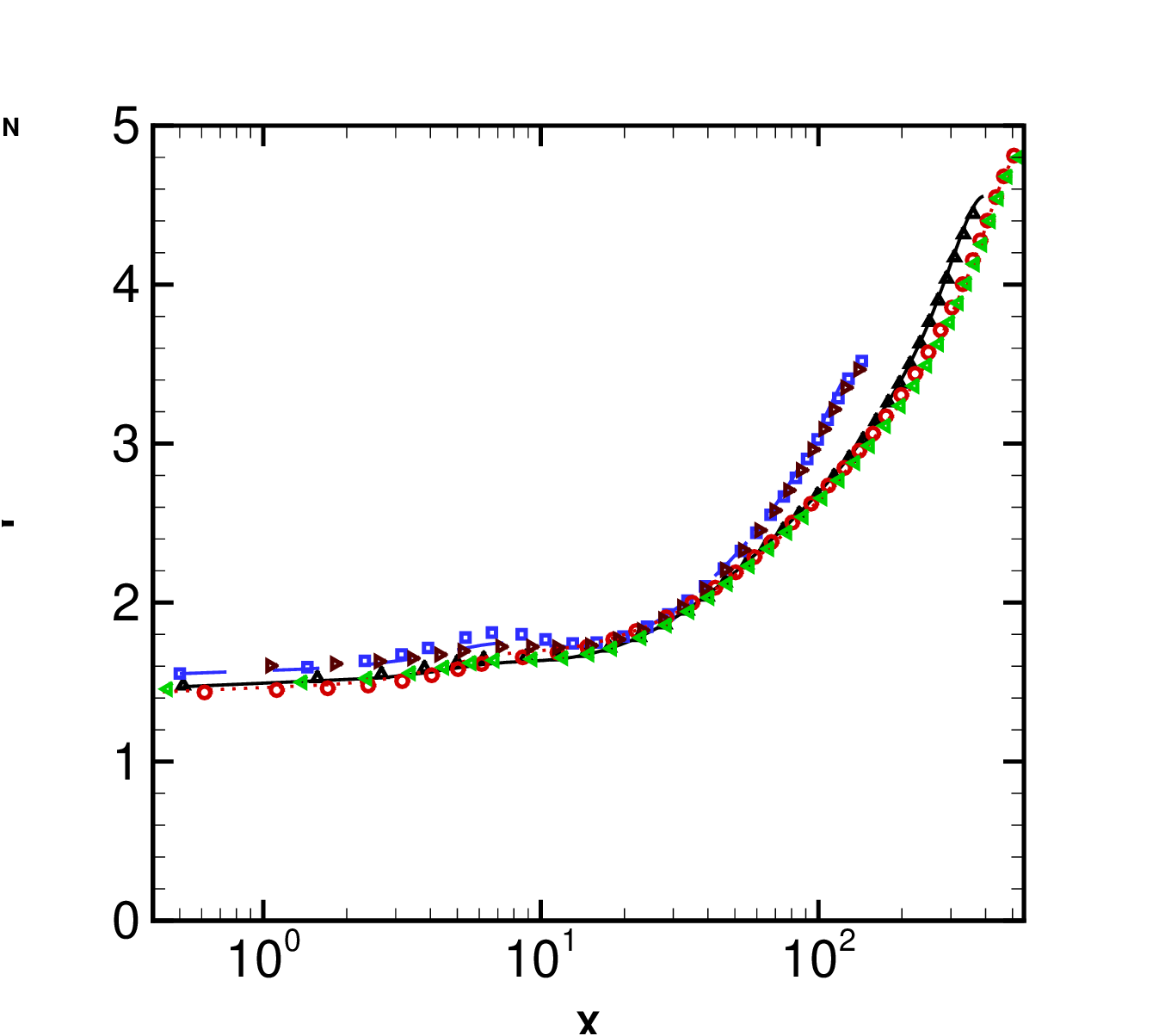}}		
   \caption{{(Colour online)} {Kolmogorov length scale normalized by and plotted using, (a) outer scales, (b) wall-based viscous length scales and (c) semi-local viscous length scales.}}   
  \label{fig:kol}
  \vspace*{-1em}
\end{figure} 

The adequacy of the mesh resolution to resolve the smallest scales is evaluated by means of wall-normal distributions of Kolmogorov length scales, $\eta=((\overline{\mu}/\overline{\rho})^3\overline{\rho}/\epsilon)^{0.25}$ ($\epsilon$ is the turbulent kinetic energy dissipation per unit volume).
Figure~\ref{fig:kol} shows $\eta$, normalized by, and plotted using outer scales (figure~\ref{fig:kolos}), wall-based viscous length scales (figure~\ref{fig:kolvs}) and semi-local viscous length scales (figure~\ref{fig:kolsl}) for all cases. Cases with similar $Re_\tau^\star$ profiles show similar distributions of Kolmogorov scales, indicating that similar to turbulence statistics \citep{patel2015semi}, the smallest scales are also strongly governed by $Re_\tau^\star$. The Kolmogorov scale normalized using the half-channel height  increases for cases with decreasing $Re_\tau^\star$, while it decreases for cases with increasing $Re_\tau^\star$. The cases with $dRe_\tau^\star/dy<0$, for which $Re_\tau^\star$ varies from 395 at the wall to approximately 140 at the channel centre, have a Kolmogorov length scale that transitions from values of case CP395 at the wall to CP150 at the channel centre. A similar observation can be made for cases with $dRe_\tau^\star/dy>0$, for which  $Re_\tau^\star$ varies from 150 at the wall to approximately 540 at centre. The Kolmogorov scale, when normalized by and plotted using classical wall-based viscous units, shows strong deviations for cases with $dRe_\tau^\star/dy\neq0$. On the other hand, using semi-local length scales provides a good collapse of $\eta$ for all cases in the near-wall region. Hence, the semi-local scaling also provides a good measure to assess the mesh spacing requirements for simulations with strong fluid property variations. 
Table \ref{tab:resol} lists the maximum grid spacing in terms of $\eta$ for all cases, and the values are within the resolution requirements of \dl$x<12\eta$, \dl$y<2\eta$, \dl$z<6\eta$, as also reported in other DNS studies \citep{zonta2012modulation,lee2013effect}. 

\begin{table}
  \begin{center}
\def~{\hphantom{0}}
  \begin{tabular}{lcccc}
Case         &$($\dl$x/\eta)_{max}$ &$($\dl$(y)_{min}/\eta)_{max}$ &$($\dl$(y)_{max}/\eta)_{max}$  &$($\dl$z/\eta)_{max}$\\ [3pt]
CP395        &$6.98$ &$0.69$  & $0.89$    &  $3.49$\\
CRe$^\star_\tau$ &$7$    &$0.7$   & $0.88$    &  $3.5$\\
SRe$^\star_{\tau GL}$ &$10.78$ &$0.64$  & $0.43$    &  $4.31$\\
GL          	      &$10.8$  &$0.64$  & $0.41$    &  $4.3$\\
LL                &$5.02$  &$0.27$  & $1.15$    &  $2.51$\\
SRe$^\star_{\tau LL}$ &$5.18$  &$0.27$  & $1.14$    &  $2.59$\\
CP150	     	  &$7.76$  &$0.44$  & $0.68$    &  $3.55$\\
CP550 			  &$7.6$   &$0.62$  & $1$       &  $3.8$\\
\end{tabular}
\caption{{Maximum spatial resolution normalized using the Kolmogorov scale $\eta$.}}
\label{tab:resol}
  \end{center}
\end{table}

\begin{figure}
  \centering
    \subfigure{\label{fig:pmsx}
    \psfrag{O}[c][][0.9]{(a)}
  	\psfrag{P}[c][90][0.9]{$k_xE_{\rho{u}^{\prime\prime}{u}^{\prime\prime}}/\overline{\rho{u}^{\prime\prime}{u}^{\prime\prime}}$}
  	\psfrag{X}[c][][0.9]{$\lambda_x/h$}
  	\psfrag{Q}[c][][0.9]{$y^\star=0.5$}
  	\psfrag{R}[c][][0.9]{$y^\star=20$}	
  	\psfrag{A}[l][][0.65]{CP395}
  	\psfrag{B}[l][][0.65]{CRe$^\star_\tau$}
  	\psfrag{C}[l][][0.65]{SRe$^\star_{\tau GL}$}
  	\psfrag{D}[l][][0.65]{GL}
  	\psfrag{E}[l][][0.65]{LL}
  	\psfrag{F}[l][][0.65]{SRe$^\star_{\tau LL}$}
  	\psfrag{G}[l][][0.65]{CP150}
  	\psfrag{H}[l][][0.65]{CP550}
	\psfrag{S}[c][][0.9]{$y/h=0.3$}
  	\psfrag{T}[c][][0.9]{$y/h=0.5$}
  	\includegraphics[width=0.5\textwidth]{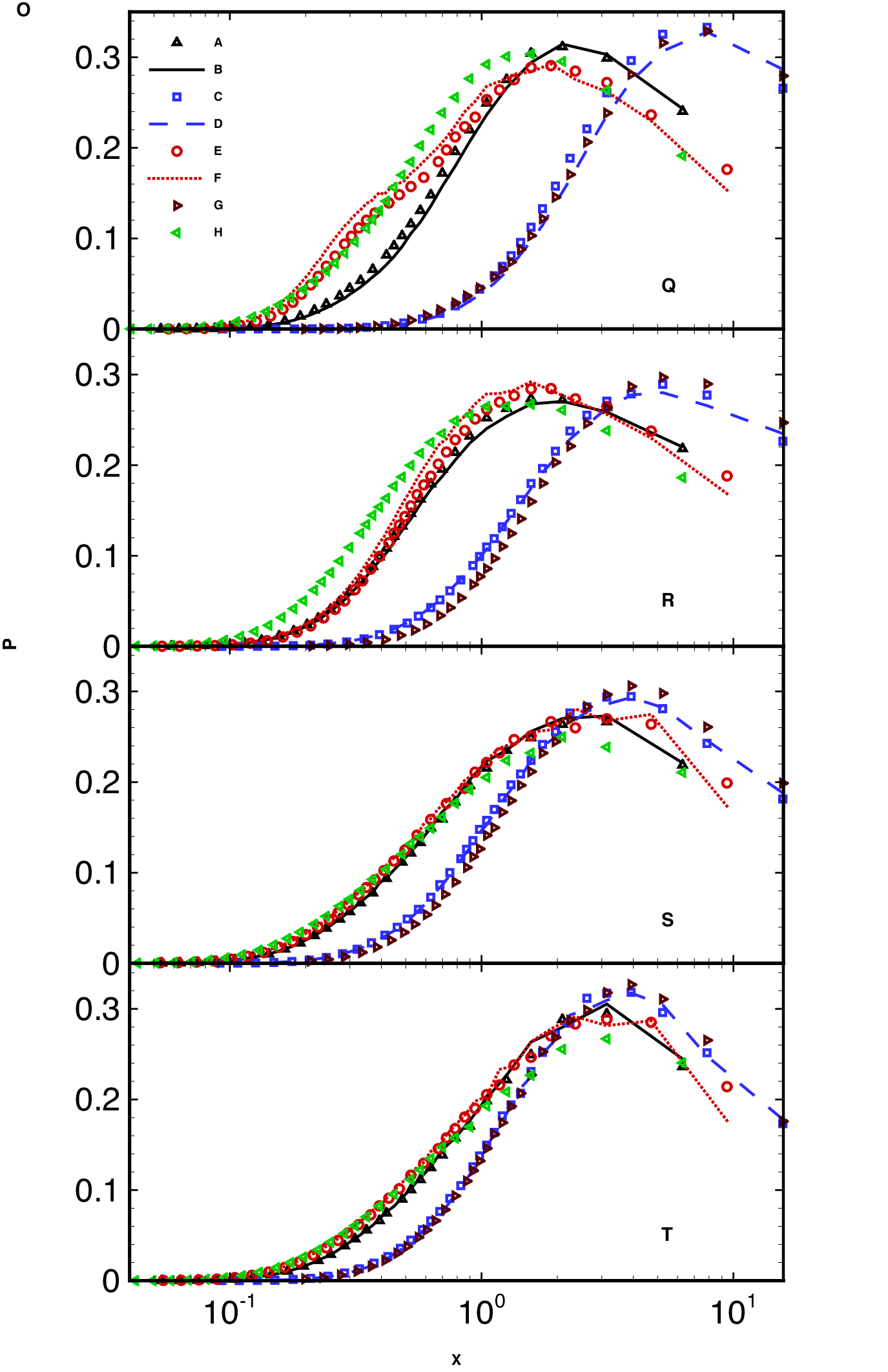}}~
  \subfigure{\label{fig:pmsz}
    \psfrag{O}[c][][0.9]{(b)}
  	\psfrag{P}[c][][0.9]{$k_zE_{\rho{u}^{\prime\prime}{u}^{\prime\prime}}/\overline{\rho{u}^{\prime\prime}{u}^{\prime\prime}}$}
  	\psfrag{X}[c][][0.9]{$\lambda_z/h$}
  	\psfrag{Q}[c][][0.9]{$~~~~~~y^\star=0.5$}
  	\psfrag{R}[c][][0.9]{$~~~~~~y^\star=20$}
  	\psfrag{A}[l][][0.65]{CP395}
  	\psfrag{B}[l][][0.65]{CRe$^\star_\tau$}
  	\psfrag{C}[l][][0.65]{SRe$^\star_{\tau GL}$}
  	\psfrag{D}[l][][0.65]{GL}
  	\psfrag{E}[l][][0.65]{LL}
  	\psfrag{F}[l][][0.65]{SRe$^\star_{\tau LL}$}
  	\psfrag{G}[l][][0.65]{CP150}
  	\psfrag{H}[l][][0.65]{CP550}
	\psfrag{S}[c][][0.9]{$~~~~~~y/h=0.3$}
  	\psfrag{T}[c][][0.9]{$~~~~~~y/h=0.5$}
  	\includegraphics[width=0.5\textwidth]{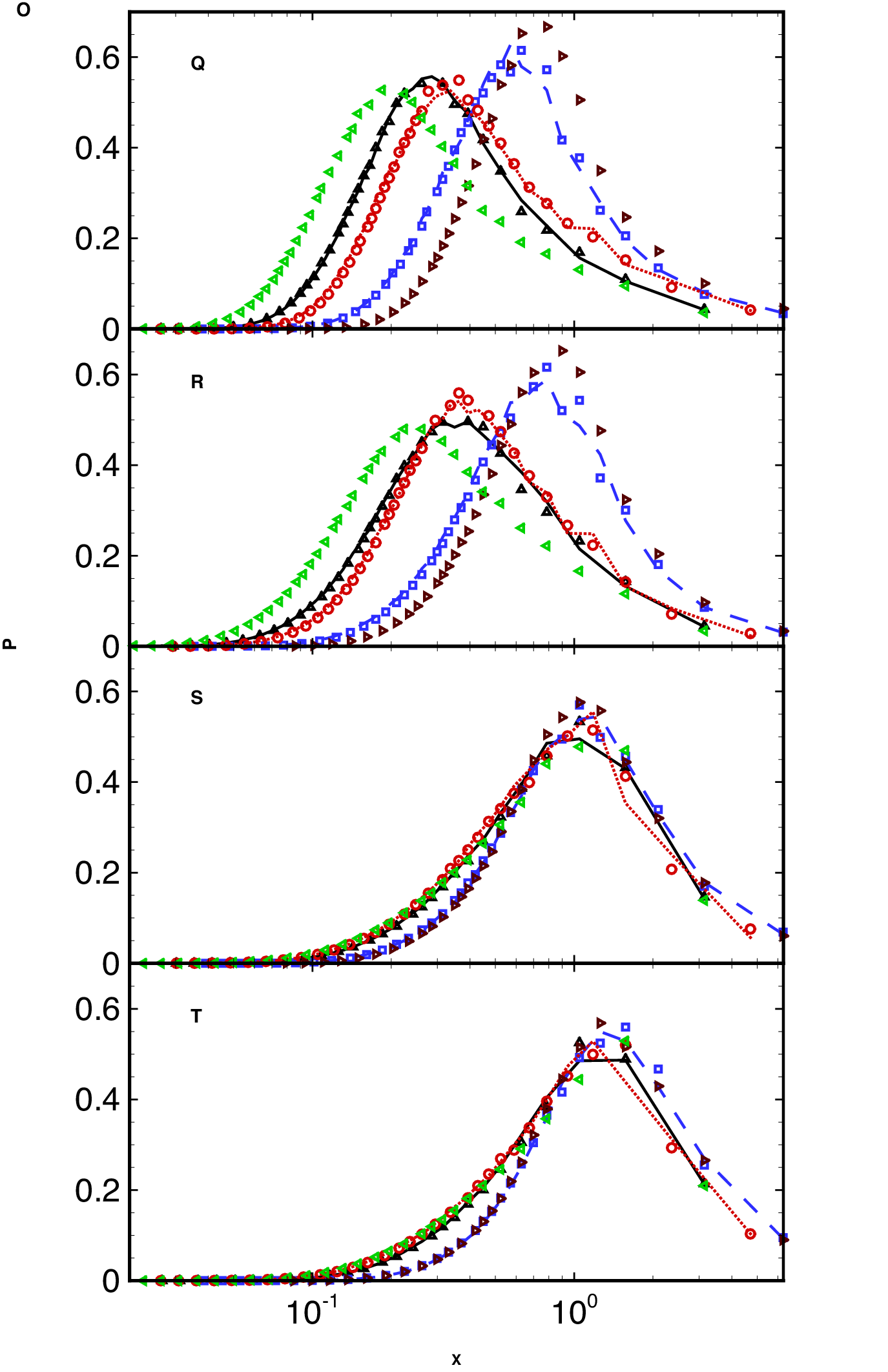}}
   \caption{{(Colour online)} {(a) Normalised pre-multiplied streamwise spectra $k_xE_{\rho{u}^{\prime\prime}{u}^{\prime\prime}}/\overline{\rho{u}^{\prime\prime}{u}^{\prime\prime}}$ as a function of streamwise wavelength $\lambda_x/h$ and (b) normalised pre-multiplied spanwise spectra $k_zE_{\rho{u}^{\prime\prime}{u}^{\prime\prime}}/\overline{\rho{u}^{\prime\prime}{u}^{\prime\prime}}$ as a function of spanwise wavelength $\lambda_z/h$, at different wall-normal locations.}}
  \label{fig:pms}
  \vspace*{-1em}
\end{figure}

The adequacy of the box size to accommodate the large-scale structures can be studied using the one-dimenisonal pre-multiplied energy spectra, defined as 
\begin{equation}
\phi_{\rho {u}^{\prime\prime}{u}^{\prime\prime}}\left(k_\alpha\right)=k_\alpha\left[\mathcal{F}\left(\sqrt{\rho}u^{\prime\prime}\right)\mathcal{F}\left(\sqrt{\rho}u^{\prime\prime}\right)^*\right]=k_\alpha\left[E_{\rho {u}^{\prime\prime}{u}^{\prime\prime}}\left(k_\alpha\right)\right]. 
\end{equation}
In the above equation, $\mathcal{F}(\psi)$ is the Fourier coefficient of $\psi$, $\mathcal{F}(\psi)^*$ is the complex conjugate, $E_{\rho {u}^{\prime\prime}{u}^{\prime\prime}}=\mathcal{F}\left(\sqrt{\rho}u^{\prime\prime}\right)\mathcal{F}\left(\sqrt{\rho}u^{\prime\prime}\right)^*$ represents the energy spectra of the density weighted streamwise velocity fluctuation, the double prime denotes Favre averaged fluctuations, and $k_\alpha$ is the wavenumber in streamwise $k_x$ or spanwise $k_z$ direction. The streamwise Reynolds stress is related to $\phi_{\rho {u}^{\prime\prime}{u}^{\prime\prime}}$ and  $E_{\rho {u}^{\prime\prime}{u}^{\prime\prime}}$ as
\begin{equation}
\overline{\rho {u}^{\prime\prime}{u}^{\prime\prime}}=\overline{\rho}\widetilde{{u}^{\prime \prime} {u}^{\prime \prime}}=\int_0^\infty \! E_{\rho {u}^{\prime\prime}{u}^{\prime\prime}}\left(k_\alpha\right) \, \mathrm{d}k_\alpha=\int_0^\infty \! \phi_{\rho {u}^{\prime\prime}{u}^{\prime\prime}}\left(k_\alpha\right) \, \mathrm{d} \left(\log\lambda_\alpha\right),
\end{equation}
where tilde denotes Favre averaging and $\lambda_\alpha=2\pi/k_\alpha$ is the wavelength. The area under the pre-multiplied spectra, when plotted in lin-log scale, represents the Reynolds stress. A plot of $\phi_{\rho {u}^{\prime\prime}{u}^{\prime\prime}}$, normalised by $\overline{\rho {u}^{\prime\prime}{u}^{\prime\prime}}$ and plotted as a function of streamwise and spanwise wavelength at different wall-normal planes, is shown in figure~\ref{fig:pms}. While the spanwise spectra indicates that the spanwise box size is sufficient, it can be argued that the length of the box in streamwise direction could be increased to resolve all the large-scale contributions. However, as noted in previous studies \citep{abe2004very,lozano2014effect} the influence of these unresolved large scales structures (due to the moderate box size) on turbulence statistics is negligible, and resolving the peak of the streamwise premultiplied spectra (e.g. DNS data of \citet{MKM590}) was considered sufficient for the present work. The spanwise premultiplied spectra $\phi_{\rho {u}^{\prime\prime}{u}^{\prime\prime}}\left(k_z\right)$ will be discussed in more detail in section~\ref{sec:str} to highlight its scaling characteristics.

\section{Mean velocity scaling}
This  section  discusses  the  effect  of  near-wall gradients in density and viscosity on scaling of mean velocity profiles. We first highlight shortcomings of the van Driest transformation for cases with strong near-wall gradients in $Re_\tau^\star$. We then derive a semi-locally scaled stress balance equation and show that the viscous stresses for the investigated cases perfectly collapse, if they are plotted as a function of semi-local wall distance $y^\star$. We will utilise this collapse to derive a velocity scaling that will be applied for the turbulent channel flows discussed above, and for adiabatic supersonic turbulent boundary layers.

\begin{figure}
  \centering
  \subfigure{\label{fig:uvdyp}
    \psfrag{T}[c][][0.9]{(a)} 
    \psfrag{V}[c][][0.9]{(c)}  	    
  	\psfrag{Z}[c][][0.9]{$\overline{u}^{\mathrm{vD}}$}
  	\psfrag{Y}[c][][0.9]{$yd \overline{u}^{\mathrm{vD}}/dy$}
  	\psfrag{X}[c][][0.9]{$y^+$}
  	\psfrag{A}[l][][0.65]{CP395}
  	\psfrag{B}[l][][0.65]{CRe$^\star_\tau$}
  	\psfrag{C}[l][][0.65]{SRe$^\star_{\tau GL}$}
  	\psfrag{D}[l][][0.65]{GL}
  	\psfrag{E}[l][][0.65]{LL}
  	\psfrag{F}[l][][0.65]{SRe$^\star_{\tau LL}$}
  	\psfrag{G}[l][][0.65]{CP150}
  	\psfrag{H}[l][][0.65]{CP550}
  	\psfrag{S}[l][][0.55]{$y^+$}
  	\psfrag{L}[l][][0.55]{$\frac{ln(y^+)}{0.41}+5.2$}
  	\includegraphics[width=0.45\textwidth]{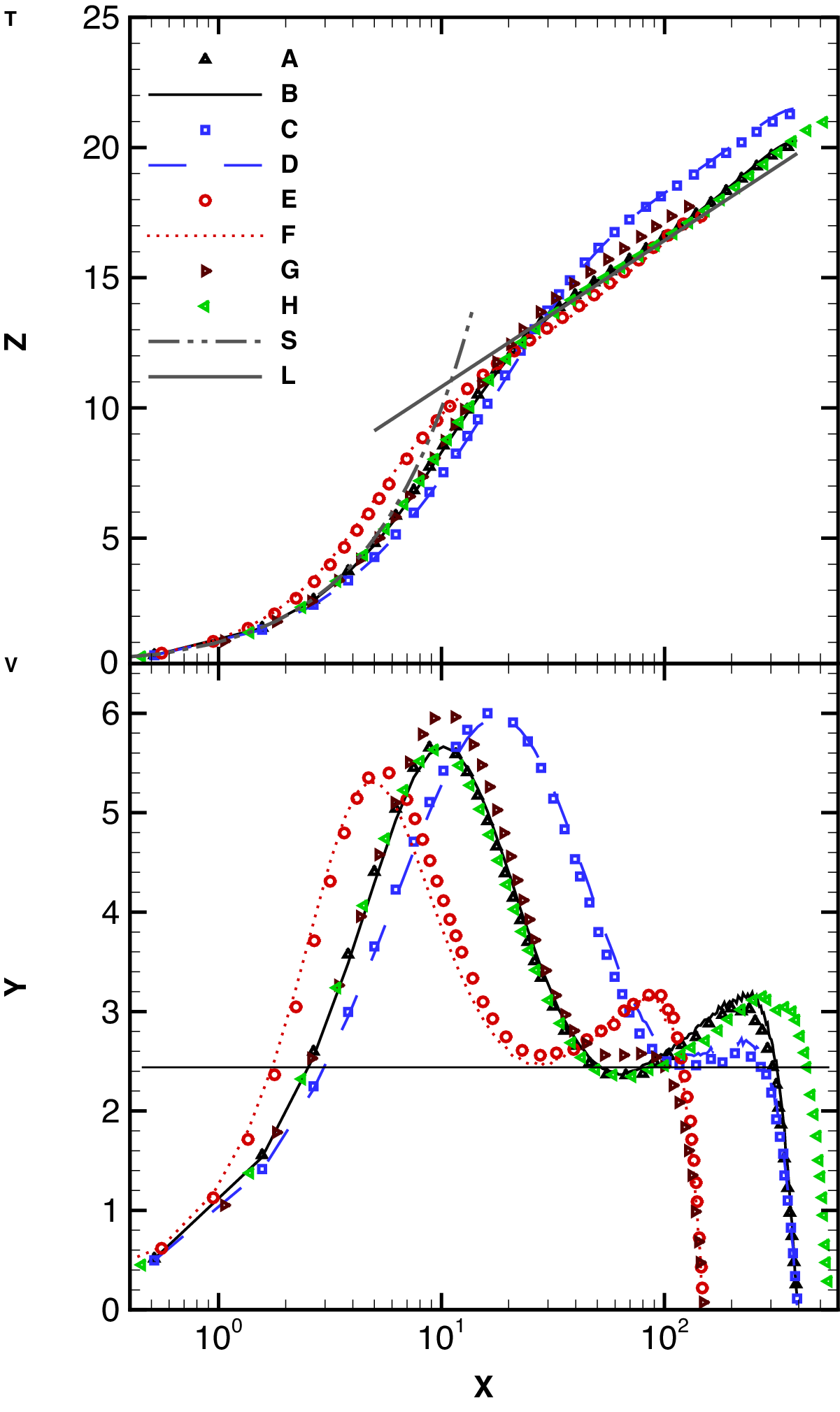}}~~~~~~~~
  \subfigure{\label{fig:uvdys}
    \psfrag{U}[c][][0.9]{(b)} 
    \psfrag{W}[c][][0.9]{(d)}
    \psfrag{Z}[c][][0.9]{$\overline{u}^{\mathrm{vD}}$}
  	\psfrag{Y}[c][][0.9]{$yd \overline{u}^{\mathrm{vD}}/dy$}
  	\psfrag{X}[c][][0.9]{$y^\star$}
  	\includegraphics[width=0.45\textwidth]{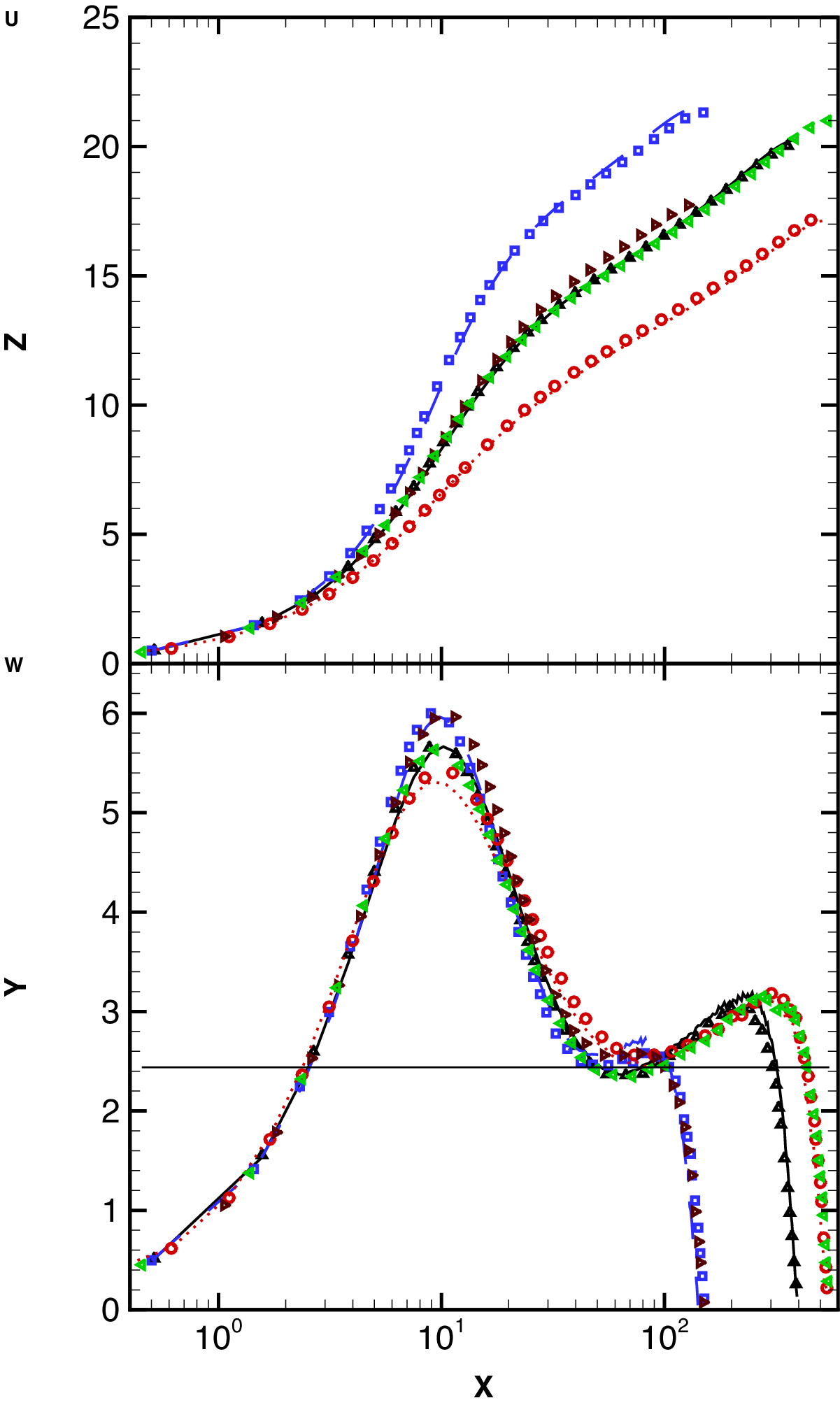}}
   \caption{{(Colour online)} {(a) and (b) van Driest velocity $\overline{u}^{\mathrm{vD}}$, (c) and (d) diagnostic function  $yd \overline{u}^{\mathrm{vD}}/dy$ shown as a function of $y^+$ (left side) and $y^\star$ (right side). The solid horizontal line in (c) and (d) has a value of 1/0.41.}}
  \label{fig:uvd}
  \vspace*{-1em}
\end{figure}

The van Driest transformed velocity $\overline{u}^{\mathrm{vD}}$ and the diagnostic function $yd \overline{u}^{\mathrm{vD}}/dy$ as a function of $y^+$ and $y^\star$ for all cases are shown in figure~\ref{fig:uvd}. As noted in our earlier study, cases with quasi-similar $Re_\tau^\star$ profiles (symbols and lines with same colour {- colour online}) exhibit similar $\overline{u}^{\mathrm{vD}}$ profiles, irrespective of their individual density and viscosity profiles. However, when comparing cases with different $Re_\tau^\star$ gradients, deviations can be observed. For example, in figure~\ref{fig:uvd}(a), the slope of the linear viscous sublayer ($\overline{u}^{\mathrm{vD}}=y^+$) increases for cases with $d {Re_\tau^\star}/dy > 0$ (red line and symbols {- colour online}) and decreases for cases with $d{Re_\tau^\star}/dy < 0$ (blue line and symbols {- colour online}). Also an increase in the log-law additive constant (commonly $B=5.2$) can be seen for cases with $d{Re_\tau^\star}/dy < 0$, while the opposite is observed for cases with $d{Re_\tau^\star}/dy > 0$. Additionally, the log-layer shifts outwards for cases with $d{Re_\tau^\star}/dy  < 0$ and inwards for cases with $d{Re_\tau^\star}/dy  > 0$, as can be seen from the profiles of the diagnostic function $yd \overline{u}^{\mathrm{vD}}/dy$ in figure~\ref{fig:uvd}(c). Figures~\ref{fig:uvd}(b) and figure~\ref{fig:uvd}(d) show $\overline{u}^{\mathrm{vD}}$ and the diagnostic function for all cases as a function of $y^\star$. The deviation of $\overline{u}^{\mathrm{vD}}$ for cases with $d{Re_\tau^\star}/dy  \neq0$ is even more prominent if plotted as a function of $y^\star$. On the other hand, the diagnostic function collapses the onset of the log-layer if plotted as function of $y^\star$. The reason for this collapse is investigated further using the streamwise stress-balance equation.

The stress-balance relation for the turbulent and viscous stresses can be obtained by integrating the mean streamwise momentum equation, which, for a fully developed turbulent channel flow (neglecting viscosity fluctuations), gives 
\begin{equation}\label{eq:vdret1_0}
-  \frac{\overline{\rho}\widetilde{{u}^{\prime \prime} {v}^{\prime \prime}}}{\overline{\rho}_w u_\tau^2} +\frac{h}{Re_\tau} \left(\frac{\overline{\mu}}{\overline{\mu}_w}\right) \frac{d\left({\overline{u}}/{u_\tau}\right)}{dy}
\approx \frac{\tau}{\tau_w} =\left(1-\frac{y}{h}\right), 
\end{equation}
where $\tau$ and $\tau_w$ are the total and wall shear stress, respectively. Analogous to \citet{patel2015semi}, 
equation~(\ref{eq:vdret1_0}) can be written in terms of semi-locally scaled velocity fluctuations (using the semi-local friction velocity $u_\tau^\star=\sqrt{\tau_w/\overline\rho}$)
\begin{equation}\label{eq:uhat}
\hat{u}_i^{\prime}\approx \frac{u_i^{\prime \prime}}{u_\tau^\star}= \sqrt{\frac{\overline{\rho}}{\overline{\rho}_w}}\left(\frac{u_i^{\prime \prime}}{u_\tau} \right)
\end{equation}
and the van Driest mean velocity increment 
\begin{equation}\label{eq:uvd}
d \overline{u}^{\mathrm{vD}}= \sqrt{\frac{\overline{\rho}}{\overline{\rho}_w}} d \left(\frac{\overline{u}}{u_\tau}\right), 
\end{equation}
to obtain the semi-locally scaled stress balance equation 
\begin{equation}\label{eq:vdret2}
-  \widetilde{\hat{u}^{\prime} \hat{v}^{\prime}} +\frac{h}{Re_\tau^\star} \frac{d\overline{u}^{\mathrm{vD}}}{dy~~} \approx \frac{\tau}{\tau_w} =\left(1-\frac{y}{h}\right). 
\end{equation}
Note, the viscous stress term is a function of the van Driest mean velocity gradient, scaled by the inverse of the semi-local Reynolds number. The reason for writing equation~(\ref{eq:vdret1_0}) in terms of $\overline{u}^{\mathrm{vD}}$ and $\widetilde{\hat{u}^{\prime} \hat{v}^{\prime}}$, originates from \citet{patel2015semi}, where the Navier-Stokes equations were rescaled using the semi-local friction velocity and local mean properties in order to obtain  governing equations for the semi-locally scaled mean, $\overline{u}^{\mathrm{vD}}$, and fluctuating velocities, $\hat{u}_i^{\prime}$, with $Re_\tau^\star$ as a strong parameter. In other words, the combined influence of density and viscosity variations on turbulence statistics can be characterised using a single parameter, $Re_\tau^\star$.

\begin{figure}
  \centering
  \subfigure{\label{fig:stressyp}
    \psfrag{T}[c][][0.9]{(a)} 
    \psfrag{V}[c][][0.9]{(c)}  
  	\psfrag{P}[c][][0.9]{$-\widetilde{\hat{u}^{\prime} \hat{v}^{\prime}}$}
  	\psfrag{Q}[c][][0.9]{$({h}/{Re_\tau^\star})d \overline{u} ^{\mathrm{vD}}/dy$}
  	\psfrag{Y}[c][][0.9]{$y^+$}
  	\psfrag{A}[l][][0.65]{CP395}
  	\psfrag{B}[l][][0.65]{CRe$^\star_\tau$}
  	\psfrag{C}[l][][0.65]{SRe$^\star_{\tau GL}$}
  	\psfrag{D}[l][][0.65]{GL}
  	\psfrag{E}[l][][0.65]{LL}
  	\psfrag{F}[l][][0.65]{SRe$^\star_{\tau LL}$}
  	\psfrag{G}[l][][0.65]{CP150}
  	\psfrag{H}[l][][0.65]{CP550}
  	\includegraphics[width=0.45\textwidth]{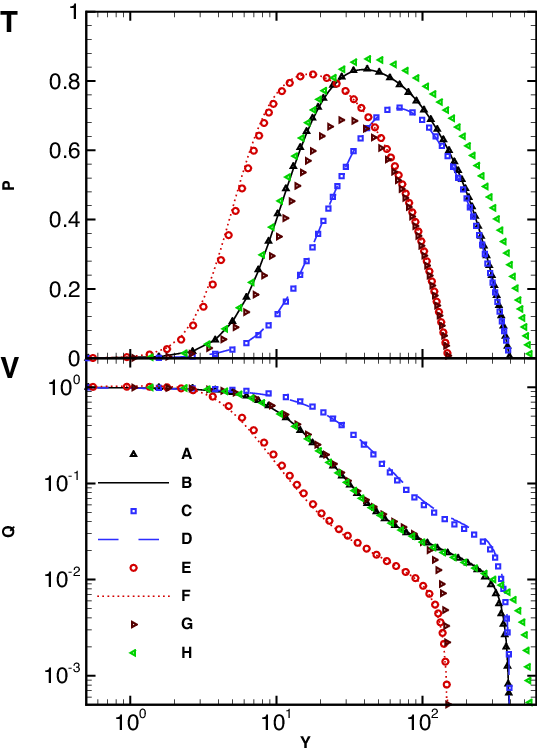}}~~~~~~~~
  \subfigure{\label{fig:stressys}
    \psfrag{T}[c][][0.9]{(b)} 
    \psfrag{V}[c][][0.9]{(d)}  
  	\psfrag{P}[c][][0.9]{$-\widetilde{\hat{u}^{\prime} \hat{v}^{\prime}}$}
  	\psfrag{Q}[c][][0.9]{$({h}/{Re_\tau^\star})d \overline{u} ^{\mathrm{vD}}/dy$}
  	\psfrag{X}[c][][0.9]{$y^\star$}
  	\psfrag{A}[c][][0.65]{}
  	\psfrag{B}[c][][0.65]{}
  	\psfrag{C}[c][][0.65]{}
  	\psfrag{D}[c][][0.65]{}
  	\psfrag{E}[c][][0.65]{}
  	\psfrag{F}[c][][0.65]{}
  	\psfrag{G}[c][][0.65]{}
  	\psfrag{H}[c][][0.65]{}
  	\includegraphics[width=0.45\textwidth]{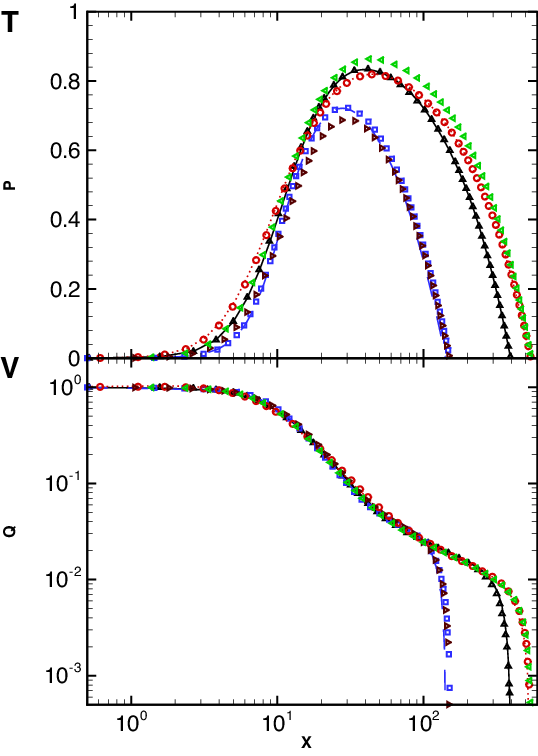}}
   \caption{{(Colour online)} {(a) and (b) Reynolds shear stress $-\widetilde{\hat{u}^{\prime} \hat{v}^{\prime}}$, (c) and (d) viscous shear stress $\left({h}/{Re_\tau^\star}\right)d \overline{u} ^{\mathrm{vD}}/dy$ shown as a function of $y^+$ (left side) and $y^\star$ (right side).}}
  \label{fig:stress}
\end{figure}

The Reynolds shear stress $\widetilde{\hat{u}^{\prime} \hat{v}^{\prime}}$ and viscous stress term $\left({h}/{Re_\tau^\star}\right) d \overline{u} ^{\mathrm{vD}}/dy$ in equation~(\ref{eq:vdret2}) are plotted as a function of $y^+$ and $y^\star$ in figure~\ref{fig:stress}. It is apparent that both stresses do not collapse if shown as a function of $y^+$ (figure~\ref{fig:stress}(a) and (c)) for cases with different $Re_\tau^\star$ gradients. Similar observations have been reported for compressible non-adiabatic flows in \citet{coleman1995numerical,foysi2004,morinishi2004} and in \citet{patel2015semi} for low-Mach number channel flows with gradients in $Re_\tau^\star$. In comparison to a constant property case with a similar $Re_\tau$ value, the Reynolds shear stress decreases for cases with $d{Re_\tau^\star}/dy < 0$ and increases for cases with $d{Re_\tau^\star}/dy > 0$. On the other hand, if the stresses are plotted as a function of $y^\star$ (figure~\ref{fig:stress}(b) and (d)), two observations can be made. First, the Reynolds shear stress profiles for all cases collapse closely in the inner layer, except in the region very close to the wall where small differences can be seen. Second, and more importantly, the viscous stresses for all cases perfectly collapse over the entire $y^\star$ range. 
Given the last observation, it is evident that the viscous stress term, expressed as $\left({h}/{Re_\tau^\star}\right) d \overline{u} ^{\mathrm{vD}}/dy$, forms the basis to develop a proper scaling law for the mean velocity in flows with large density and viscosity gradients. We thus propose that the transformation must be based on the viscous stress 
\begin{equation}\label{eq:visst}
\frac{h}{Re_\tau^\star}\frac{d\overline{u}^{\mathrm{vD}}}{dy}=\Phi(y^\star)~, 
\end{equation}
where $\Phi$ is an unknown function of $y^\star$. Using the chain rule $d/dy=(dy^\star/dy) d/dy^\star$, we can write equation~(\ref{eq:visst}) as
\begin{equation}\label{eq:visst1}
\frac{h}{Re_\tau^\star} \left(\frac{dy^\star}{dy~}\right)  \frac{d\overline{u} ^{\mathrm{vD}}}{d{y^\star}}=\Phi(y^\star)~. 
\end{equation}
This allows us to replace $dy^\star/dy$ in equation~(\ref{eq:visst1}) by taking the derivative of $y^\star=yRe_\tau^\star/h$ with respect to $y$, to obtain 
\begin{equation}\label{eq:utr0}
\left(1+\frac{y}{Re_\tau^\star}\frac{dRe_\tau^\star}{dy} \right) \frac{d \overline{u} ^{\mathrm{vD}}}{d{y^\star}}=\Phi(y^\star). 
\end{equation}
Equation~(\ref{eq:utr0}) thus provides the scaling law for the mean streamwise velocity, which we will denote as $\overline{u}^{\star}$. Hence, $d\overline{u} ^{\star}$ and $d\overline{u} ^{\mathrm{vD}}$ are related through   
\begin{equation}\label{eq:utr}
d \overline{u} ^{\star}=\left(1+\frac{y}{Re_\tau^\star}\frac{dRe_\tau^\star}{dy} \right) d \overline{u} ^{\mathrm{vD}}=\Phi(y^\star)d{y^\star}. 
\end{equation}

{Note, it can be shown that the transformation proposed by \citet{trettel2016} is equivalent  when substituting the definitions of $Re_\tau^\star$ and $d\overline{u} ^{\mathrm{vD}}$ into equation~(\ref{eq:utr}). Their transformation, which is expressed in terms of density and viscosity gradients, was obtained by equating the transformed log-law velocity gradient with the velocity gradient obtained from the stress-balance equation, assuming that the Reynolds shear stress is similar for compressible and constant property cases. The present derivation follows an alternative route. It is based on rescaling the Navier--Stokes equations using the local mean properties and semi-local friction velocity, which naturally suggests that the viscous terms are scaled by the semi-local Reynolds number $Re_\tau^\star$ to account for changes in viscous scales due to property variations. The transformation expressed in terms of equation~(\ref{eq:utr}) thus emphasises the fact that, similar to turbulence statistics, the transformation is also governed by the $Re_\tau^\star$ profiles, rather than individual density and viscosity profiles.} 

\begin{figure}
  \centering
  \subfigure{\label{fig:velu}
    \psfrag{N}[c][][0.9]{(a)}
	\psfrag{A}[l][][0.65]{CP395}
  	\psfrag{B}[l][][0.65]{CRe$^\star_\tau$}
  	\psfrag{C}[l][][0.65]{SRe$^\star_{\tau GL}$}
  	\psfrag{D}[l][][0.65]{GL}
  	\psfrag{E}[l][][0.65]{LL}
  	\psfrag{F}[l][][0.65]{SRe$^\star_{\tau LL}$}
  	\psfrag{G}[l][][0.65]{CP150}
  	\psfrag{H}[l][][0.65]{CP550}
	\psfrag{Y}[c][][0.9]{$\overline{u}^{\star}$}
  	\psfrag{X}[c][][0.9]{$y^\star$}
  	\includegraphics[width=0.5\textwidth]{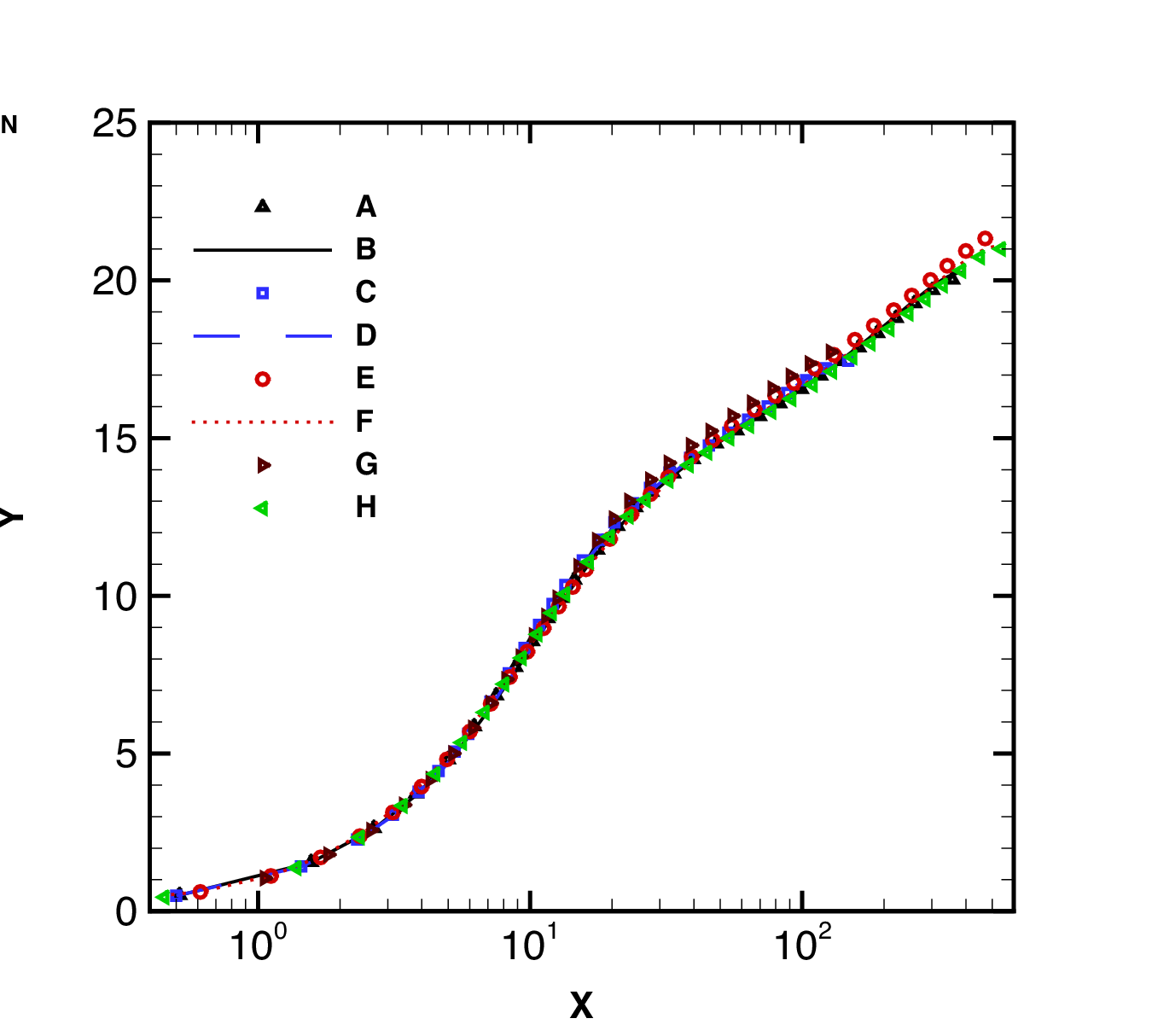}}~~~
     \subfigure{\label{fig:lmys}
     \psfrag{N}[c][][0.9]{(b)}
  	\psfrag{Y}[c][][0.9]{$l_m Re_\tau^\star/h$}
  	\psfrag{X}[c][][0.9]{$y^\star$}
  	\includegraphics[width=0.5\textwidth]{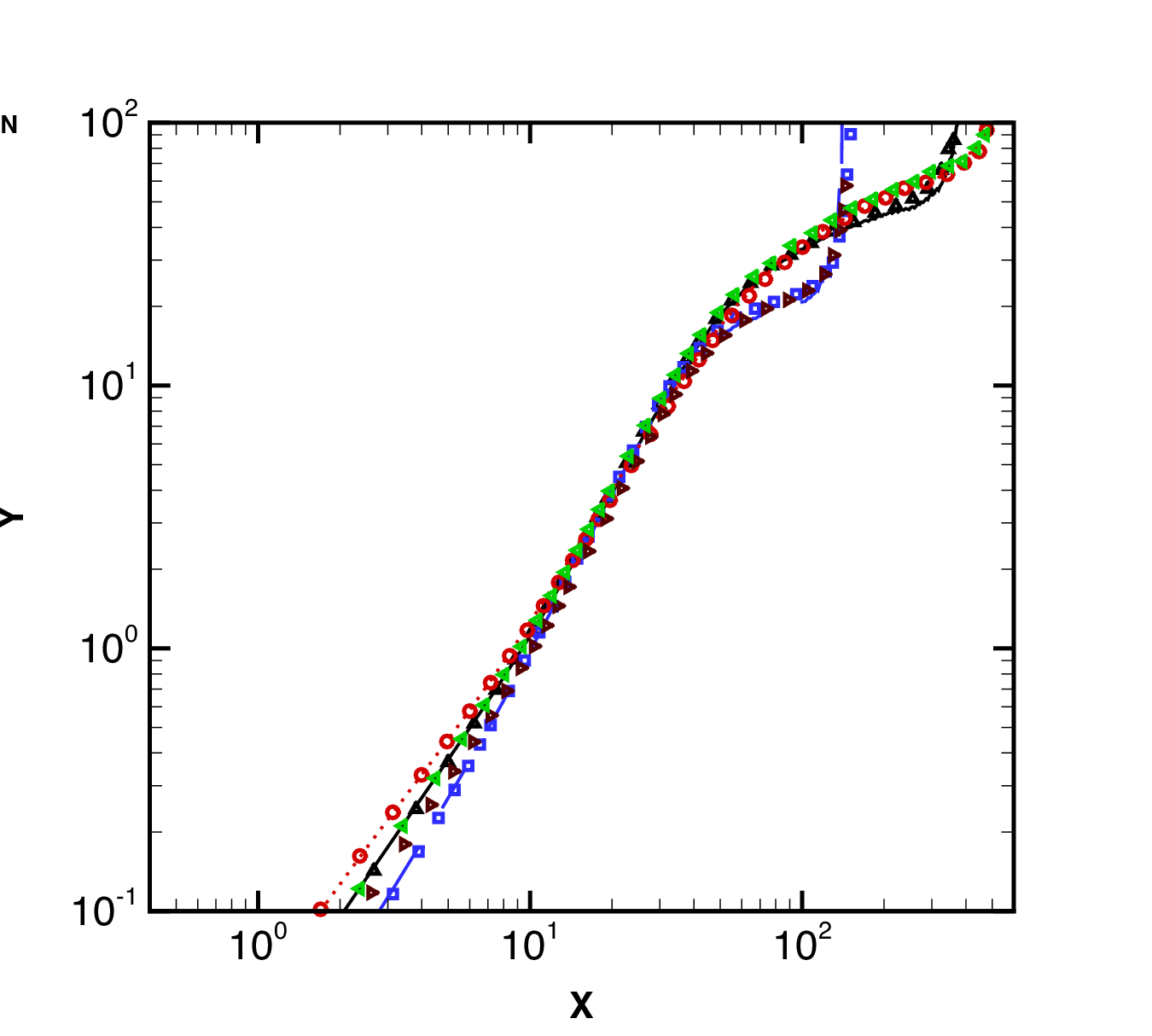}}
   \caption{{(Colour online)} {(a) Transformed velocity $\overline{u}^{\star}$ and (b) mixing length $l_m Re_\tau^\star/h$ as a function of $y^\star$.}}
  \label{fig:lmys_velu}
  \vspace*{-1em}
\end{figure}

The derived velocity transformation 
\begin{equation}\overline{u} ^{\star}=\int^{\overline{u}^{\mathrm{vD}}}_0\left(1+\frac{y}{Re_\tau^\star}\frac{dRe_\tau^\star}{dy} \right) d \overline{u} ^{\mathrm{vD}}, 
\end{equation}
is shown in figure~\ref{fig:velu} as a function of $y^\star$. It can be seen that $\overline{u} ^{\star}$ is perfectly able to collapse the velocity profiles for all cases over the entire $y^\star$ range. In our previous work \citep{patel2015semi} we showed that similar turbulence statistics and van Driest mean velocity profiles $\overline{u} ^{\mathrm{vD}}$ are obtained for cases with similar $Re_\tau^\star$ distributions, even if their individual mean density and viscosity profiles substantially differ. Here, we derive a transformation that accounts for gradients in $Re_\tau^\star$ in order to extend the van Driest transformation. 

Using the derived velocity transformation it is also possible to explain the collapse of the diagnostic function as a function of $y^\star$, as mentioned earlier. The definitions in equations~(\ref{eq:visst}) and (\ref{eq:utr}) can be expressed as 
\begin{equation}\label{eq:utr_relation1}
\frac{h}{Re_\tau^\star}\frac{d \overline{u} ^{\mathrm{vD}}}{dy} =\frac{d \overline{u} ^{\star}}{dy^\star}=\Phi(y^\star)~. 
\end{equation}
Substituting $h/Re_\tau^\star=y/y^\star$ in equation~(\ref{eq:utr_relation1}) we can obtain the correlation between the diagnostic function and the newly derived velocity scale $u^\star$ as 
\begin{equation}\label{eq:utr_relation2}
y\frac{d \overline{u} ^{\mathrm{vD}}}{dy} =y^\star\frac{d \overline{u} ^{\star}}{dy^\star}=y^\star\Phi(y^\star)~, 
\end{equation} 
which explains the collapse of the diagnostic function in figure~\ref{fig:uvd}(d). 

The invariance of $\Phi$ as a function of $y^\star$ can further be quantified by expressing it in terms of turbulence mixing length. Following \citet{huang1994van}, the Reynolds shear stress can be written in terms of the velocity gradient using the mixing-length theory as 
\begin{equation}\label{eq:mixl}
l_m^2 =  \frac{-\widetilde{{u}^{\prime\prime} {v}^{\prime\prime}}}{\left(d \overline{u}/dy\right)^2} = \frac{-\widetilde{\hat{u}^{\prime} \hat{v}^{\prime}}}{\left(d\overline{u} ^{\mathrm{vD}}/dy\right)^2}~.
\end{equation}
Substituting the Reynolds shear stress in equation~(\ref{eq:vdret2}) using equation~(\ref{eq:mixl}) results in a quadratic equation for $d\overline{u} ^{\mathrm{vD}}/dy$, which, when solved and simplified, gives 
\begin{equation}\label{eq:vdml}
\Phi(y^\star) = \frac{2 {\tau}/{\tau_w} }{1+\sqrt{1+4{\tau}/{\tau_w}\left({l_m Re_\tau^\star}/{h}\right)^2 }}~.
\end{equation}
The above expression naturally suggests that $l_m$ scales with the semi-local length scale $h/Re_\tau^\star$. Figure~\ref{fig:lmys} shows $l_m Re_\tau^\star/h$ as a function of $y^\star$. A satisfactory collapse is obtained for almost the entire inner layer, except very close to wall where small deviations occur. These deviations stem from turbulence modulation caused by strong $Re_\tau^\star$ gradients, which we will discuss in detail in section~\ref{sec:mod} and \ref{sec:str}. However, close to the wall the viscous stress dominates and small values of mixing length do not alter the velocity scaling.

\begin{figure}
  \centering
  \subfigure{\label{fig:blret}
    \psfrag{N}[c][][0.9]{(a)}
  	\psfrag{Y}[c][][0.9]{$Re^\star_\tau$}
  	\psfrag{X}[c][][0.9]{$y^\star$}
  	\psfrag{A}[c][][0.65]{~~~Ma=2}
  	\psfrag{B}[c][][0.65]{~~~Ma=3}
  	\psfrag{C}[c][][0.65]{~~~Ma=4}
  	\psfrag{G}[c][][0.65]{~~~Ma=0}
  	\includegraphics[width=0.5\textwidth]{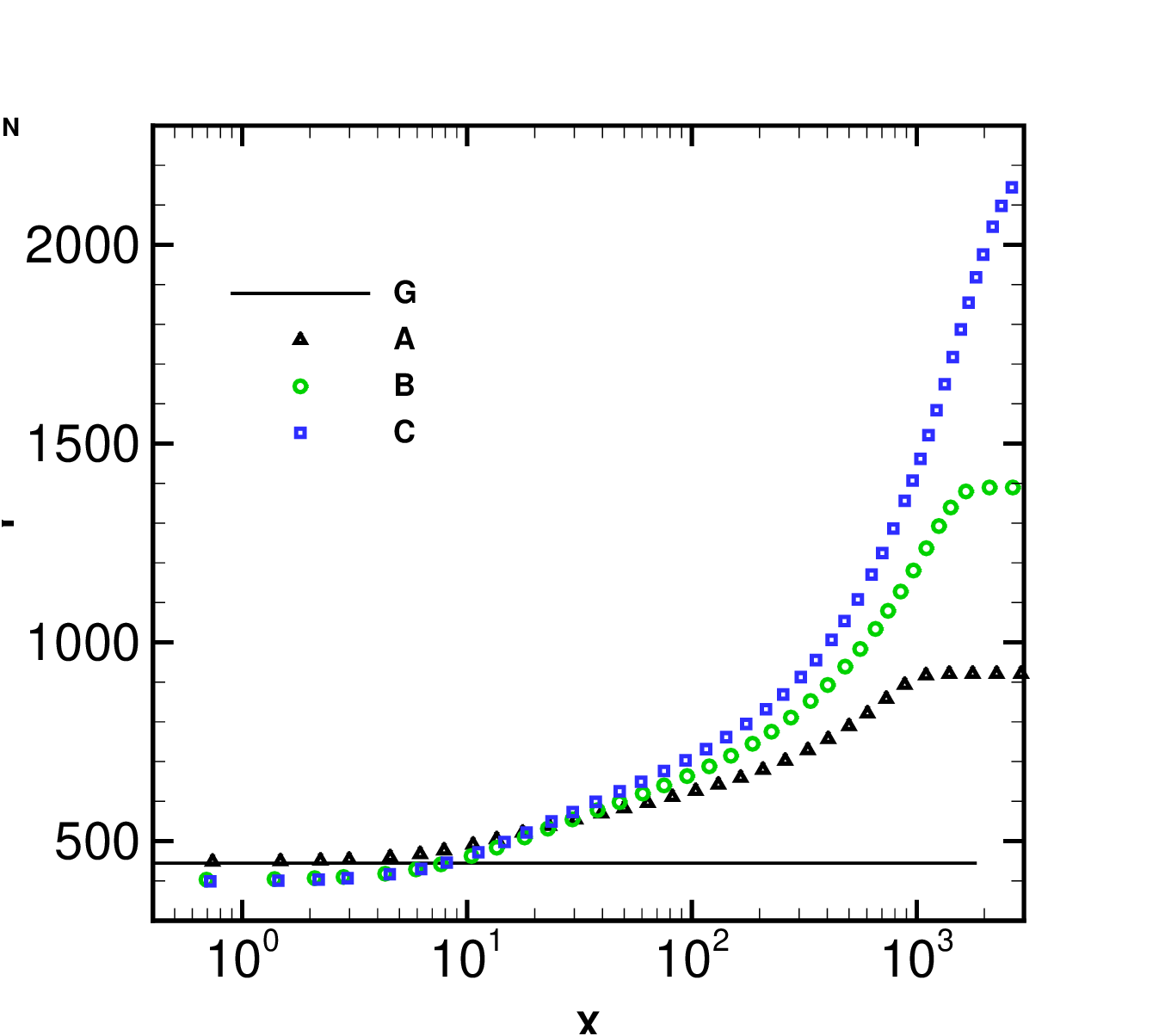}}
  \subfigure{\label{fig:blvelyp}
    \psfrag{N}[c][][0.9]{(b)}
  	\psfrag{Y}[c][][0.9]{$\overline{u}^{\mathrm{vD}}$}
  	\psfrag{X}[c][][0.9]{$y^+$}
  	\includegraphics[width=0.5\textwidth]{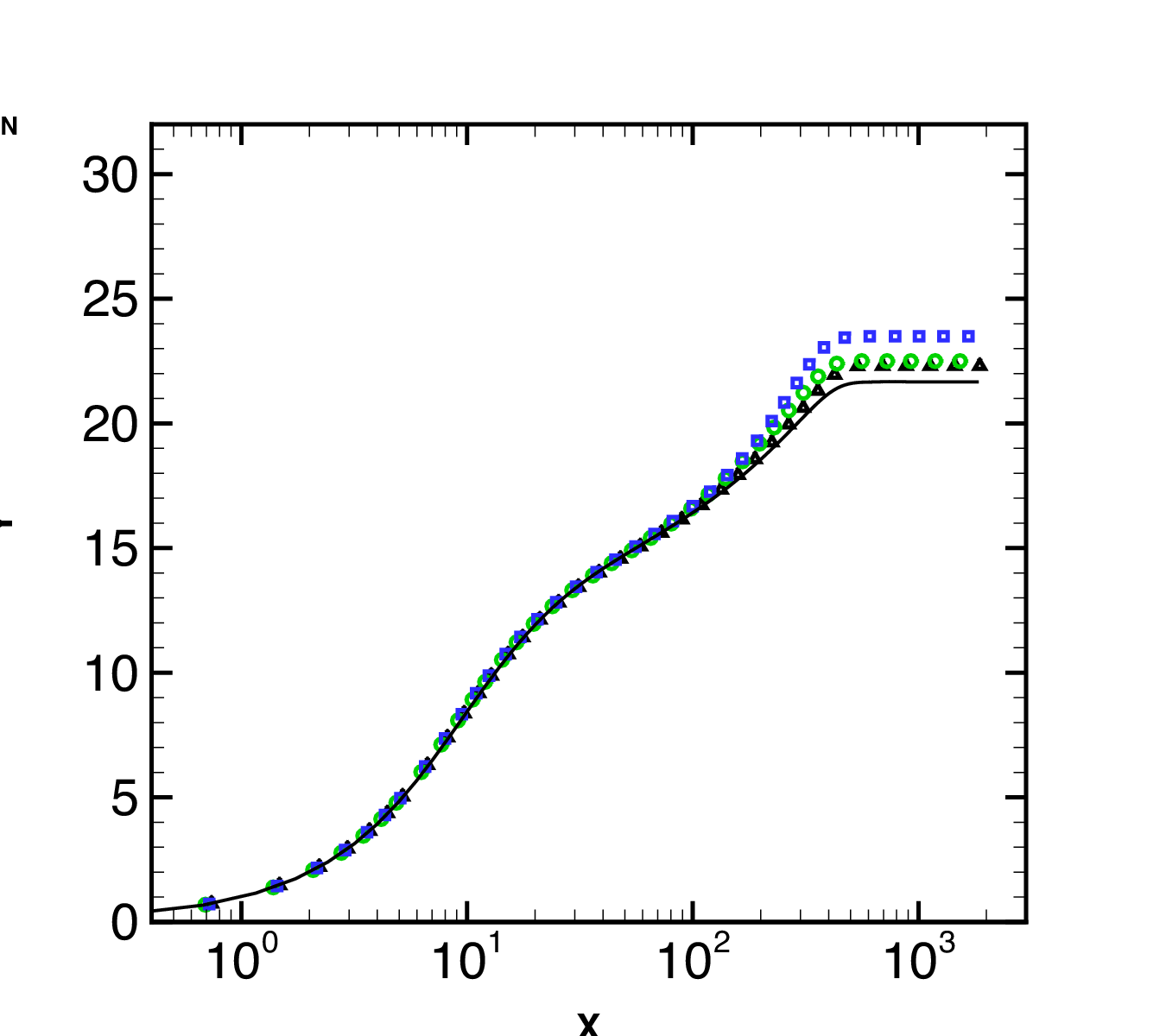}}~~~
  \subfigure{\label{fig:blvelys}
    \psfrag{N}[c][][0.9]{(c)}
  	\psfrag{Y}[c][][0.9]{$\overline{u}^\star$}
  	\psfrag{X}[c][][0.9]{$y^\star$}
  	\includegraphics[width=0.5\textwidth]{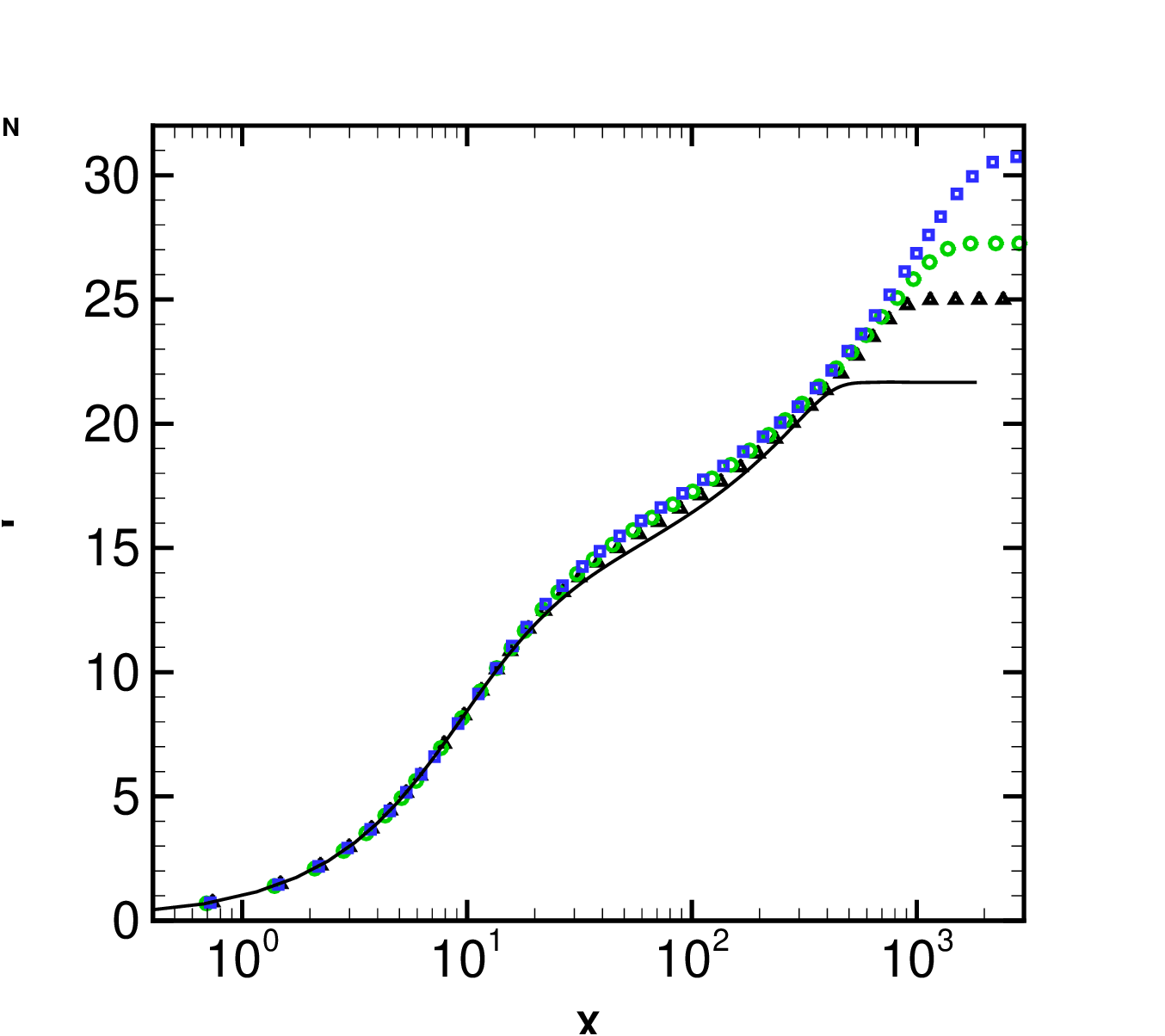}}
   \caption{{(Colour online)} {(a) $Re^\star_\tau$ shown as a function of $y^\star$, (b) van Driest velocity $\overline{u}^{\mathrm{vD}}$ shown as a function of $y^+$, (c) transformed velocity $\overline{u}^{\star}$ shown as a function of $y^\star$ for adiabatic supersonic boundary layers obtained from \citet{pirozzoli2011turbulence,bernardini2011wall}, compared with reference boundary layer data from \citet{jimenez2010turbulent}.}}
  \label{fig:blvel}
  \vspace*{-1em}
\end{figure}

\begin{figure}
  \centering
  \subfigure{\label{fig:bl_stressyp}
    \psfrag{T}[c][][0.9]{(a)} 
    \psfrag{V}[c][][0.9]{(c)} 
  	\psfrag{P}[c][][0.9]{$-\widetilde{\hat{u}^{\prime} \hat{v}^{\prime}}$}
  	\psfrag{Q}[c][][0.9]{$({h}/{Re_\tau^\star})d \overline{u} ^{\mathrm{vD}}/dy$}
  	\psfrag{Y}[c][][0.9]{$y^+$}
  	\psfrag{A}[c][][0.65]{~~~Ma=2}
  	\psfrag{B}[c][][0.65]{~~~Ma=3}
  	\psfrag{C}[c][][0.65]{~~~Ma=4}
  	\psfrag{G}[c][][0.65]{~~~Ma=0}
  	\includegraphics[width=0.45\textwidth]{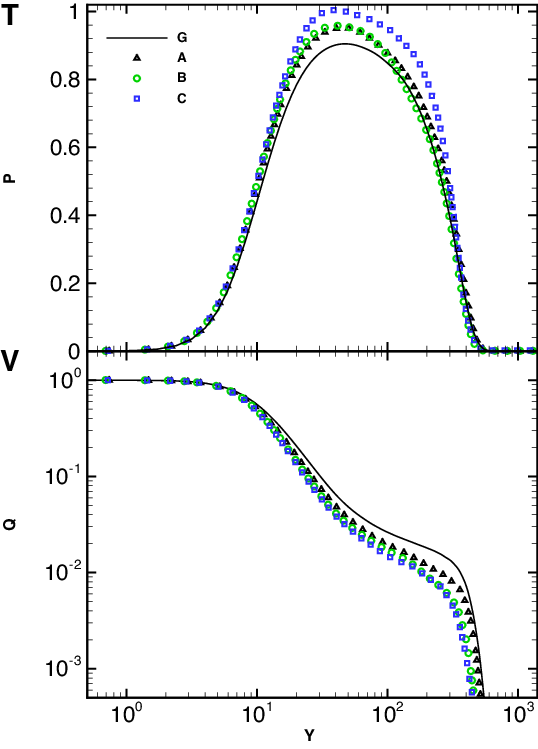}}~~~~~~~~
  \subfigure{\label{fig:bl_stressys}
    \psfrag{U}[c][][0.9]{(b)} 
    \psfrag{W}[c][][0.9]{(d)} 
  	\psfrag{P}[c][][0.9]{$-\widetilde{\hat{u}^{\prime} \hat{v}^{\prime}}$}
  	\psfrag{Q}[c][][0.9]{$({h}/{Re_\tau^\star})d \overline{u} ^{\mathrm{vD}}/dy$}
  	\psfrag{Y}[c][][0.9]{$y^\star$}
  	\psfrag{A}[c][][0.65]{~~~Ma=2}
  	\psfrag{B}[c][][0.65]{~~~Ma=3}
  	\psfrag{C}[c][][0.65]{~~~Ma=4}
  	\psfrag{G}[c][][0.65]{~~~Ma=0}
  	\includegraphics[width=0.45\textwidth]{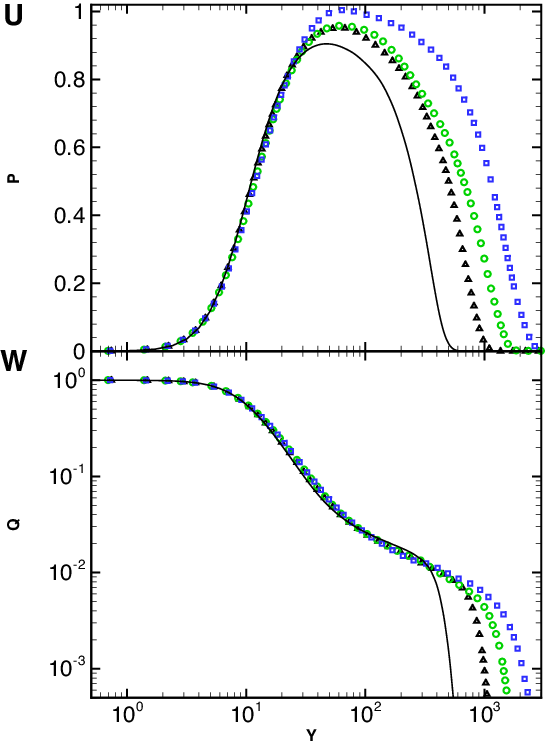}}
   \caption{{(Colour online)} {(a) and (b) Reynolds shear stress $-\widetilde{\hat{u}^{\prime} \hat{v}^{\prime}}$, (c) and (d) viscous shear stress $({h}/{Re_\tau^\star})d \overline{u} ^{\mathrm{vD}}/dy$ shown as a function of $y^+$ (left side) and $y^\star$ (right side) for adiabatic supersonic boundary layers obtained from \citet{pirozzoli2011turbulence,bernardini2011wall}, compared with  boundary layer data from \citet{jimenez2010turbulent}.}}
  \label{fig:bl_stress}
\end{figure}

In the following, we will investigate if the scaling can also be successfully applied for supersonic adiabatic flows, for which the van Driest scaling already shows a satisfying collapse. For this we will use a DNS database for adiabatic supersonic boundary layers from \citet{pirozzoli2011turbulence,bernardini2011wall} and data for an incompressible boundary layer from \citet{jimenez2010turbulent}. The investigated compressible cases are with Mach numbers $Ma$=2, 3 and 4 and corresponding Reynolds numbers are $Re_\tau$=450, 400 and 400, respectively. The incompressible boundary layer is at $Re_\tau$=450. Figure~\ref{fig:blret} shows the $Re_\tau^\star$ distribution for these cases as a function of $y^\star$. Note, in contrast to the heated and cooled channel flows, the gradient of $Re_\tau^\star$ at the wall is negligible. The largest $Re_\tau^\star$ variation is obtained for the $Ma$=4 case, where $Re_\tau^\star$ at the edge of the boundary layer is more than four times as large than at the wall. The van Driest mean velocity $\overline{u} ^{\mathrm{vD}}$ is plotted as a function of $y^+$ in figure~\ref{fig:blvelyp}. A satisfying collapse for the compressible and incompressible boundary layer is obtained, except in the wake region, which has been studied in more detail by \citet{zhang2012mach}. The transformed velocity $\overline{u}^{\star}$ as a function of $y^\star$ is shown in figure~\ref{fig:blvelys}. Also here, a reasonable collapse for the $\overline{u}^{\star}$ velocity profiles is obtained. However, small deviations with respect to the incompressible boundary layer occur in the buffer and logarithmic region. Interestingly, also \citet{trettel2016} have found a similar disagreement for cooled turbulent boundary layers.

{The implication of the above mean velocity scaling characteristics on the stress balance relation for these turbulent boundary layers is discussed next.} The Reynolds shear and viscous stresses are  shown in figure~\ref{fig:bl_stress} as a function of $y^+$ and $y^\star$. Both stresses show a superior collapse if they are plotted as a function of $y^\star$. Analogous to the plot of $\overline{u}^{\star}$ vs $y^\star$ in figure~\ref{fig:blvelys}, also the viscous stresses of the compressible cases slightly deviate from the incompressible case, as can be seen in figure~\ref{fig:bl_stress}(d). On the other hand, if the stresses are plotted as a function of $y^+$, it can be seen that the turbulent shear stress profiles increase and the viscous stress profiles decrease with Mach number, see figure~\ref{fig:bl_stress}(a) and figure~\ref{fig:bl_stress}(c). This Mach number dependence of the stresses as a function of $y^+$ can be shown mathematically by writing the stress-balance equation (\ref{eq:vdret2}) for the incompressible and  compressible boundary layer as, 
\begin{equation}\label{eq:blsb1Incomp} 
  \left. \left( \widetilde{\hat{u}^{\prime} \hat{v}^{\prime}} 
  + \frac{\tau}{\tau_w} \right) \right|_{\mathrm{incomp}}=\left.\frac{h}{Re_\tau} \frac{d\overline{u}^{\mathrm{vD}}}{dy}\right|_{\mathrm{incomp}}~\mathrm{for~the~incompressible~BL},
\end{equation}
\begin{equation}\label{eq:blsb1Comp} 
 \left.\frac{Re_\tau^\star}{Re_\tau} \left( \widetilde{\hat{u}^{\prime} \hat{v}^{\prime}} 
  + \frac{\tau}{\tau_w} \right) \right|_{\mathrm{comp}}=\left.\frac{h}{Re_\tau} \frac{d\overline{u}^{\mathrm{vD}}}{dy}\right|_{\mathrm{comp}}~\mathrm{for~the~compressible~BL}.
\end{equation}
In the constant stress layer $(\tau/\tau_w)_{\mathrm{comp}}=(\tau/\tau_w)_{\mathrm{incomp}}=1$ and due to the collapse of $\overline{u}^{\mathrm{vD}}$ as a function of $y^+$, equations~(\ref{eq:blsb1Incomp}) and (\ref{eq:blsb1Comp}) can be equated as, 
\begin{equation}\label{eq:blsb1equating} 
  \left. \left( \widetilde{\hat{u}^{\prime} \hat{v}^{\prime}} 
  + 1 \right) \right|_{\mathrm{incomp}}=
  \left.\frac{Re_\tau^\star}{Re_\tau} \left( \widetilde{\hat{u}^{\prime} \hat{v}^{\prime}} 
  + 1 \right) \right|_{\mathrm{comp}}~.
\end{equation}
Therefore, if ${Re_\tau^\star} > {Re_\tau}$ it follows that $\left.\left(-\widetilde{\hat{u}^{\prime} \hat{v}^{\prime}}\right) \right|_{\mathrm{comp}}>\left.\left(-\widetilde{\hat{u}^{\prime} \hat{v}^{\prime}}\right) \right|_{\mathrm{incomp}}$, which explains the Mach number dependence of the stresses. 
The increase in shear stress for the supersonic cases also explains the corresponding increase in turbulence intensities when compared with  incompressible cases \citep{pirozzoli2004direct,bernardini2011wall}, due to the increase in the turbulent kinetic energy production. The partial success of both $y^+$ and $y^\star$ in adiabatic supersonic boundary layers, thus warrants future studies. The semi-local scaling utilizes the invariance of $({h}/{Re_\tau^\star}) {d\overline{u}^{\mathrm{vD}}}/{dy}$ as a function of $y^\star$ to obtain $\overline{u}^{\star}$, while adiabatic boundary layers exhibit invariance of a more fundamental variable, namely the wall-normalized van Driest transformed mean spanwise vorticity $\overline{\omega}_z^{\mathrm{vD}}/(\overline{\omega}_z^{\mathrm{vD}})_w  = {d\overline{u}^{\mathrm{vD}}}/{dy^+}=({h}/{Re_\tau}) {d\overline{u}^{\mathrm{vD}}}/{dy}$ as a function of $y^+$. More discussion on the significance of mean vorticity is provided in the next section.

\section{Influence of Re$_\tau^\star$ gradients on near-wall turbulence statistics}\label{sec:mod}
Turbulence statistics are strongly influenced by near-wall modifications of turbulence. Near-wall turbulence modifications for variable property flows can be classified into two main mechanisms: (i) changes in viscous scales, and (ii) structural changes of turbulence. The semi-local scaling is successful in collapsing turbulence statistics, such as mean velocity (see figure~\ref{fig:velu}), Reynolds and viscous shear stress (see figure~\ref{fig:stress}(b) and (d)), because it accommodates the change in viscous scales using local quantities, and because these quantities are not significantly affected by structural changes and non-local effects very close to the wall. On the other hand, the semi-local scaling fails for turbulence statistics that are sensitive to strong non-local interactions of the buffer layer vortical structures with the sub-layer region, e.g., statistics of vorticity fluctuations. Furthermore, the semi-local scaling also fails for turbulence statistics that contain direct information of turbulence structure, e.g., turbulence anisotropy \citep{foysi2004} and low-speed streaks \citep{patel2015semi}. 

We first study the gradient of the van Driest transformed mean velocity $\overline u^\mathrm{vD}$, which we will refer to as the van Driest transformed mean spanwise vorticity $\overline{\omega}_z^{\mathrm{vD}}$. $\overline u^\mathrm{vD}$ is the analogue to the mean velocity in a variable density flow, and therefore a study of its gradients allows us to draw conclusions on the stability of the flow. We then discuss the turbulent vorticity fluctuations and the turbulence anisotropy to emphasise the failure of the semi-local scaling due to non-locality and changes in turbulence structure, respectively.

\subsection{Mean spanwise vorticity}\label{sec:mod1}
\begin{figure}
  \centering
  \subfigure{\label{fig:mvortyp}
    \psfrag{N}[c][][0.9]{(a)}  
  	\psfrag{Y}[c][][0.9]{$\overline{\omega}_z^{\mathrm{vD}}/(\overline{\omega}_z^{\mathrm{vD}})_w$}
  	\psfrag{X}[c][][0.9]{$y^+$}
  	\psfrag{A}[l][][0.65]{CP395}
  	\psfrag{B}[l][][0.65]{CRe$^\star_\tau$}
  	\psfrag{C}[l][][0.65]{SRe$^\star_{\tau GL}$}
  	\psfrag{D}[l][][0.65]{GL}
  	\psfrag{E}[l][][0.65]{LL}
  	\psfrag{F}[l][][0.65]{SRe$^\star_{\tau LL}$}
  	\psfrag{G}[l][][0.65]{CP150}
  	\psfrag{H}[l][][0.65]{CP550}
  	\includegraphics[width=0.45\textwidth]{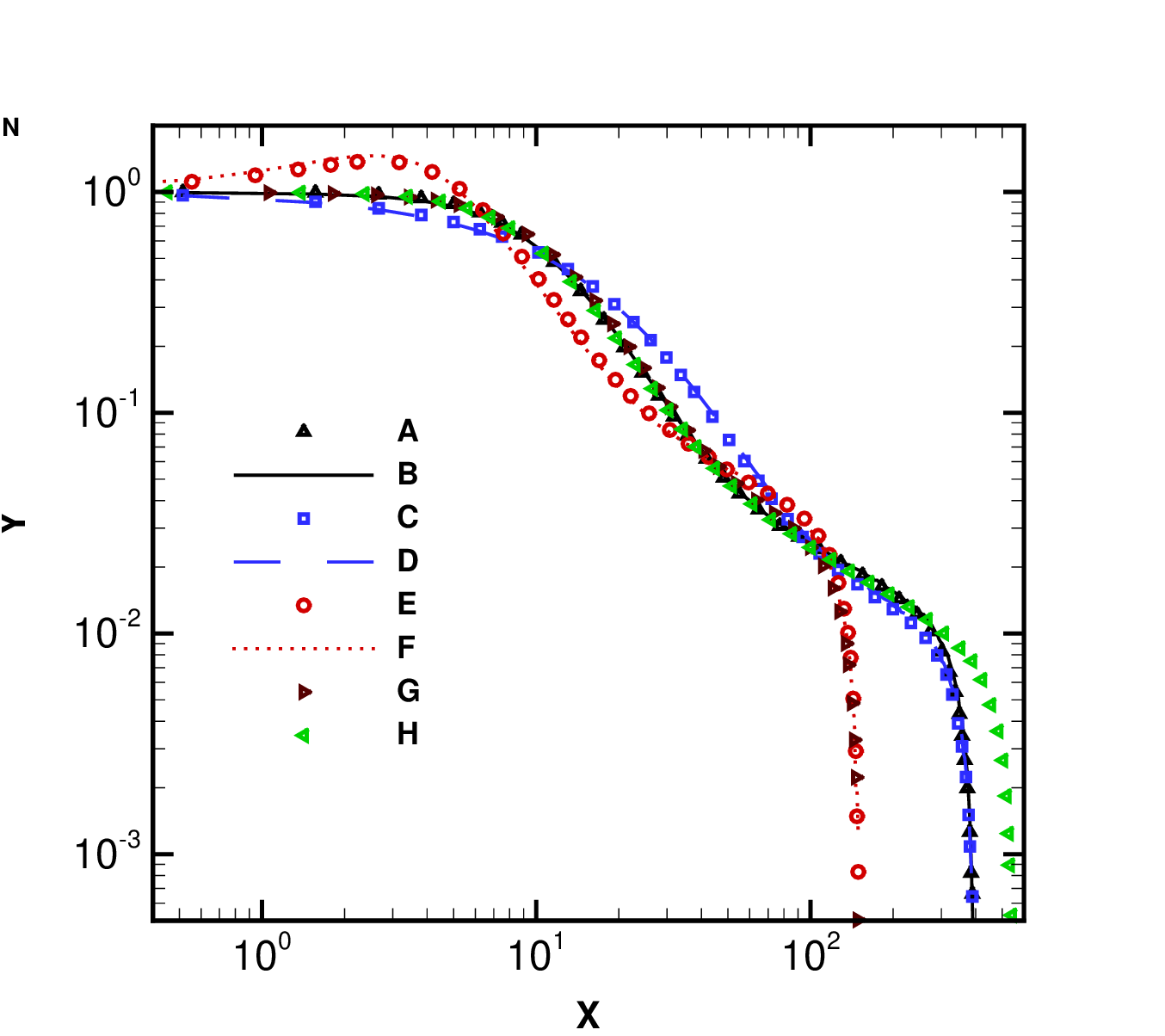}}~~~
  \subfigure{\label{fig:mvortys}
    \psfrag{N}[c][][0.9]{(b)}  
  	\psfrag{Y}[c][][0.9]{$\overline{\omega}_z^{\mathrm{vD}}/(\overline{\omega}_z^{\mathrm{vD}})_w$}
  	\psfrag{X}[c][][0.9]{$y^\star$}
  	\psfrag{A}[c][][0.65]{}
  	\psfrag{B}[c][][0.65]{}
  	\psfrag{C}[c][][0.65]{}
  	\psfrag{D}[c][][0.65]{}
  	\psfrag{E}[c][][0.65]{}
  	\psfrag{F}[c][][0.65]{}
  	\psfrag{G}[c][][0.65]{}
  	\psfrag{H}[c][][0.65]{}
  	\includegraphics[width=0.45\textwidth]{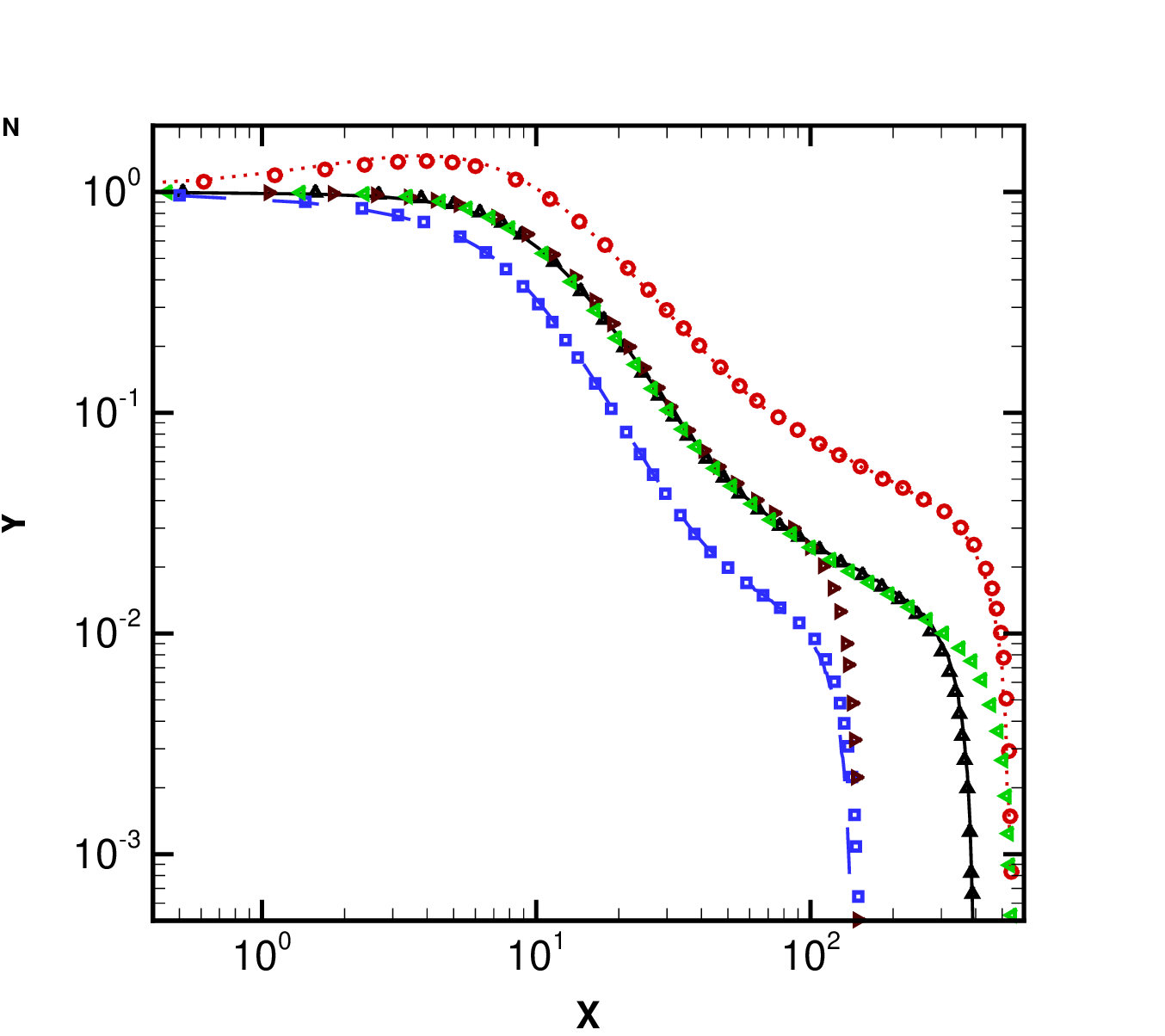}}
   \caption{{(Colour online)} Wall-normalized van Driest transformed mean spanwise vorticity $\overline{\omega}_z^{\mathrm{vD}}/(\overline{\omega}_z^{\mathrm{vD}})_w$ shown as a function of (a) $y^+$ and (b) $y^\star$.}
\label{fig:mvort}
\end{figure}
The wall-normalized van Driest transformed mean spanwise vorticity $\overline{\omega}_z^{\mathrm{vD}}/(\overline{\omega}_z^{\mathrm{vD}})_w  = {d\overline{u}^{\mathrm{vD}}}/{dy^+}=({h}/{Re_\tau}) {d\overline{u}^{\mathrm{vD}}}/{dy}$ is plotted as a function of $y^+$ and $y^\star$ in figure~\ref{fig:mvortyp} and figure~\ref{fig:mvortys}, respectively. 
Unlike for adiabatic walls, where $\overline{\omega}_z^{\mathrm{vD}}/(\overline{\omega}_z^{\mathrm{vD}})_w$ is invariant as a function of $y^+$ in the inner layer \citep[e.g., ][]{pirozzoli2011turbulence}, $\overline{\omega}_z^{\mathrm{vD}}/(\overline{\omega}_z^{\mathrm{vD}})_w$  is neither a function of $y^+$ nor $y^\star$ for non-adiabatic walls. 
Based on \citet{eyink2008turbulent} vorticity generated at the wall is transported outward, first diffused by viscosity and subsequently advected by turbulence. Strong gradients in viscosity in the viscous  dominated region thus change the vorticity transport and consequently the mean velocity profile. The change in mean velocity profile then directly influences turbulence. Equivalently, in variable density flows, strong gradients in $Re_\tau^\star$ (since $Re_\tau^\star$ characterises the viscous scale), influence the van Driest transformed mean velocity profile. It is clearly visible in figure~\ref{fig:mvort} that $\overline{\omega}_z^{\mathrm{vD}}/(\overline{\omega}_z^{\mathrm{vD}})_w$ shows a local maximum for cases with $d{Re_\tau^\star}/dy > 0$ (red line and symbols {- colour online}). This maximum implies an inflection point in the velocity profile, which also indicates a more unstable flow condition. The reverse happens for cases with $d {Re_\tau^\star}/dy < 0$ (blue line and symbols {- colour online}) where the velocity profile becomes fuller (higher negative curvature of $d^2 \overline{u} ^{\mathrm{vD}}/dy^2$), causing the flow to become more stable \citep{gad1990control}. This effect is similar to flow control techniques that change the curvature of the velocity profile at the wall by introducing, e.g., wall motion, suction/injection, streamwise pressure-gradients, or wall-normal viscosity-gradients \citep{gad1990control}. {For example, in an adverse-pressure-gradient region (decelerated flow) the appearance of an inflection point is known to increase the wall-burst rate \citep{bushnell1989turbulence}. The opposite is observed for flows with favorable-pressure-gradient, where the base velocity state mitigates the formation of localised near-wall inflections \citep{bushnell1989turbulence}. \citet{marquillie2011instability} performed a linear stability analysis about the mean base profile of a turbulent boundary layer with adverse-pressure-gradient, and showed a higher streak instability for cases with a pronounced wall-normal inflection point.} This fact has an important implication on near-wall turbulent structures and is discussed in more detail in section~\ref{sec:str}.

\subsection{Turbulent vorticity fluctuations}
\begin{figure}
  \centering
  \subfigure{\label{fig:vortws}
    \psfrag{N}[c][][0.9]{(a)}
    \psfrag{M}[c][][0.9]{(c)}
    \psfrag{O}[c][][0.9]{(e)}
  	\psfrag{P}[c][][0.9]{$(h/Re_\tau)({\hat{\omega}^{\prime}_x})_{rms}$}
  	\psfrag{Q}[c][][0.9]{$({h}/{Re_\tau})({\hat{\omega}^{\prime}_y})_{rms}$}
  	\psfrag{R}[c][][0.9]{$({h}/{Re_\tau})({\hat{\omega}^{\prime}_z})_{rms}$}
  	\psfrag{X}[c][][0.9]{$y^+$}
  	\psfrag{A}[c][][0.65]{~~CP395}
  	\psfrag{B}[c][][0.65]{~~~~CRe$^\star_\tau$}
  	\psfrag{C}[c][][0.65]{~~~~~~SRe$^\star_{\tau GL}$}
  	\psfrag{D}[c][][0.65]{~GL}
  	\psfrag{E}[c][][0.65]{~LL}
  	\psfrag{F}[c][][0.65]{~~~~~~SRe$^\star_{\tau LL}$}
  	\psfrag{G}[c][][0.65]{~~CP150}
  	\psfrag{H}[c][][0.65]{~~CP550}
  	\includegraphics[width=0.5\textwidth]{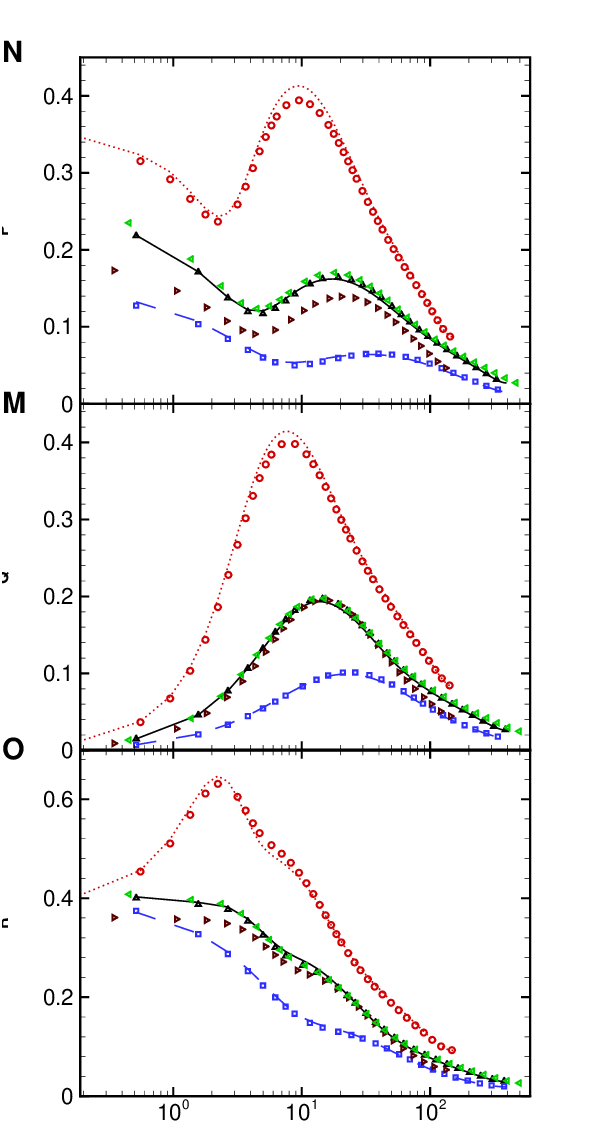}}~~~
  \subfigure{\label{fig:vortsl}
    \psfrag{N}[c][][0.9]{(b)}
    \psfrag{M}[c][][0.9]{(d)}
    \psfrag{O}[c][][0.9]{(f)}
  	\psfrag{P}[c][][0.9]{$({h}/{Re_\tau^\star})({\hat{\omega}^{\prime}_x})_{rms}$}
  	\psfrag{Q}[c][][0.9]{$({h}/{Re_\tau^\star})({\hat{\omega}^{\prime}_y})_{rms}$}
  	\psfrag{R}[c][][0.9]{$({h}/{Re_\tau^\star})({\hat{\omega}^{\prime}_z})_{rms}$}
  	\psfrag{X}[c][][0.9]{$y^\star$}
  	\psfrag{A}[l][][0.65]{CP395}
  	\psfrag{B}[l][][0.65]{CRe$^\star_\tau$}
  	\psfrag{C}[l][][0.65]{SRe$^\star_{\tau GL}$}
  	\psfrag{D}[l][][0.65]{GL}
  	\psfrag{E}[l][][0.65]{LL}
  	\psfrag{F}[l][][0.65]{SRe$^\star_{\tau LL}$}
  	\psfrag{G}[l][][0.65]{CP150}
  	\psfrag{H}[l][][0.65]{CP550}	
  	\includegraphics[width=0.5\textwidth]{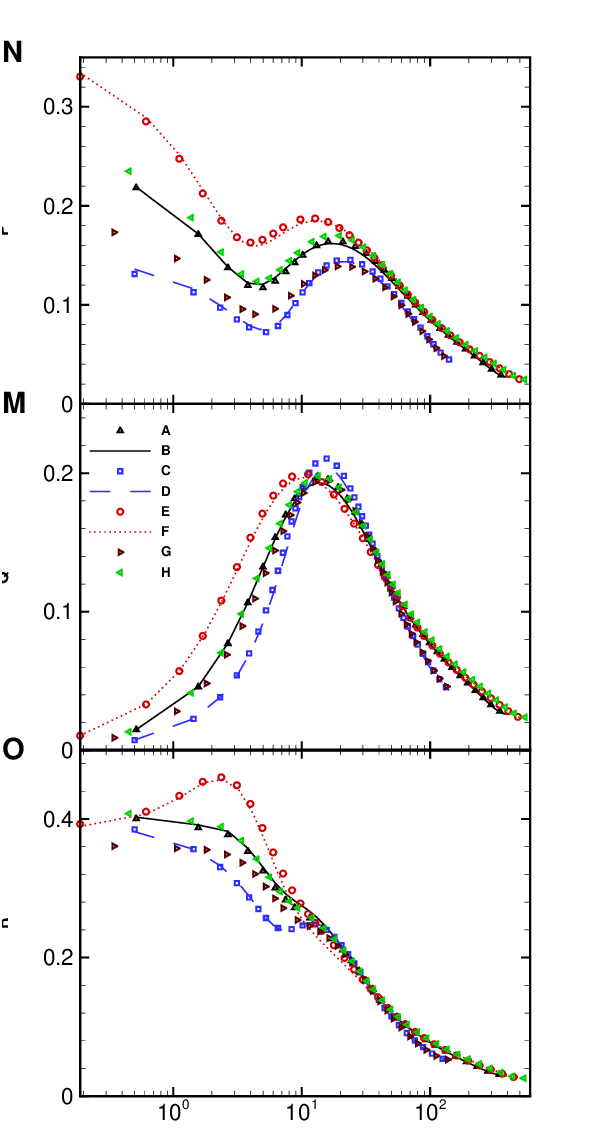}}
   \caption{{(Colour online)} {Root mean square $rms$ vorticity fluctuations of (a) and (b) streamwise $({\hat{\omega}^{\prime}_x})$, (c) and (d) wall-normal $({\hat{\omega}^{\prime}_y})$, (e) and (f) spanwise components $({\hat{\omega}^{\prime}_z})$, normalized by and plotted using wall based scales (left side) and semi-local scales (right side).}}
  \label{fig:vort}
\end{figure}
The $rms$ of the semi-locally scaled turbulent vorticity fluctuations $\pmb{\hat{\omega}}^{\prime}=\nabla \times \mathbf{\hat{u}}^{\prime}$ for streamwise ($\hat{\omega}^{\prime}_x$), wall-normal ($\hat{\omega}^{\prime}_y$) and spanwise ($\hat{\omega}^{\prime}_z$) directions are shown in figure \ref{fig:vort}. Figure~\ref{fig:vort}(a), \ref{fig:vort}(c) and \ref{fig:vort}(e) shows the vorticity normalised by the wall-based viscous scale $h/Re_\tau$ as a function of $y^+$, whereas the vorticity normalised by the semi-local viscous scale $h/Re_\tau^\star$ as a function of $y^\star$ is shown in figure~\ref{fig:vort}(b), \ref{fig:vort}(d) and \ref{fig:vort}(f). According to our previous work \citep{patel2015semi}, quasi-similar vorticity statistics are obtained for cases with quasi-similar ${Re_\tau^\star}$ profiles, even though the individual density and viscosity profiles differ. Comparing cases with different ${Re_\tau^\star}$ gradients in figure~\ref{fig:vort}(a), \ref{fig:vort}(c) and \ref{fig:vort}(e) shows considerable differences both in terms of magnitude and wall-normal location of peak values. On the other hand, the semi-local normalisation (figure~\ref{fig:vort}(b), \ref{fig:vort}(d) and \ref{fig:vort}(f)) provides a reasonable collapse in regions away from the wall for comparable ${Re_\tau^\star}_c$,  and it is also able to preserve the wall-normal locations of the peaks. 
However, in the near-wall region the semi-local scaling with $h/Re_\tau^\star$ fails and does not provide a collapse of the vorticity profiles, even though the van Driest transformed mean spanwise vorticity scales well with $h/Re_\tau^\star$ (see figure~\ref{fig:stress}(d)). In the following, we will make several comments on statistics of semi-locally normalised vorticity fluctuations. 
\begin{enumerate}
\item It is known that even for constant property cases the $x$ and $z$ vorticity fluctuation components increase with Reynolds number, whereas the $y$ component shows $Re_\tau$ independence \citep{antonia1994low}. This $Re_\tau$ dependence is particularly prominent for the low Reynolds number case CP150 (brown triangles {- colour online}). 
\item For the variable property cases, it is interesting to see that the cases with $d {Re_\tau^\star}/dy > 0$ have the highest magnitude of the semi-locally scaled $\hat{\omega}^{\prime}_x$ in the near-wall region, although their $Re_\tau$ values are the lowest ($Re_\tau=150$). The opposite occurs for cases with decreasing $Re_\tau^\star$ away from the wall. There, the streamwise vorticity fluctuations are the lowest, even for comparatively high values of $Re_\tau=395$. This is clearly a variable property effect that is in contrast to the Reynolds number dependence as discussed in point (i).  
Similar observations can be made for the $y$ and $z$ components. The semi-locally scaled rms values of $\hat{\omega}^{\prime}_y$ and $\hat{\omega}^{\prime}_z$ are the highest in the near-wall region for cases with $d {Re_\tau^\star}/dy > 0$ and vice versa. 
\item For cases with $d {Re_\tau^\star}/dy > 0$, the $z$ component has a maximum at approximately the same location where the van Driest transformed mean spanwise vorticity also shows a maximum. This is in agreement to the inflection point of the van Driest transformed mean velocity, which causes the flow to become more unstable as mentioned in section~\ref{sec:mod1}. For cases with $d {Re_\tau^\star}/dy < 0$ the fluctuations fall off rapidly away from the wall. 
\end{enumerate}

The failure of the semi-local scaling is presumably caused by both structural changes in turbulence and strong non local interactions between structures in the buffer- and the viscous sub-layer that will be discussed in detail in section~\ref{sec:str}.

\subsection{Turbulent stress anisotropy}
\begin{figure}
  \centering
  \subfigure{\label{fig:bii}
    \psfrag{N}[c][][0.9]{(a)}
  	\psfrag{Y}[c][][0.9]{${\widetilde{\hat{u}_i^{\prime} \hat{u}_i^{\prime}}}/{2k}-{1}/{3}$}
  	\psfrag{X}[c][][0.9]{$y^\star$}
  	\psfrag{A}[c][][0.65]{~~CP395}
  	\psfrag{B}[c][][0.65]{~~~~CRe$^\star_\tau$}
  	\psfrag{C}[c][][0.65]{~~~~~~SRe$^\star_{\tau GL}$}
  	\psfrag{D}[c][][0.65]{~GL}
  	\psfrag{E}[c][][0.65]{~LL}
  	\psfrag{F}[c][][0.65]{~~~~~~SRe$^\star_{\tau LL}$}
  	\psfrag{G}[c][][0.65]{~~CP150}
  	\psfrag{H}[c][][0.65]{~~CP550}
  	\psfrag{P}[c][][0.8]{~~$b_{11}$}
  	\psfrag{R}[c][][0.8]{~~$b_{22}$}
  	\psfrag{Q}[c][][0.8]{~~$b_{33}$}
  	\includegraphics[width=0.5\textwidth]{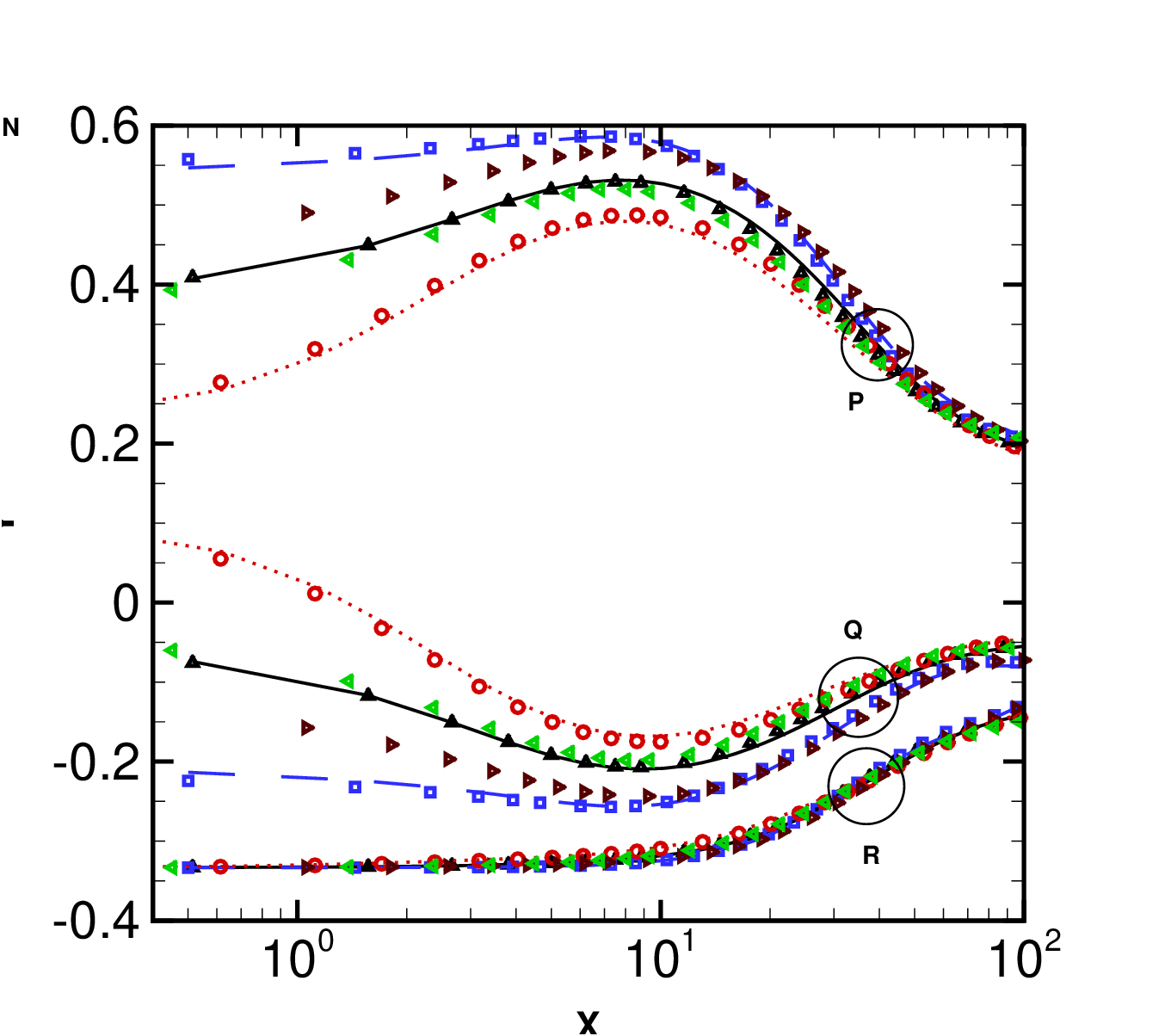}}~~~
  \subfigure{\label{fig:buw}
    \psfrag{N}[c][][0.9]{(b)}
  	\psfrag{Y}[c][][0.9]{${\widetilde{\hat{u}^{\prime} \hat{v}^{\prime}}}/{2k}$}
  	\psfrag{X}[c][][0.9]{$y^\star$}
  	\psfrag{A}[l][][0.65]{CP395}
  	\psfrag{B}[l][][0.65]{CRe$^\star_\tau$}
  	\psfrag{C}[l][][0.65]{SRe$^\star_{\tau GL}$}
  	\psfrag{D}[l][][0.65]{GL}
  	\psfrag{E}[l][][0.65]{LL}
  	\psfrag{F}[l][][0.65]{SRe$^\star_{\tau LL}$}
  	\psfrag{G}[l][][0.65]{CP150}
  	\psfrag{H}[l][][0.65]{CP550}
  	\includegraphics[width=0.5\textwidth]{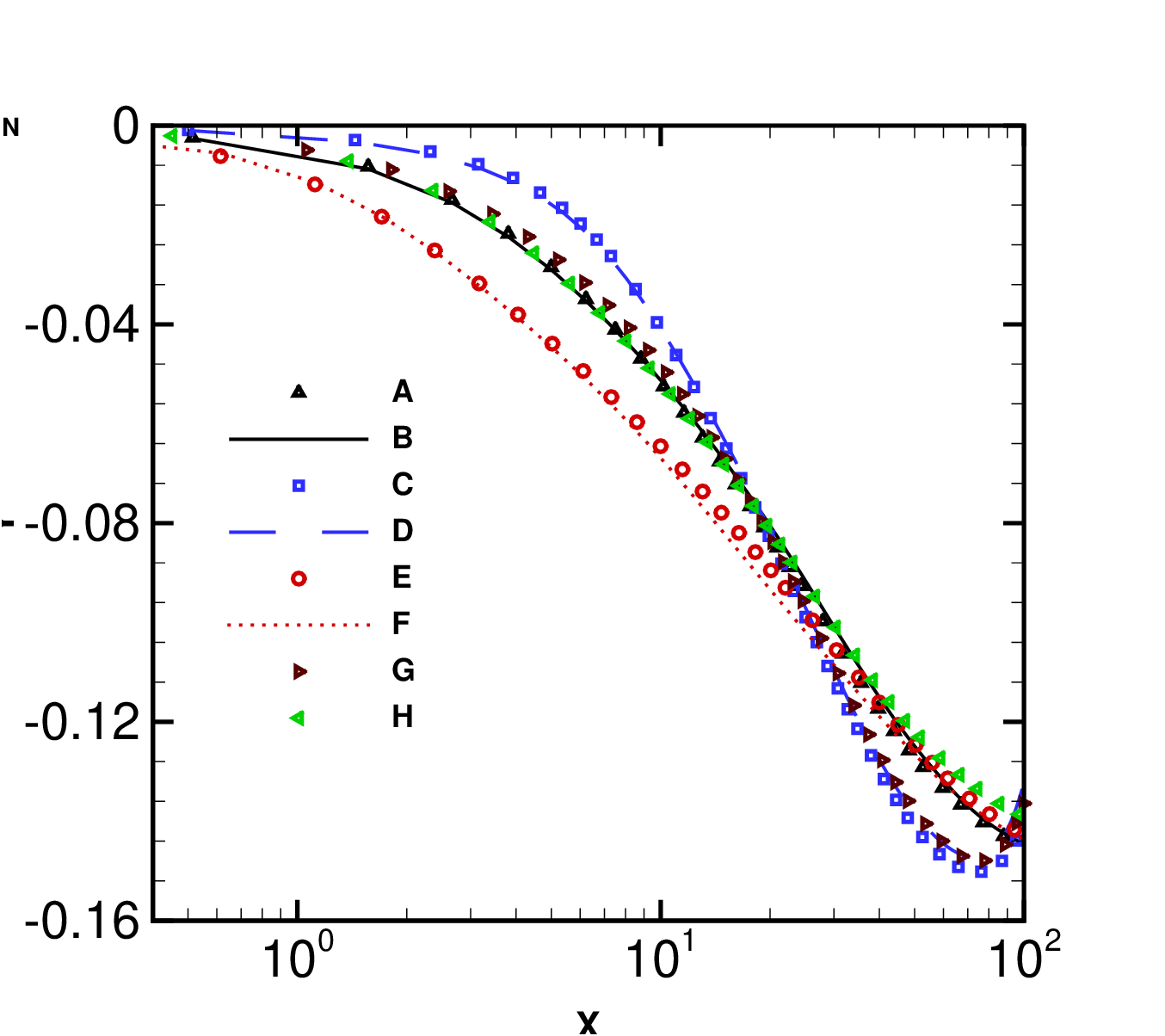}}
   \caption{{(Colour online)} {(a) Normal Reynolds stress anisotropies, (b) Reynolds shear stress anisotropy as a function of $y^\star$.}}
  \label{fig:bii_buw}
  \vspace*{-1em}
\end{figure}

\begin{figure}
  \centering
  \subfigure{\label{fig:bl_bii}
    \psfrag{N}[c][][0.9]{(a)}
  	\psfrag{Y}[c][][0.9]{${\widetilde{\hat{u}_i^{\prime} \hat{u}_i^{\prime}}}/{2k}-{1}/{3}$}
  	\psfrag{X}[c][][0.9]{$y^+$}
  	\psfrag{A}[c][][0.65]{~~CP395}
  	\psfrag{B}[c][][0.65]{~~~~CRe$^\star_\tau$}
  	\psfrag{C}[c][][0.65]{~~~~~~SRe$^\star_{\tau GL}$}
  	\psfrag{D}[c][][0.65]{~GL}
  	\psfrag{E}[c][][0.65]{~LL}
  	\psfrag{F}[c][][0.65]{~~~~~~SRe$^\star_{\tau LL}$}
  	\psfrag{G}[c][][0.65]{~~CP150}
  	\psfrag{H}[c][][0.65]{~~CP550}
  	\psfrag{P}[c][][0.8]{~~$b_{11}$}
  	\psfrag{R}[c][][0.8]{~~$b_{22}$}
  	\psfrag{Q}[c][][0.8]{~~$b_{33}$}
  	\includegraphics[width=0.5\textwidth]{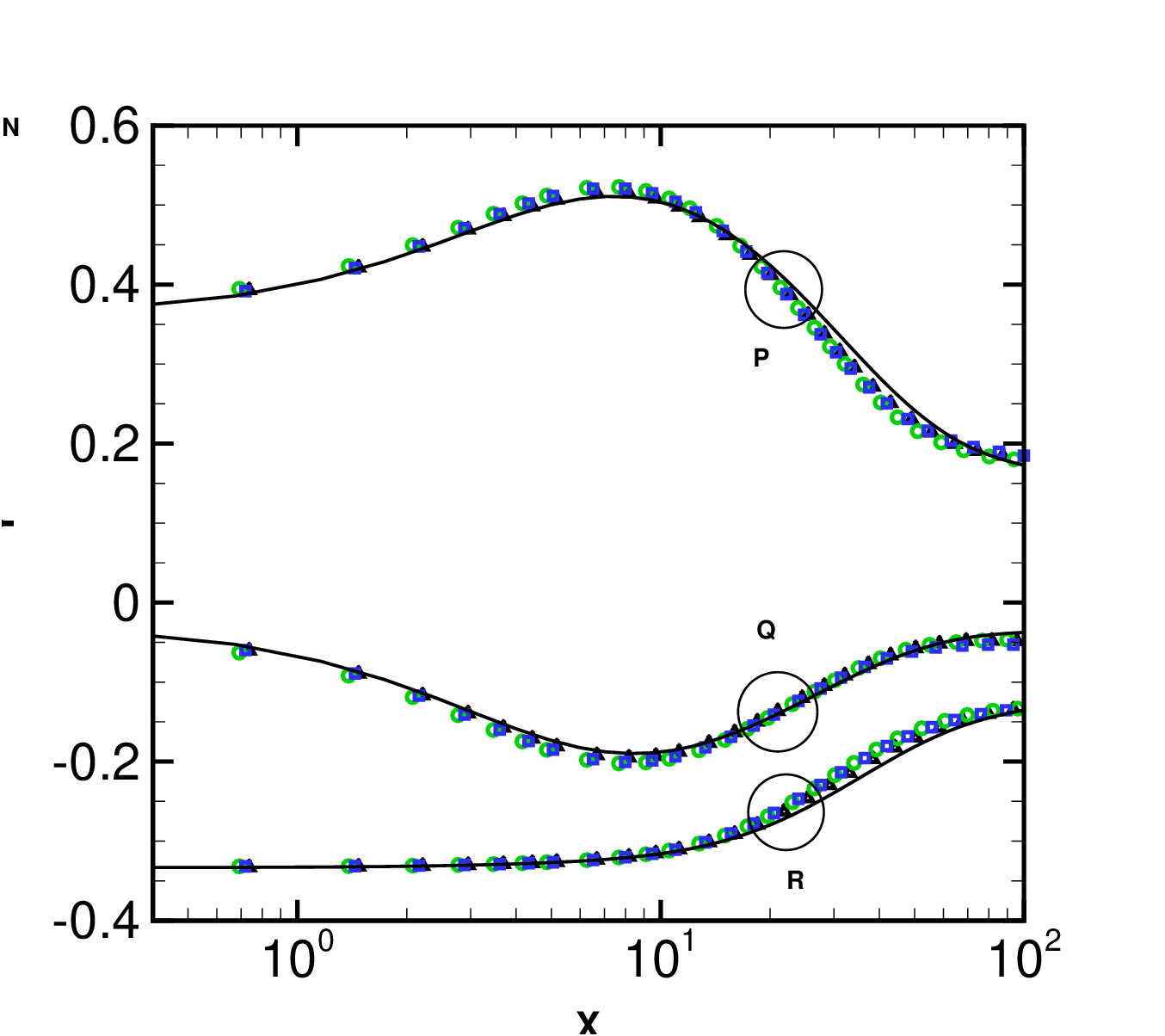}}~~~
  \subfigure{\label{fig:bl_buw}
    \psfrag{N}[c][][0.9]{(b)}
  	\psfrag{Y}[c][][0.9]{${\widetilde{\hat{u}^{\prime} \hat{v}^{\prime}}}/{2k}$}
  	\psfrag{X}[c][][0.9]{$y^+$}
  	\psfrag{A}[c][][0.65]{~~~Ma=2}
  	\psfrag{B}[c][][0.65]{~~~Ma=3}
  	\psfrag{C}[c][][0.65]{~~~Ma=4}
  	\psfrag{G}[c][][0.65]{~~~Ma=0}
  	\includegraphics[width=0.5\textwidth]{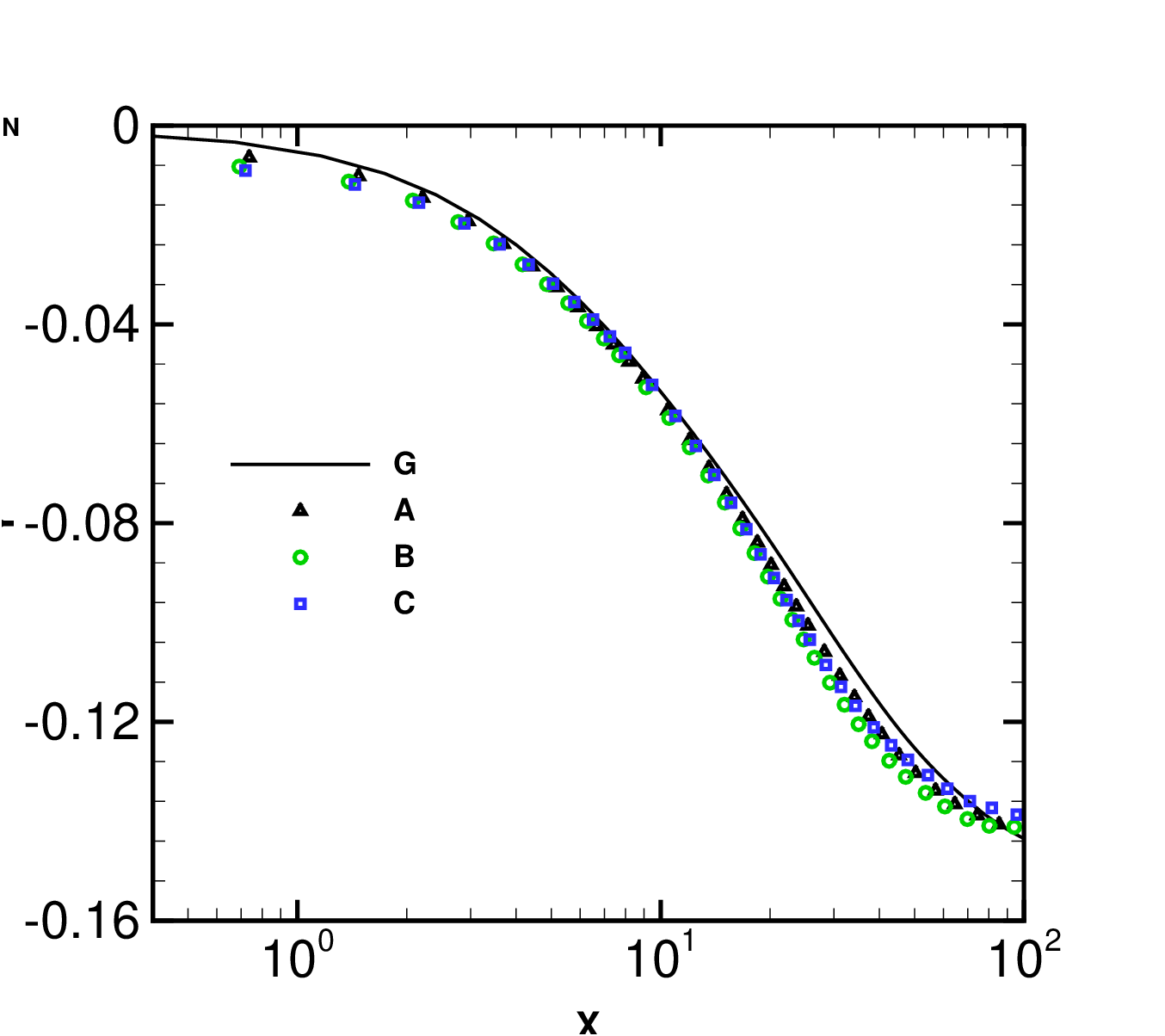}}
   \caption{{(Colour online)} {(a) Normal Reynolds stress anisotropies, (b) Reynolds shear stress anisotropy as a function of $y^+$ for adiabatic supersonic boundary layers obtained from \citet{pirozzoli2011turbulence,bernardini2011wall} and compared with reference boundary layer data from \citet{jimenez2010turbulent}.}}
  \label{fig:bl_bii_buw}
  \vspace*{-1em}
\end{figure}

As described earlier, and as discussed in our previous work \citep{patel2015semi}, gradients of $Re_\tau^\star$ have a large effect on inter-component energy transfer and thus on turbulence anisotropy. The anisotropy tensor is defined as,
\begin{equation}
b_{ij}=\frac{\widetilde{\hat{u}_i^{\prime} \hat{u}_j^{\prime}}}{2k}-\delta_{ij}\frac{1}{3}~,
\end{equation}
with the turbulent kinetic energy $k=\widetilde{\hat{u}_k^{\prime} \hat{u}_k^{\prime}}/2$ and  $\delta_{ij}$ the Kronecker delta. Figure~\ref{fig:bii} shows that for cases with $d {Re_\tau^\star}/dy < 0$ an increase in streamwise component $b_{11}$ and a decrease in spanwise component $b_{33}$ occurs in the near-wall region if compared to cases with $d {Re_\tau^\star}/dy = 0$. The opposite is seen for cases with $d{Re_\tau^\star}/dy>0$. The wall-normal component $b_{22}$, however, is not influenced  by gradients in $Re_\tau^\star$. The increase in $b_{11}$ for cases with $d {Re_\tau^\star}/dy < 0$ indicates  a decrease in the redistribution of turbulent energy from the streamwise direction to the other two directions. The component $b_{12}$, which is the ratio of turbulent shear stress and turbulent kinetic energy, is shown in figure~\ref{fig:buw}. It decreases for cases with $d {Re_\tau^\star}/dy < 0$, indicating a reduced momentum transfer in spite of higher turbulent kinetic energy, while the reverse is true for cases with $d {Re_\tau^\star}/dy > 0$.  
The same quantities are also shown in figure~\ref{fig:bl_bii_buw} for the adiabatic supersonic turbulent boundary layer cases from \citet{pirozzoli2011turbulence,bernardini2011wall} and compared with reference boundary layer data from \citet{jimenez2010turbulent}. No significant changes in anisotropy are noticeable since $d {Re_\tau^\star}/dy \approx 0$ in the near-wall region. The change in anisotropy for cases with $d {Re_\tau^\star}/dy \neq 0$ is linked to modifications in turbulent structures that will be discussed next.

\section{Turbulent structures}\label{sec:str}
Here we investigate the influence of near-wall $Re_\tau^\star$ gradients on the characteristics of near-wall streaks and quasi-streamwise vortices, which are both known to be the dominant structures in near-wall turbulence, in order to provide a mechanistic description for the modulated statistics. 

\subsection{Near-wall streaks}
\begin{figure}
  \centering
  \subfigure{\label{fig:specya}
    \psfrag{O}[c][][0.9]{(a)}
    \psfrag{P}[c][][0.9]{}
  	\psfrag{Y}[c][][0.9]{$k_zE_{\hat{u}^{\prime} \hat{u}^{\prime}}/\overline{\hat{u}^{\prime} \hat{u}^{\prime}}$}
  	\psfrag{Z}[c][][0.9]{$k_zE_{\hat{u}^{\prime} \hat{u}^{\prime}}/\overline{\hat{u}^{\prime} \hat{u}^{\prime}}$}
  	\psfrag{X}[c][][0.9]{$\lambda_z^\star$}
  	\psfrag{A}[l][][0.65]{CP395}
  	\psfrag{B}[l][][0.65]{CRe$^\star_\tau$}
  	\psfrag{C}[l][][0.65]{SRe$^\star_{\tau GL}$}
  	\psfrag{D}[l][][0.65]{GL}
  	\psfrag{E}[l][][0.65]{LL}
  	\psfrag{F}[l][][0.65]{SRe$^\star_{\tau LL}$}
  	\psfrag{G}[l][][0.65]{CP150}
  	\psfrag{H}[l][][0.65]{CP550}
  	\includegraphics[width=0.5\textwidth]{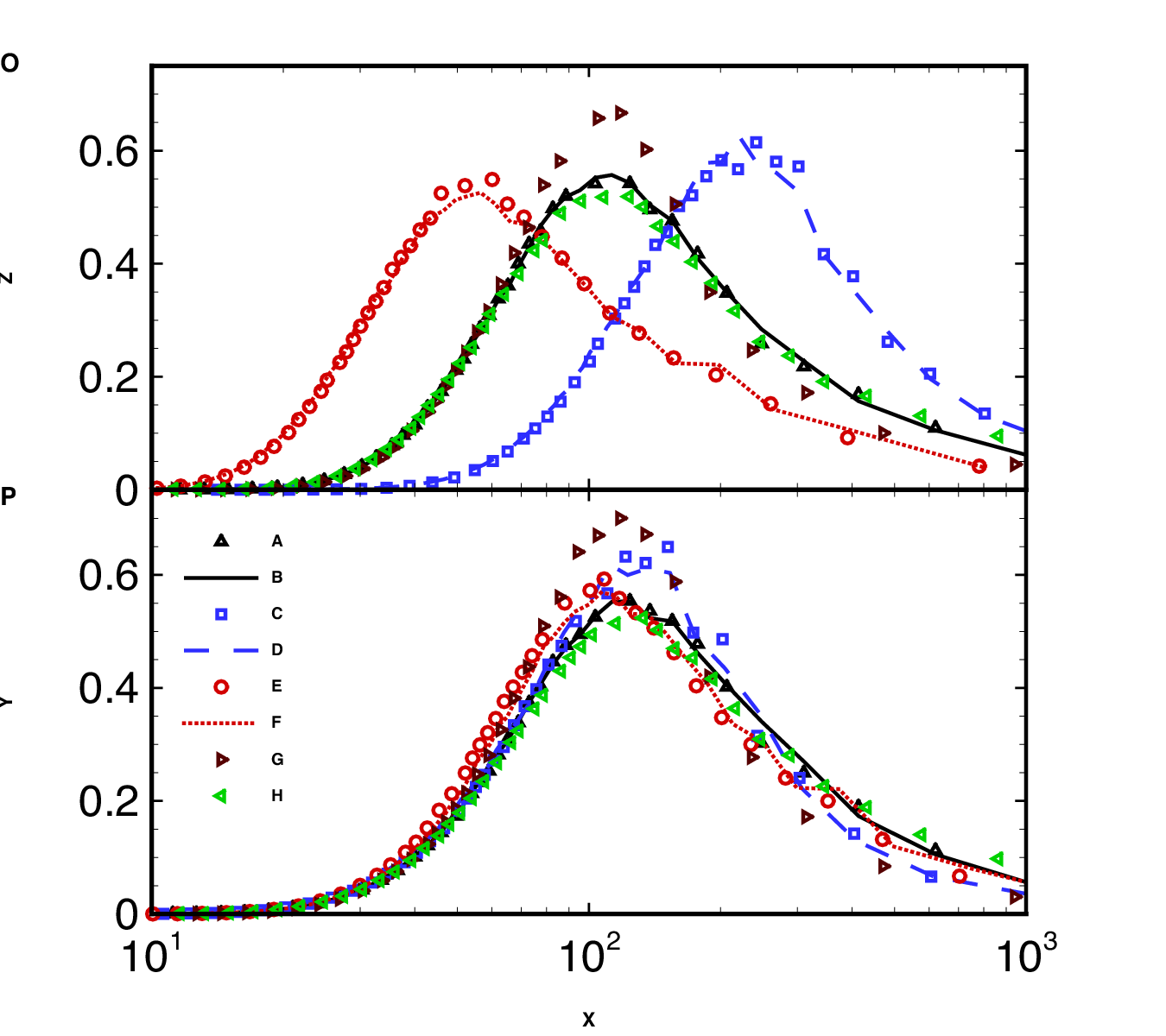}}~\hspace*{-1.2em}
  \subfigure{\label{fig:specyc}
    \psfrag{O}[c][][0.9]{~~~$d Re_\tau^\star/dy<0$}
    \psfrag{P}[c][][0.9]{~~~$d Re_\tau^\star/dy=0$}
    \psfrag{R}[c][][0.9]{~~~$d Re_\tau^\star/dy>0$}  
    \psfrag{N}[c][][0.9]{(b)}
  	\psfrag{Y}[c][][0.9]{$\lambda_z^\star$}
  	\psfrag{X}[c][][0.9]{$y^\star$}		
  	\includegraphics[width=0.525\textwidth]{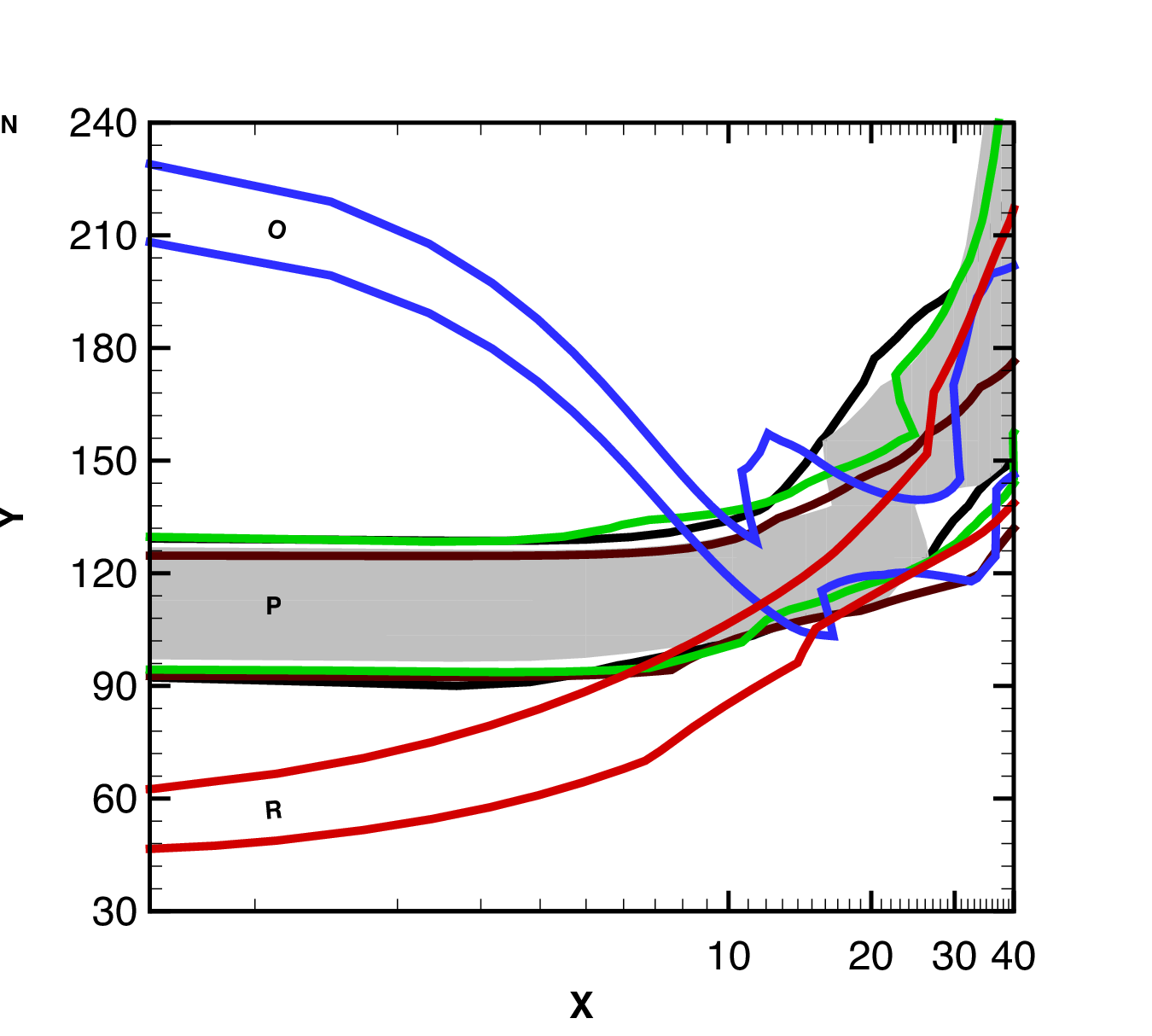}}
\caption{{(Colour online)} (a) Normalised pre-multiplied spanwise spectra $k_zE_{\hat{u}^{\prime} \hat{u}^{\prime}}/\overline{\hat{u}^{\prime} \hat{u}^{\prime}}$ as a function of $\lambda_z^\star$ at $y^\star\approx0.5$ (top) and $y^\star\approx13$ (bottom). (b) {$\lambda_z^\star$ as a function of $y^\star$ obtained using bands of  $k_zE_{\hat{u}^{\prime} \hat{u}^{\prime}}/\overline{\hat{u}^{\prime} \hat{u}^{\prime}}$ with values larger than 96\% of its maximum value; black, brown, green, blue and red lines correspond to case CP395, CP150, CP550, GL and LL, respectively; the grey region corresponds to case CRe$^\star_\tau$.}}
  \label{fig:specy}
  \vspace*{-1em}
\end{figure}

For constant property cases, the near-wall streak spacing in the viscous sub-layer remains remarkably constant over a wide range of Reynolds numbers \citep{klewicki1995viscous}. Figure \ref{fig:specya} shows the normalised pre-multiplied spanwise spectra of the semi-locally scaled streamwise velocity fluctuation  $k_zE_{\hat{u}^{\prime} \hat{u}^{\prime}}/\overline{\hat{u}^{\prime} \hat{u}^{\prime}}$ at $y^\star\approx 0.5$ (top) and $y^\star\approx 13$ (bottom) as a function of semi-locally scaled wavelength $\lambda_z^\star=\lambda_z Re_\tau^\star/h$. {The peak location of the spectra represents the mean streak spacing and it can be seen that at $y^\star\approx 13$ the spacing is the same for all cases ($\lambda_z^\star \approx 120$ ).} This is one of the few similarities that variable property and constant property cases share. On the other hand, at $y^\star\approx 0.5$ only the constant property cases (CP395, CP150, CP550) and the variable property case CRe$^\star_\tau$ show a similar streak spacing of $\lambda_z^\star=\lambda_z^+ \approx 110$, whereas  the variable property cases with $d {Re_\tau^\star}/dy \neq 0$ show a modulation in streak spacing. The cases with $d {Re_\tau^\star}/dy <0$ (blue lines and symbols {- colour online}) show an increased streak spacing of $\lambda_z^\star \approx 220$, while $\lambda_z^\star \approx 50$ for cases with $d {Re_\tau^\star}/dy >0$ (red lines and symbols {- colour online}). Figure~\ref{fig:specyc} gives an overview of the near-wall streak spacing as a function of $y^\star$. 
The lines in the plot indicate the boundaries where $k_zE_{\hat{u}^{\prime} \hat{u}^{\prime}}/\overline{\hat{u}^{\prime} \hat{u}^{\prime}}$ is 96\% of the peak value at a certain $y^\star$ location. It can be seen that the mean streak spacing appears to become universal after $y^\star\approx 12-13$, while it deviates significantly in the viscous sub-layer for cases with $d {Re_\tau^\star}/dy \neq 0$. 
This deviation is not surprising, because of the fact that the sub-layer flow is known to be induced by advecting dominant structures in the buffer layer \citep{kim1993propagation}. The wall-normal location of the dominant structures is also found to be universal in semi-local units, as shown by \citet{pei2013new}, using a vorticity-velocity correlation in supersonic channel flows with isothermal walls at different Mach numbers. The strong non-local influence of turbulence structures in the buffer layer on the near-wall region creates a disparity between the semi-local scales and the actual turbulent scales in the sub-layer. 

\subsection{Vortical structures}
The three-dimensional swirling strength $\Lambda_{ci}(x,y,z)$, which is based on the imaginary part of the complex eigenvalue of the velocity gradient tensor \citep{zhou1999mechanisms}, is used to identify the near-wall vortical structures. The swirling strength separates swirling- from shearing motion, and it can be evaluated using the gradient tensor of the instantaneous- or the fluctuating velocity field. Note, it has been analytically shown that it is not possible to decouple the mean shear from the instantaneous field \citep{chen2014analytic}. In our investigations we evaluate the swirling strength using the instantaneous velocity gradient tensor, because the local shear depends on the instantaneous field. Similar to \citet{wu2006population}, we normalise the local swirling strength with its corresponding wall-normal root-mean-square $\Lambda^{rms}_{ci}(y)$ value, such that $\acute{\Lambda}_{ci}(x,y,z)=\Lambda_{ci}(x,y,z)/\Lambda^{rms}_{ci}(y)$.   

\begin{figure}
  \centering
  \subfigure{\label{fig:strtop}
  \psfrag{A}[c][][0.9]{(a)}
  	\includegraphics[trim = 90 0 90 0, clip,width=0.55\textwidth,angle=-90]{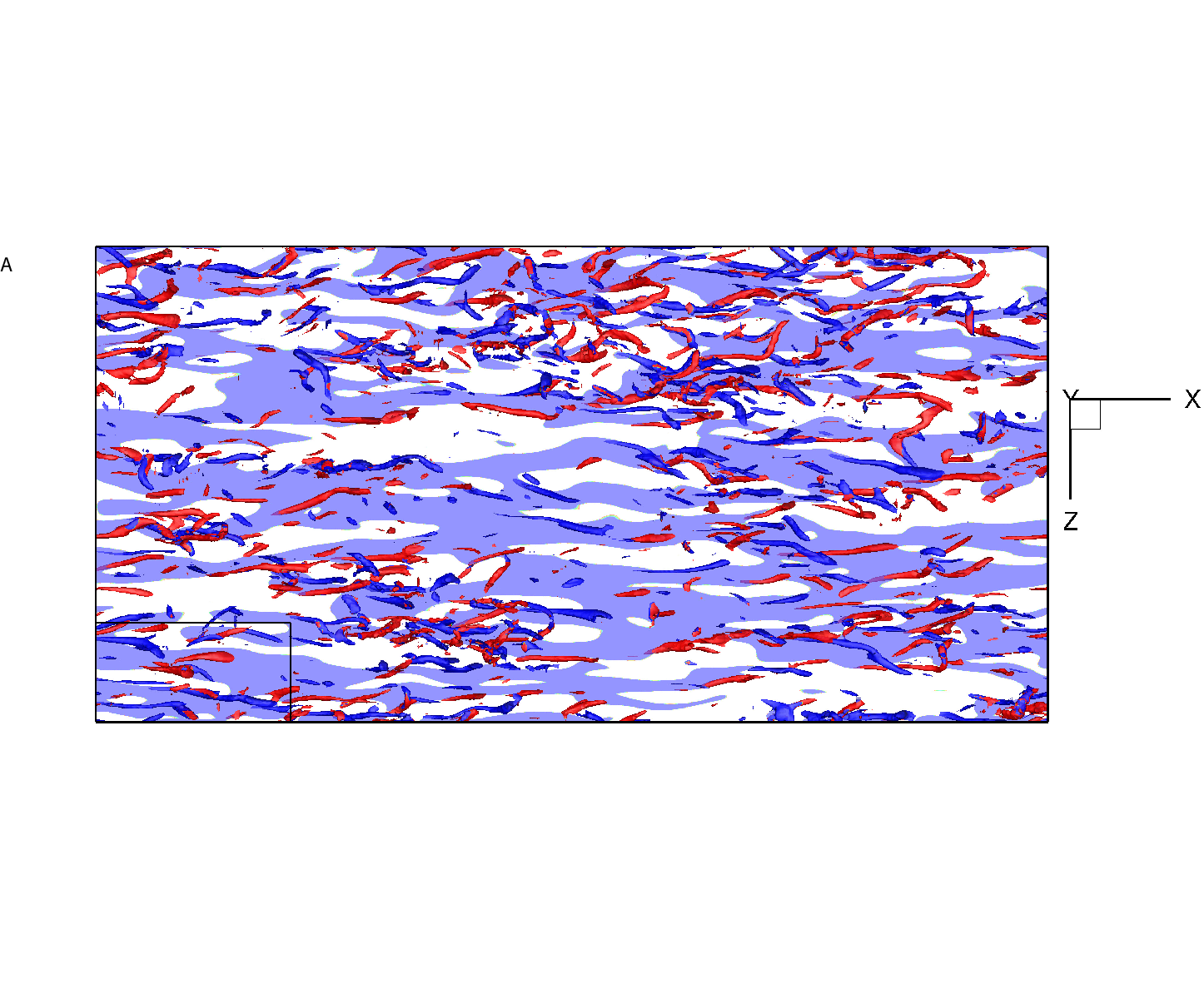}}
  	\vspace*{-4.5em}
  \subfigure{\label{fig:striso}
    \psfrag{B}[c][][0.9]{(b)}
  	\includegraphics[trim = 120 10 0 30, clip,width=0.65\textwidth,angle=-90]{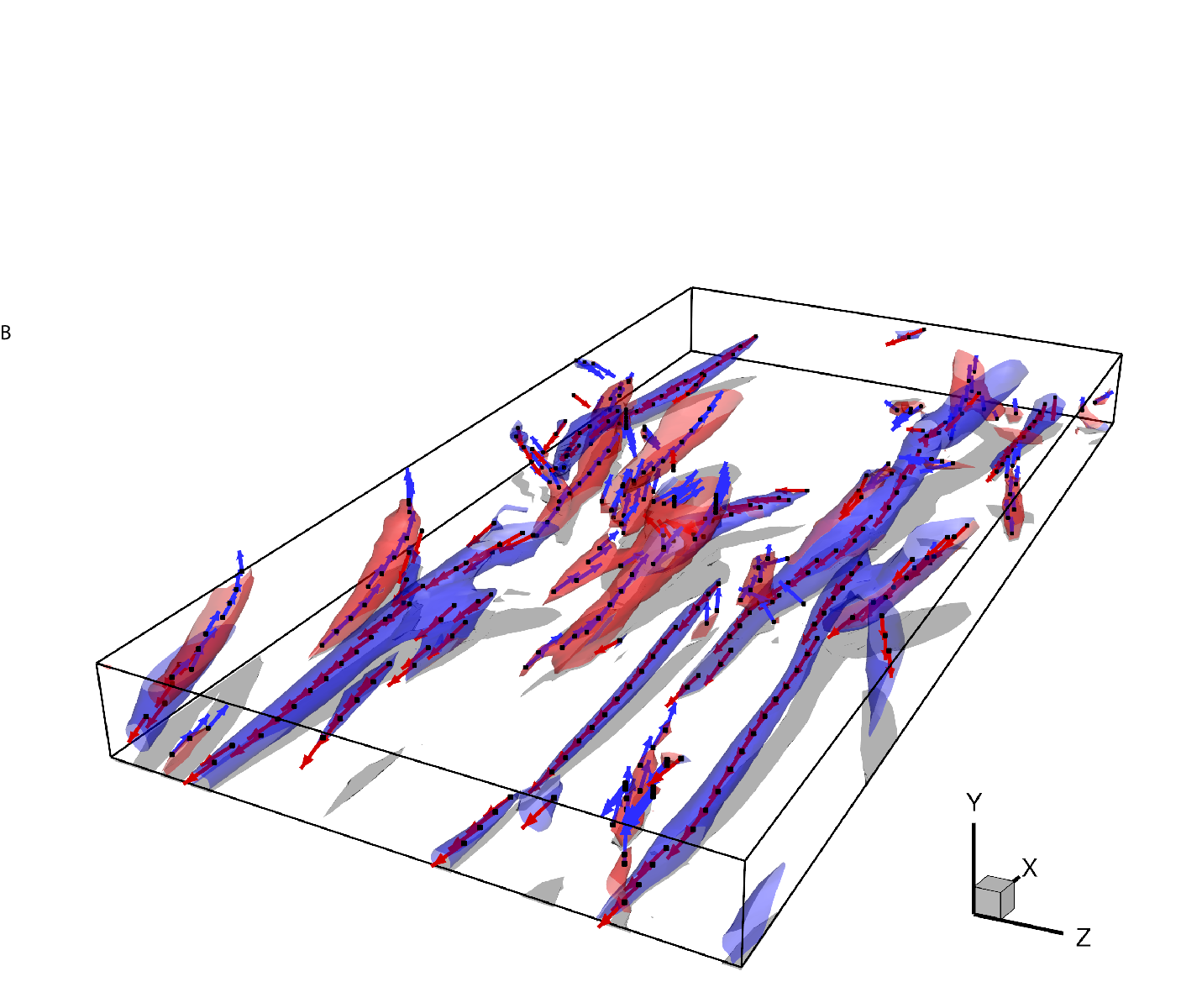}}
  \vspace*{-2em}
   \caption{{(Colour online)} (a) Iso-surfaces of swirling strength seen from top, red iso-surfaces correspond to positive voriticity while blue denotes negative vorticity; contours show the low speed-streaks; rectangular box  corresponds to figure b  (b) isometric view of structures with real eigen-vectors at the vortex centre- the grey color corresponds to projection of the structures at wall, vectors corresponding to positive vorticity shown as blue while red denotes negative vorticity.}
  \label{fig:str}
  \vspace*{-1em}
\end{figure}

Figure \ref{fig:strtop} shows the top-view of an iso-surface of $\acute{\Lambda}_{ci}=1.5$ for case CP395 in the near-wall region up to $y^+\approx50$. As stated in previous studies \citep{jeong1997coherent,robinson1991coherent}, the near-wall region is mostly populated by quasi-streamwise vortices. The iso-surfaces are coloured by the sign of their streamwise vorticity, where red denotes positive and blue negative vorticity (colour online). The figure also shows a slice of low-speed streaks at $y^+=13$ (seen as light blue - colour online). A small section (black box) of the same instantaneous flow field is shown in an isometric view in figure \ref{fig:striso}, in order to show the inclination and elevation of these structures with respect to the wall (the grey shades are projections of the iso-surfaces onto the wall). As noted by \citet{jeong1997coherent}, the structures with positive vorticity tend to tilt in negative direction, and those with negative vorticity tend to tilt in positive direction with respect to the streamwise direction within the x-z plane. This tilting is {correlated} with the waviness of streaks and it is {associated} with transferring of streamwise turbulent energy to spanwise and wall-normal components \citep{jeong1997coherent}. Therefore, studying the orientation of these structures can clarify the changes in anisotropy that occur in flows with  gradients in $Re_\tau^\star$. 

The orientation of the vortical structures can be characterised by the real eigenvector $\nu_r$ of the velocity gradient tensor, which is able to differentiate the swirling direction from the vorticity direction \citep{gao2011analysis, pirozzoli2008characterization}. However, before evaluating the vortex orientation we must first find the vortex centre. This is done by performing the following steps. First, all the grid points that correspond to the local maxima of the swirling strength $\Lambda_{ci}$ are flagged in all $y-z$ planes of the numerical domain. Next the real eigenvector $\nu_r$ of these points is used to check if the projection angles in the perpendicular planes, which define the lift and the tilt and are denoted as $\theta_{xy}$ and $\theta_{xz}$, are within $\pm 45^{\circ}$. If this condition is met, the points are retained, otherwise they are discarded. These steps are repeated to also find the vortex centres in the $z-x$ and $x-y$ planes of the computational domain. Finally, only the points with $\acute{\Lambda}_{ci} \geq 1.5$ are kept and used as the vortex centres. Note, most of vortex centres were found in the $y-z$ plane, showing the dominance of quasi-streamwise vortices. 
An outcome of this eduction procedure is given in figure \ref{fig:striso}, where the vortex centres with  their corresponding eigenvectors are shown. Since eigenvectors can have either of the two opposing directions, its positive direction is chosen such that the dot product with the the vorticity vector is positive. As evident, the eigenvectors $\nu_r$ provide an excellent measure of the orientation axis of the vortical structures. We will now apply this procedure to obtain statistics of the inclination and tilting angle to draw conclusions on turbulence modulation for all constant and variable property cases investigated in this work. 

\begin{figure}
  \centering
  \subfigure{\label{fig:pdfik}
    \psfrag{N}[c][][0.9]{(a)}
  	\psfrag{Y}[c][][0.9]{PDF}
  	\psfrag{X}[c][][0.9]{$\theta_{xy}$}
  	\psfrag{A}[l][][0.65]{CP395}
  	\psfrag{B}[l][][0.65]{CRe$^\star_\tau$}
  	\psfrag{C}[l][][0.65]{SRe$^\star_{\tau GL}$}
  	\psfrag{D}[l][][0.65]{GL}
  	\psfrag{E}[l][][0.65]{LL}
  	\psfrag{F}[l][][0.65]{SRe$^\star_{\tau LL}$}
  	\psfrag{G}[l][][0.65]{CP150}
  	\psfrag{H}[l][][0.65]{CP550}
  	\includegraphics[width=0.52\textwidth]{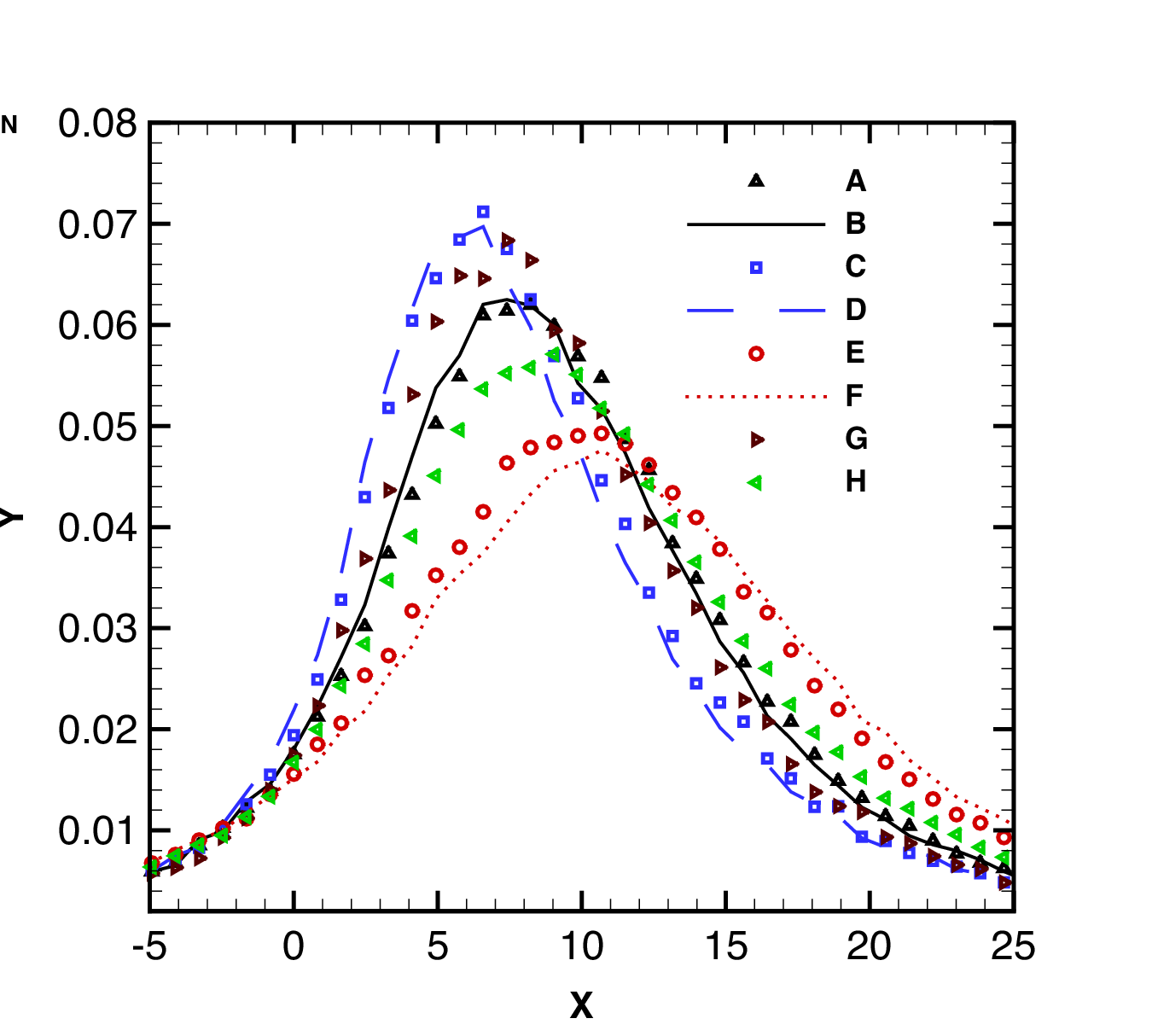}}~\hspace*{-1.2em}
  \subfigure{\label{fig:pdfjk}
    \psfrag{N}[c][][0.9]{(b)}
  	\psfrag{Y}[c][][0.9]{PDF}
  	\psfrag{X}[c][][0.9]{$\theta_{xz}$}
  	\includegraphics[width=0.52\textwidth]{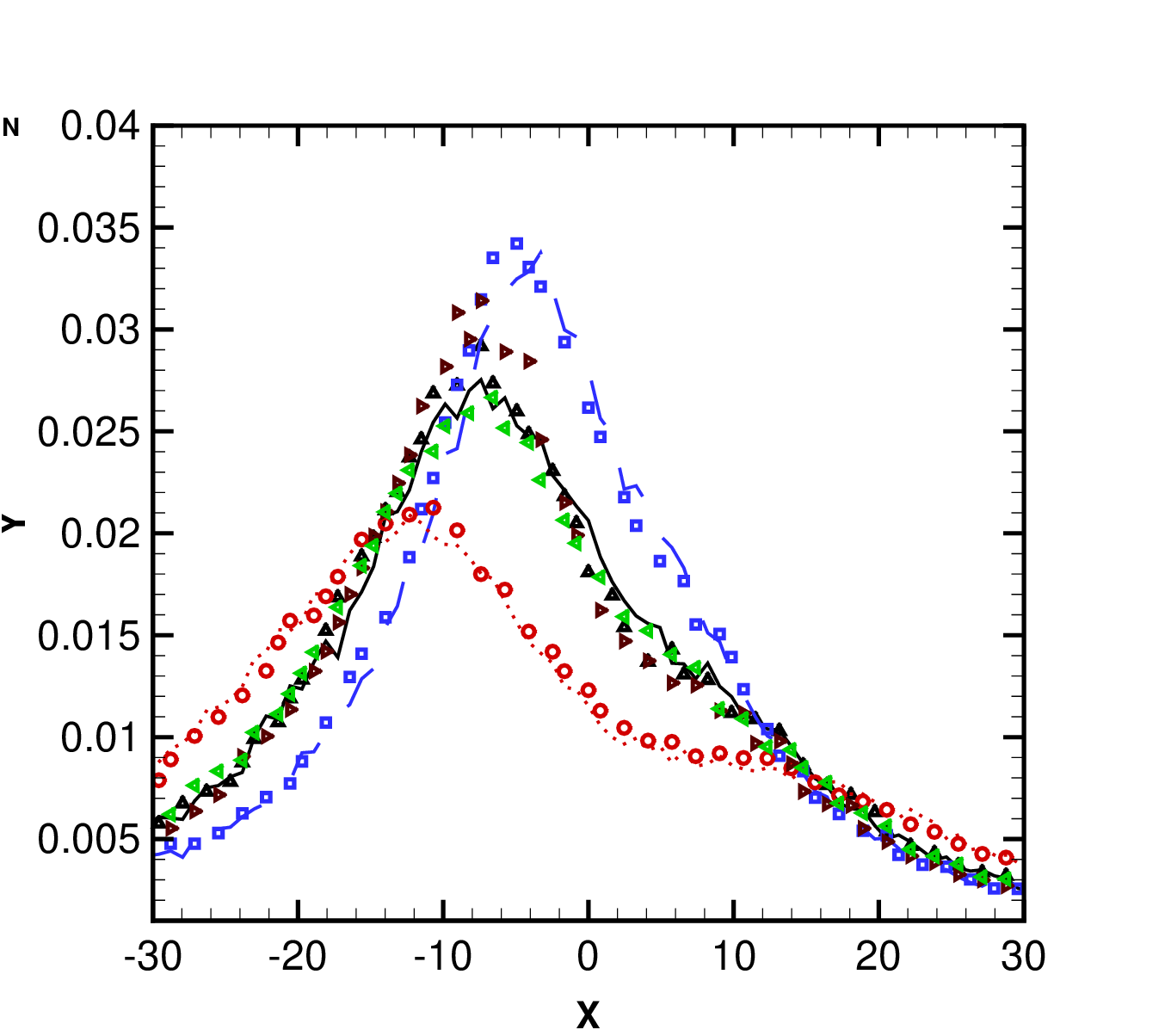}}
   \caption{{(Colour online)} Probability density function of projection angle of eigenvector $\nu_r$ at $y^\star=13$ (a) inclination angle $\theta_{xy}$ (b) tilting angle $\theta_{xz}$.}
  \label{fig:pdf}
  \vspace*{-1em}
\end{figure}

The probablity-density-functions ($pdf$) of the inclination $\theta_{xy}$ and tilting angle $\theta_{xz}$ of the vortical structures at $y^\star=13$ are shown in figure~\ref{fig:pdf}. Similar to turbulence statistics, the orientation of vortical structures for cases with quasi-similar Re$^\star_\tau$ profiles also shows quasi-similarity and is therefore independent of individual density and viscosity profiles. As shown in figure~\ref{fig:pdfik}, an increase in Re$_\tau$ from 150 to 550 for the constant property cases, increases the mode of the $pdf$ from 7$^{\circ}$ to 8$^{\circ}$ and broadens the $pdf$ showing an increase in standard deviation. The variable property cases with $dRe_\tau^\star/d y < 0$ (blue line and symbols {- colour online}) show a decrease in the mode of the $pdf$ (6$^{\circ}$) and a decrease in standard deviation. The opposite is seen for cases with $d Re_\tau^\star/d y > 0$ (red line and symbols {- colour online}), which show an increase in the mode (10$^{\circ}$) and a broadening of the distribution. These results show that lifting of coherent vortical structures reduces for cases with $d Re_\tau^\star/d y < 0$, and increases for $d Re_\tau^\star/d y > 0$. This is in agreement with our previous findings \citep{patel2015semi}, where we observed that the streaks were stabilised and do not lift as intensely for cases with $d Re_\tau^\star/d y < 0$, while they lift more intensely for cases with $d Re_\tau^\star/d y > 0$. 

The tilting of the structures is indicated in figure \ref{fig:pdfjk}. Because the direction of the tilting is coupled to the sign of the streamwise vorticity, and the events related to positive and negative streamwise vorticity are symmetric, we only show tilting angles for positive streamwise vorticity. Only small changes can be seen for the constant property cases with the mode at approximately $\pm$ 8$^{\circ}$. For cases with $d Re_\tau^\star/d y < 0$, the mode of the $pdf$'s occurs at $\pm$ 5$^{\circ}$ and the distributions narrow.  The reverse happens for cases with $d Re_\tau^\star/d y > 0$, with the mode at $\pm$ 11$^{\circ}$ and broader distributions. 

We will now summarise the effects related to flows with variable properties and provide a mechanistic description to explain the observed turbulence modulations. In order to avoid switching between different cases, we will only use cases with $d Re_\tau^\star/d y > 0$ (the effects for cases with $d Re_\tau^\star/d y < 0$ are exactly opposite). The increase in magnitude of van Driest transformed mean spanwise vorticity $-d \overline{u} ^{\mathrm{vD}}/dy$ increases the mean forcing in  spanwise direction, causing the increased tilting of vortical structures. The increased tilting results in stronger streak waviness, which then leads to an increased turbulence activity, enabling the evolution of near-wall structures and the generation of stronger shear layers \citep{johansson1991evolution}. The increased lifting of both streaks and vortical structures is therefore closely associated with an increase in tilting of the structures. This also explains the increase in shear stress anisotropy, which increases momentum transfer in spite of lower turbulent kinetic energy. Furthermore, the increase in tilting angle also provides a structural interpretation for the increase in the pressure-strain correlation, which acts as a sink in the budget of the streamwise energy equation and therefore redistributes the streamwise turbulent energy in the other two directions. From \citet{jeong1997coherent} it is known that the preferential alignment of structures with respect to streamwise vorticity produces positive values of $\partial u^{\prime}/\partial x$ within the structure and since $p^{\prime}$ is negative within the structure, $p^{\prime}\partial u^{\prime}/\partial x$ is also negative. Additionally, the spanwise asymmetry results in internal shear layers, where high-speed fluid collides with low-speed fluid, causing a positive $p^{\prime}$ and a negative $\partial u^{\prime}/\partial x$, and hence in  $p^{\prime}\partial u^{\prime}/\partial x<0$. The modulation in tilting of the structures therefore provides a physical interpretation of the modulated turbulence statistics.

\section{Conclusion}
Effects of strong near-wall gradients in density and viscosity on the mean velocity scaling, near-wall turbulence statistics and turbulent structures are studied by performing DNS of a fully developed channel flow under the low Mach number approximation of the Navier--Stokes equation. Five variable property cases (CRe$^\star_\tau$, SRe$^\star_{\tau GL}$, GL, LL, SRe$^\star_{\tau LL}$) with different relations for density and viscosity as a function of temperature are investigated. The different density and viscosity profiles are parametrised by means of the semi-local Reynolds number $Re_\tau^\star \equiv Re_\tau\sqrt{(\overline{\rho}/{\overline{\rho}_w})}/(\overline{\mu}/{\overline{\mu}_w})$, which is known from \citet{patel2015semi} to be the governing parameter for turbulence statistics. The case CRe$^\star_\tau$ corresponds to a constant wall-normal Re$_\tau^\star$ profile; for cases SRe$^\star_{\tau GL}$ and GL the Re$_\tau^\star$ profile decreases towards the channel centre ($d Re_\tau^\star/d y < 0$), and for cases LL and SRe$^\star_{\tau LL}$ Re$_\tau^\star$ increases towards the channel centre  ($d Re_\tau^\star/d y > 0$). Three constant property cases (CP395,CP150, CP550) at different $Re_\tau$ values are also performed to distinguish any Reynolds number effects with effects caused by property gradients. 

Strong near-wall gradients in $Re_\tau^\star$ result in a failure to collapse the van Driest transformed mean velocity $\overline{u}^{\mathrm{vD}}$ as a function of $y^+$. {An extension of the van Driest transformation that accounts for the gradients in $Re_\tau^\star$ $(d \overline{u} ^{\star}=\left(1+\left( {y}/{Re_\tau^\star}\right){dRe_\tau^\star}/{dy} \right) d \overline{u} ^{\mathrm{vD}})$ is derived based on the compelling collapse of the viscous stresses $(h/Re_\tau^\star) d\overline{u}^{\mathrm{vD}}/dy$ for all cases when plotted as a function of the semi-local wall coordinate $y^\star$.} {A successful collapse of $\overline{u} ^{\star}$ when plotted as a function of $y^\star$ was obtained for all the investigated cases.} The applicability of the transformation was also tested on adiabatic supersonic boundary layers of \cite{bernardini2011wall,pirozzoli2011turbulence}, {which} are known to show an excellent agreement of $\overline{u}^{\mathrm{vD}}$ as a function of $y^+$. The new transformed velocity collapses supersonic cases at different Mach numbers. However, the supersonic cases showed a small increase in the log-law additive constant when compared to an incompressible turbulent boundary layer. In a recent work \citet{trettel2016} derived a similar transformation in terms of density and viscosity gradients, and applied it successfully to supersonic channel flows with isothermal cooled walls.

Near-wall gradients in $Re_\tau^\star$ also result in turbulence modification when compared to constant property cases with similar $Re_\tau$ values. Partial success in accounting for this change in turbulence is obtained using the semi-local scaling, which accommodates the changes in viscous scales using local fluid properties. The success of the semi-local scaling is evident for the profiles of Reynolds shear stress and viscous shear stress. However, statistics like $rms$ of vorticity fluctuations, which are sensitive to the strong non-local interactions of the buffer layer vortical structures with the viscous sub-layer, show a poor collapse in the near-wall region using the semi-local scaling. This strong non-local interactions are evident from the semi-locally scaled streak spacing that tends to become universal after $y^\star\approx 12-13$, while it deviates significantly in the viscous sub-layer for cases with $d {Re_\tau^\star}/dy \neq 0$. Furthermore, the failure of the semi-local scaling also occurs due to structural changes in turbulence that affect lifting and tilting of quasi-streamwise vortices. These changes influence the Reynolds stress generation mechanism and the inter-component energy transfer for turbulent stresses. The influence of this failure on the scaling of the $\overline {u}^\star$ profile is negligible, since similar to the Reynolds shear stress the mixing length is modified in the viscosity dominated region and therefore has a negligible influence on the transformed velocity profile. 

The influence of $Re_\tau^\star$ gradients on vortical structures was studied using the three-dimensional swirling strength based on the instantaneous velocity gradient tensor. The orientation of these structures was determined using real eigenvector of the tensor at the vortex centre. The analysis shows that similar to constant property cases, the near-wall region in variable property flows is also mostly populated by quasi-streamwise vortices that are slightly inclined away from the the wall and tilted towards spanwise direction. Similar to turbulence statistics, the near-wall turbulent structures are also strongly governed by $Re_\tau^\star$ profile and their dependence on individual density and viscosity profile is negligible. Cases with $d {Re_\tau^\star}/dy > 0$ show an inflection point in $\overline{u}^{\mathrm{vD}}$, causing a higher strain with respect to the wall. 
This increase in strain increases the mean forcing in spanwise direction, which results in an increase in tilting of quasi-streamwise vortices. The increased tilting of the structures increases the asymmetry of the streaks, which are known to play an important role in maintaining the near-wall cycle and generation of strong shear layers \citep{johansson1991evolution}. This increased turbulence activity also causes an increased lifting of the structures and explains why cases with $d {Re_\tau^\star}/dy > 0$ show an increased momentum transfer in spite of lower turbulent kinetic energy. The increase in tilting of the structures also provides a physical interpretation for the increase in negative pressure-strain, which enables transfer of streamwise fluctuation energy towards spanwise and wall-normal components \citep{jeong1997coherent}. The opposite is true for cases with $d {Re_\tau^\star}/dy < 0$. 

\vspace*{10mm}

The authors would like to acknowledge the access to large scale computing facilities from the Netherlands Organisation for Scientific Research (NWO) through the grant with the dossier number SSH-223-13. The authors
also wish to thank Dr. S. Pirozzoli for providing additional information for his DNS data of supersonic turbulent boundary layers.

\bibliographystyle{jfm}
\bibliography{PatelEtAl2016}
\end{document}